\documentclass[%
 reprint,
superscriptaddress,
preprintnumbers,
nofootinbib,
 amsmath,amssymb,
 aps,
]{revtex4-2}

\usepackage[colorlinks]{hyperref}
\usepackage{graphicx}
\usepackage{dcolumn}
\usepackage{bm}
\usepackage{color}
\usepackage{placeins}
\usepackage{slashed}

\definecolor{red}{rgb}{0.9, 0,0}
\definecolor{cerulean}{rgb}{0., 0.62,0.9}
\definecolor{navy}{rgb}{0.05, 0.05,0.8}

\renewcommand{\eqref}[1]{Eq.~\ref{#1}}

\begin{document}

\title{ Probing naturally light singlets with a displaced vertex trigger}

\author{Yuri Gershtein}
\affiliation{Department of Physics and Astronomy, Rutgers University, Piscataway, NJ 08854, U.S.A.}
\author{Simon Knapen}
\affiliation{CERN, Theoretical Physics Department, Geneva, Switzerland.}
\author{Diego Redigolo}
\affiliation{CERN, Theoretical Physics Department, Geneva, Switzerland.}
\affiliation{INFN, Sezione di Firenze and Department of Physics and Astronomy, University of Florence,
Via G. Sansone 1, 50019 Sesto Fiorentino, Italy}

\preprint{CERN-TH-2020-207}

\date{\today}

\begin{abstract}
We investigate the physics case for a dedicated trigger on a low mass, hadronic displaced vertex at the high luminosity LHC, relying on the CMS phase II track trigger. We  estimate the trigger efficiency with a simplified simulation of the CMS track trigger and show that the L1 trigger rate from fake vertices, $B$ meson decays and secondary interactions with the detector material can likely be brought down to the kHz level with a minimal set of cuts. While it would with any doubt be a severe experimental challenge to implement, we conclude that a displaced vertex trigger could open qualitatively new parameter space for exotic Higgs decays, exotic B decays and even direct production of light resonances. We parametrize the physics potential in terms of a singlet scalar mixing with the Standard Model Higgs and an axion-like particle with a coupling to gluons, and review a number or relevant models motivated by the hierarchy and strong CP problems, dark matter and baryogenesis.
\end{abstract}

\maketitle

\section{Introduction}

With its high luminosity upgrade, the LHC will be capable of delivering up to 7 times its current luminosity to ATLAS and CMS, which implies a corresponding increase of the number pile-up vertices per bunch crossing. Both ATLAS \cite{Collaboration:2285585} and CMS \cite{collaboration:2714892} will undergo substantial upgrades to enable them to handle this increasingly challenging environment. One of the most important innovations is the introduction of tracking information into the level-1 (L1) trigger decision making. The CMS collaboration will accomplish this by making use of double layered sensors in the outer tracker, which detect track ``stubs'' rather than individual hits. By correlating both sensors, the tracker can then assign an approximate $p_T$ to each stub, such that the computationally expensive track fitting can be restricted to stubs satisfying $p_T>2$ GeV. This innovation makes track reconstruction feasible at the L1 trigger, though necessarily with a somewhat worse resolution as compared to off-line track reconstruction.

Interestingly, this method also enables the CMS track trigger to reconstruct \emph{displaced} tracks with impact parameters up to about 10 cm \cite{Gershtein:2017tsv,collaboration:2714892}, which can be used to search for displaced jets \cite{Gershtein:2017tsv,CMS-PAS-FTR-18-018,Bhattacherjee:2020nno} and low mass, displaced dimuon resonances \cite{Gershtein:2019dhy}. Both signatures are supported by very strong theory motivation, as low mass, long-lived particles (LLP) are rather generic ingredients for hidden sector models. Specifically, so far the displaced jet topology has been studied for exotic Higgs decays \cite{Gershtein:2017tsv} and axion-like particles (ALPs)~\cite{Hook:2019qoh}. A displaced dimuon vertex trigger on the other hand can be a very powerful probe of low mass extended Higgs sectors, by leveraging exotic $B$-meson decays \cite{Gershtein:2019dhy,Evans:2020aqs}.

In this work, we combine both ideas and investigate the physics potential of a hypothetical trigger on a displaced, multitrack hadronic vertex. Of the L1 track trigger applications listed above, this would clearly be the most difficult to implement, given the  combinatorical challenge when attempting to reconstruct such a vertex at the trigger level. It is therefore paramount to establish \emph{(i)} a strong physics case for this type of trigger and \emph{(ii)} that manageable background rates can be achieved, assuming the vertex reconstruction challenge can be met. Our goal with this paper is to address both questions and to motivate further experimental studies in this direction.

\begin{figure*}[t]\centering
\includegraphics[width=0.475\textwidth]{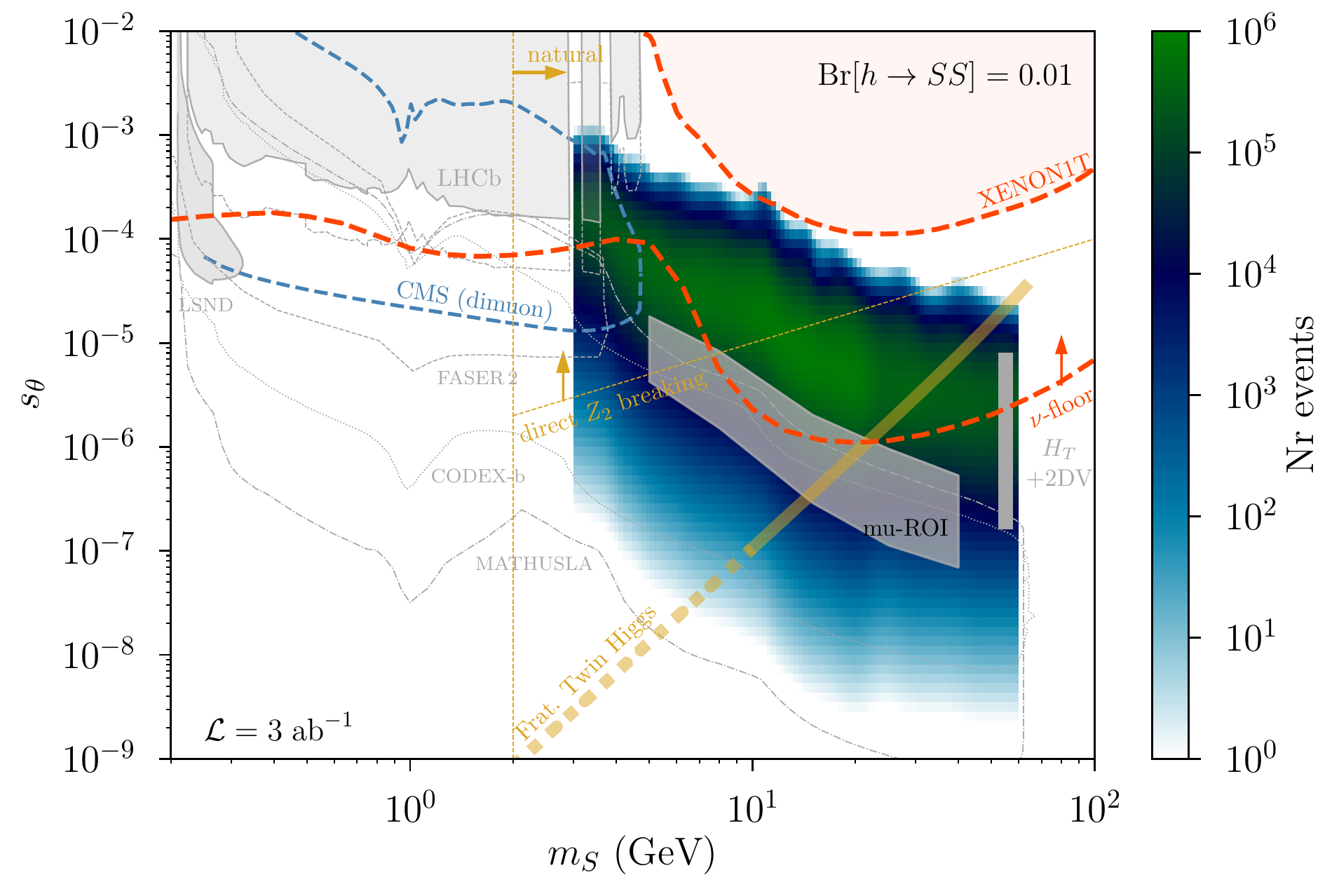}\hfill
\includegraphics[width=0.475\textwidth]{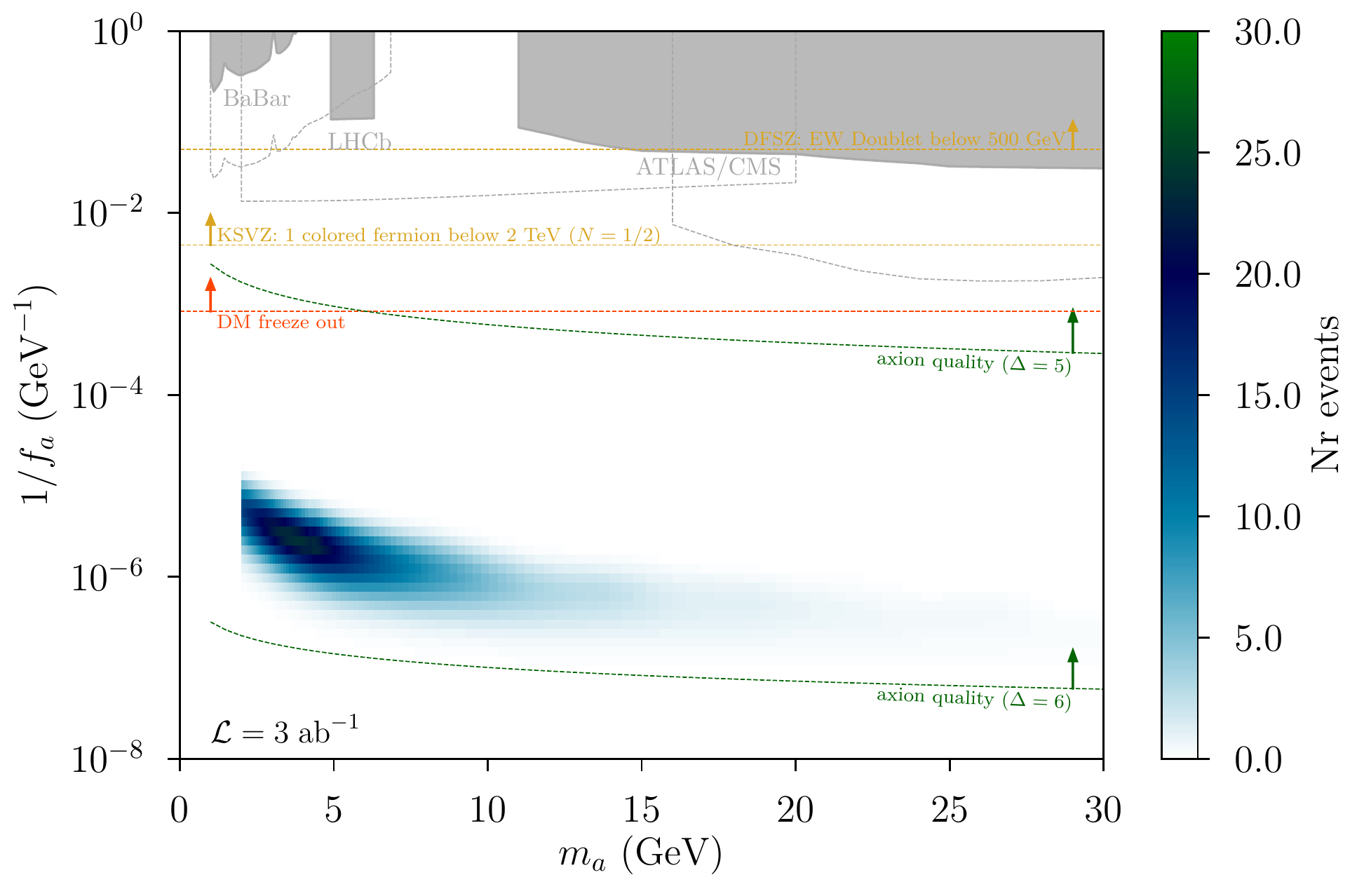}
\caption{\textbf{Left:} Projected number of events collected by a CMS L1 displaced vertex trigger at HL-LHC for the scalar model, fixing $\mathrm{Br}[h\to SS]=1\%$. {\bf shaded grey} regions represent existing limits from LSND \cite{Foroughi-Abari:2020gju}, LHCb \cite{Aaij:2016qsm}, the ATLAS  \cite{Aaboud:2018aqj}  (``mu-ROI'') and CMS (``$H_T$+2DV'') \cite{Sirunyan:2020cao}. The {\bf dashed grey} curves are projections for Belle II \cite{Kachanovich:2020yhi}, FASER 2 \cite{Feng:2017vli}, MATHUSLA \cite{Alpigiani:2020tva} and CODEX-b \cite{Aielli:2019ivi}. The {\bf blue dashed} curve is the projected reach of the CMS track trigger in the dimuon channel \cite{Evans:2020aqs}. The {\bf yellow} lines represent theory priors from naturalness (\eqref{eq:lowerbound} in Sec.~\ref{sec:singlet}), perturbativity (\eqref{eq:lowermix} in Sec.~\ref{sec:singlet}) and the Fraternal Twin Higgs scenario (\eqref{eq:twinglue} in Sec.~\ref{sec:hierarchy}). The two {\bf red} lines represent the XENON1T limit \cite{Aprile:2018dbl} and the neutrino floor in an example dark matter model \cite{Evans:2017kti}, where we fixed $m_\chi/m_S=3$ (Sec.~\ref{sec:darkmatter}).
\textbf{Right:} Projected number of events collected by a CMS L1 displaced vertex trigger at HL-LHC for the ALP model. {\bf shaded grey} regions represent existing limits, assuming $E/N=8/3$ (see Sec.~\ref{sec:ALP}), from diphoton searches at ATLAS and CMS~\cite{Aad:2014ioa}, the diphoton cross section measurements at ATLAS and CMS~\cite{Mariotti:2017vtv} and at LHCb ~\cite{CidVidal:2018blh}, the boosted dijets search at CMS~\cite{Sirunyan:2017nvi} and the $\Upsilon$ decays at Babar~\cite{Lees:2011wb}. The {\bf dashed grey} curves are projections for Belle II~\cite{CidVidal:2018blh}, and HL-LHC~\cite{Mariotti:2017vtv}. The regions above the {\bf dashed yellow} lines have colored fermions below 2 TeV in the KSVZ UV completion (\eqref{eq:KSVZ} in Sec.~\ref{sec:ALP}) or heavy Higgses below 500 GeV in the DFSZ UV completion (\eqref{eq:DFSZ} in Sec.~\ref{sec:ALP}). The region above the {\bf dashed dark green} lines shows where the ALP can solve the strong CP problem even if the PQ symmetry is broken by operators with dimension $\Delta=5,6$ (\eqref{eq:qualityPQ} in Sec.~\ref{sec:ALP}). Above the {\bf dashed dark red} line standard Dark Matter freeze-out can be realized in a perturbative model (Sec.~\ref{sec:darkmatter}). \label{fig:higgsmoney}}
\end{figure*}

To parametrize the potential of a displaced vertex trigger, we study two simple  benchmark models which encode the salient features of a vast class of beyond the Standard Model (SM) scenarios. In the first model we consider the most minimal extension of the SM Higgs sector by adding a single, real scalar field whose most general lagrangian reads
\begin{equation}
\mathcal{L}_{\text{S}}\supset-\frac{1}{2}\tilde{m}_S^2 S^2-\mu S H^\dagger H -\frac{1}{2}\lambda_{SH} S^2 H^\dagger H-V_{\text{int}}(S)\, ,\\ \label{eq:benchmark}
\end{equation}
where $V_{\text{int}}(S)$ contains the singlet self-interactions.  As we detail in Sec.~\ref{sec:benchmarks}, the spontaneous and/or explicit breaking of the approximate $\mathbb{Z}_2$-symmetry $S\to -S$ induces a mixing between the $S$ and the SM Higgs, parametrized by the mixing angle $\theta$. The latter sets the lifetime of the singlet, which has a dominant branching ratio into hadrons in most of the parameter space. The $\mathbb{Z}_2$-preserving quartic ($\lambda_{SH}$) controls the exotic Higgs decay $h\to SS$. Given that the SM Higgs is an exceptionally narrow particle, the branching ratio for this process can be large even for very small values of $\lambda_{SH}$.  

As a second benchmark, we study an axion-like particle (ALP) which couples primarily to gauge bosons
\begin{equation}
\mathcal{L}_{a}\supset -\frac{1}{2}m_a^2 a^2- \frac{\alpha_s}{8\pi}\frac{a}{f_a} \tilde{G}G+ \frac{E}{N}\frac{\alpha_{\text{em}}}{8\pi}\frac{a}{f_a} \tilde{F}F\, ,\label{eq:ALPL}
\end{equation}
where coupling to massive gauge bosons are mostly irrelevant in the ALP mass range of interest. $N$ and $E$ are the anomaly coefficients for the gluon and photon couplings respectively. To simplify the notation in the crucial second term in \eqref{eq:ALPL}, we have absorbed $N$ in the definition of the $f_a$. The rational number $E/N$ therefore sets the model dependent strength of the coupling to photon pairs relative to that to the gluons. The gluon coupling dominates the ALP width as long as $E/N\lesssim 8(\alpha_s/\alpha_{\text{em}})^2\simeq 4\times 10^{3}$, which is generically satisfied in models that are compatible with grand unification. Moreover, for low $\sqrt{\hat s}$, the gluon-gluon luminosity at the LHC is so enormous that there is still an appreciable $pp\to a j$ cross section even for extremely high values of the decay constant $f_a$. In part of the parameter space, this means that $a$ can decay through displaced hadronic vertex, despite there being no other parametric suppression for the width of $a$.

The event yield for our proposed trigger is shown in Fig.~\ref{fig:higgsmoney} for both models, where we  have fixed the quartic $\lambda_{SH}$ by choosing the branching ratio of $h\to SS$ to be 1\%. This benchmark value for the exotic Higgs branching ratio will be difficult to exclude just with precision measurements of the Higgs couplings at the HL LHC~\cite{Bechtle:2014ewa,Belanger:2013xza,Frugiuele:2018coc}. We find that up to $10^6$ events could be recorded, and the trigger could cover between 3 and 4 orders of magnitude in the mixing angle $s_\theta$. Hunting directly for exotic Higgs decays in this manner provides then an experimental avenue that is very complementary with the Higgs precision program.

The event yield for the ALP model on the other hand is probably marginal, due to the still rather short lifetime of the ALP. The model is nevertheless still a useful proxy for less minimal models, in which the lifetime and production cross section of the LLP are not controlled by a single parameter. Indeed, both models are to be viewed as straw man models for more complete scenarios which address the naturalness of the electroweak scale, the strong CP problem, the origin of the Dark Matter or the present baryonic asymmetry. In Sec.~\ref{sec:theory} we explain how the phenomenology of these more complete frameworks can be mapped onto our simple benchmark models, hereby establishing the strong theory motivation for this experimental effort. In Secs.~\ref{sec:simulation} and \ref{sec:backgrounds} we discuss respectively our simulation framework and background calculations, arguing that a rate well below 10 kHz can likely be achieved with a minimal set of cuts. We close with additional results and an outlook in Sec.~\ref{sec:results}.

\section{Benchmark models}\label{sec:benchmarks}

In this section we provide more details on both benchmark models, with a special focus on their most natural parameter space and their various production modes.

\subsection{A light real singlet\label{sec:singlet}} The first model is the most minimal extension of the SM, with only one new, real degree of freedom $S$, one mass parameter ($m_S$) and two coupling constants ($\mu,\lambda_{HS}$). The lagrangian is given in \eqref{eq:benchmark} and the resulting mass squared matrix of the Higgs-singlet system is 
\begin{equation}
\mathcal{M}=\begin{pmatrix}
m_h^2& v(\mu+\lambda_{SH} s_0) \\
v(\mu+\lambda_{SH} s_0) & \tilde{m}_S^2+\frac{1}{2}\lambda_{SH} v^2+V''_{\text{int}}(s_0)
\end{pmatrix}\ ,\label{eq:Smassmatrix}
\end{equation} 
where in the limit of small mixing,  $\mathcal{M}_{11}\simeq m_h^2=2\lambda_H v^2$ and $\mathcal{M}_{22}\simeq m_S^2$ are the mass eigenstates corresponding to the SM Higgs and the singlet $S$. $s_0$ is the singlet vacuum expectation value (VEV) and $v=246\text{ GeV}$ is the VEV of the SM Higgs.  

Here we are interested in the light singlet regime where $m_S<m_h/2$, which means that the mixed quartic coupling will induce the exotic Higgs decay $h\to SS$ with branching ratio
\begin{equation}
\mathrm{Br}[h\to SS]\approx\frac{\Gamma_{h\to SS}}{\Gamma_{h\to b\bar b}}\approx\frac{\lambda_{SH}^2}{6y_b^2 \lambda_H}\, ,
\end{equation}
where we expanded for $m_S\ll m_h$. $y_b\simeq 2\times 10^{-2}$ and $\lambda_H\simeq 0.13$ are respectively the bottom Yukawa and SM Higgs quartic couplings. 
The extremely narrow width of the SM Higgs implies that $\mathrm{Br}[h\to SS]=0.01$, as assumed in Fig.~\ref{fig:higgsmoney}, corresponds to a small quartic
\begin{equation}
\lambda_{SH}\approx1.7\times10^{-3}.
\end{equation}
The value of this quartic has two important theoretical consequences that we illustrate in turn. 

First, the quartic back-reacts on the SM Higgs potential and sets the cutoff $\Lambda$ of a possible, natural UV completion for the lagrangian in \eqref{eq:benchmark}. The most divergent contribution is the infamous mass correction to the SM Higgs mass
\begin{align}\label{eq:naturalness}
\delta m_H^2 \sim \frac{\lambda_{SH}}{16\pi^2}\Lambda^2 \ .
\end{align} 
Setting $\delta m_H^2\sim m_H^2$ implies $\Lambda\lesssim 38$ TeV which is easily outside the reach of the LHC. In other words, the hierarchy problem introduced by coupling $S$ to the Higgs is much less severe the hierarchy problem corresponding to the SM top quark. A similar cut-off dependence arises in the analogous correction to $\tilde m_S$, but it can easily be UV completed without directly observable consequences.

Second, a non-zero $\lambda_{SH}$ induces an irreducible, tree-level contribution to the singlet mass in \eqref{eq:Smassmatrix}. This means that fine tuning in the lower right block of \eqref{eq:Smassmatrix} is needed if $\frac{1}{2}\lambda_{SH} v^2\gg m_S^2$. Numerically, we can write $m_S$ as
\begin{equation}
m_S\approx 2\text{ GeV}\times\left(\frac{\mathrm{Br}[h\to SS]}{0.01}\right)^{1/4}\times\left(\frac{\Delta}{0.1}\right)^{1/2}\ ,\label{eq:lowerbound}
\end{equation}
where $\Delta\equiv m_S^2/ (\frac{1}{2}\lambda_{SH} v^2 ) $ is a measure of the degree of fine tuning in $\mathcal{M}_{22}$. In other words, for $m_S\lesssim 2$ GeV, more than 10\% fine tuning is needed given our benchmark value for $\lambda_{SH}$. This lower bound leaves most of the relevant parameter space in Fig.~\ref{fig:higgsmoney} open. 
 \begin{figure*}[t]\centering
\includegraphics[width=0.48\textwidth]{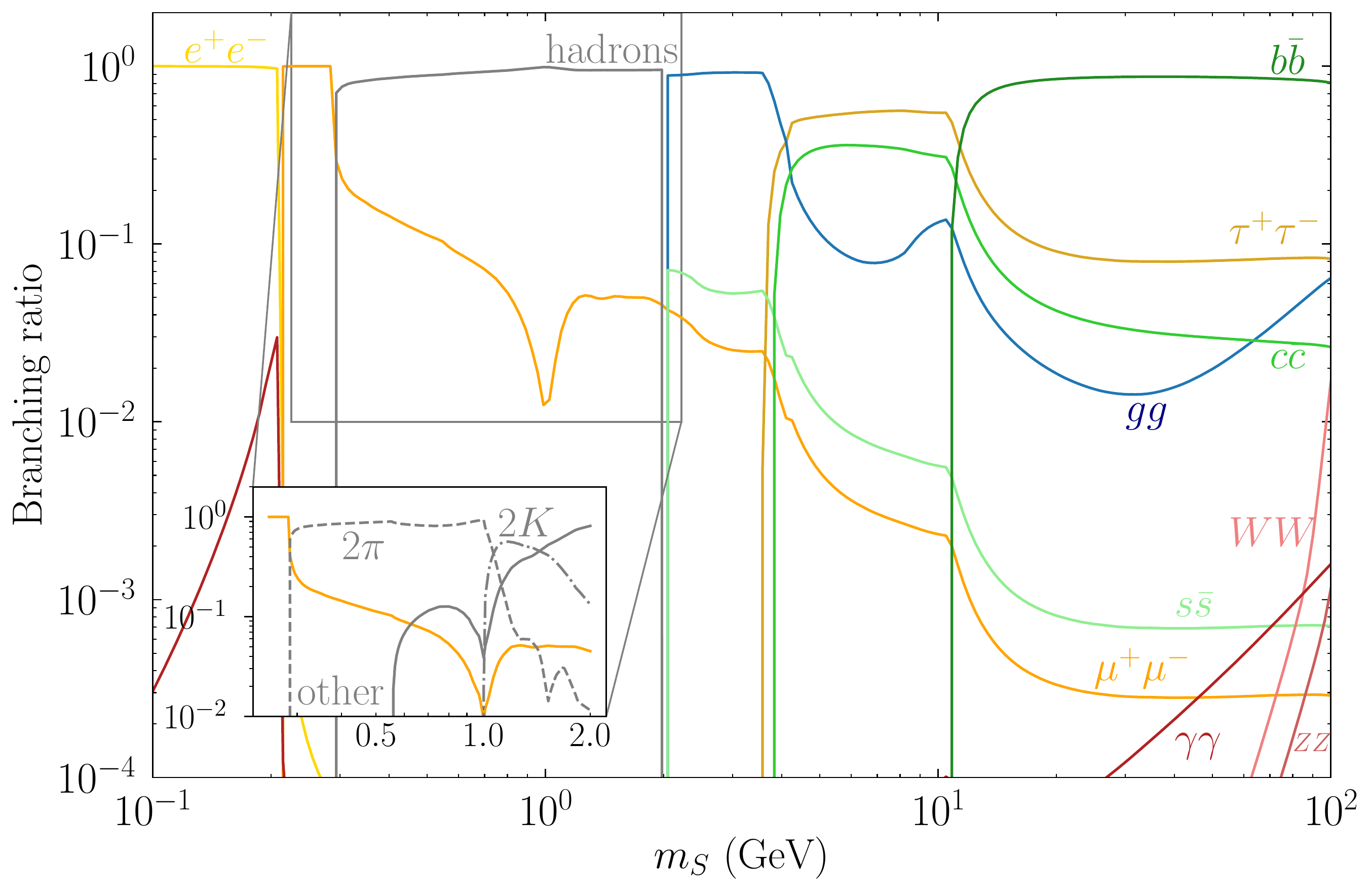}\hfill
\includegraphics[width=0.458\textwidth]{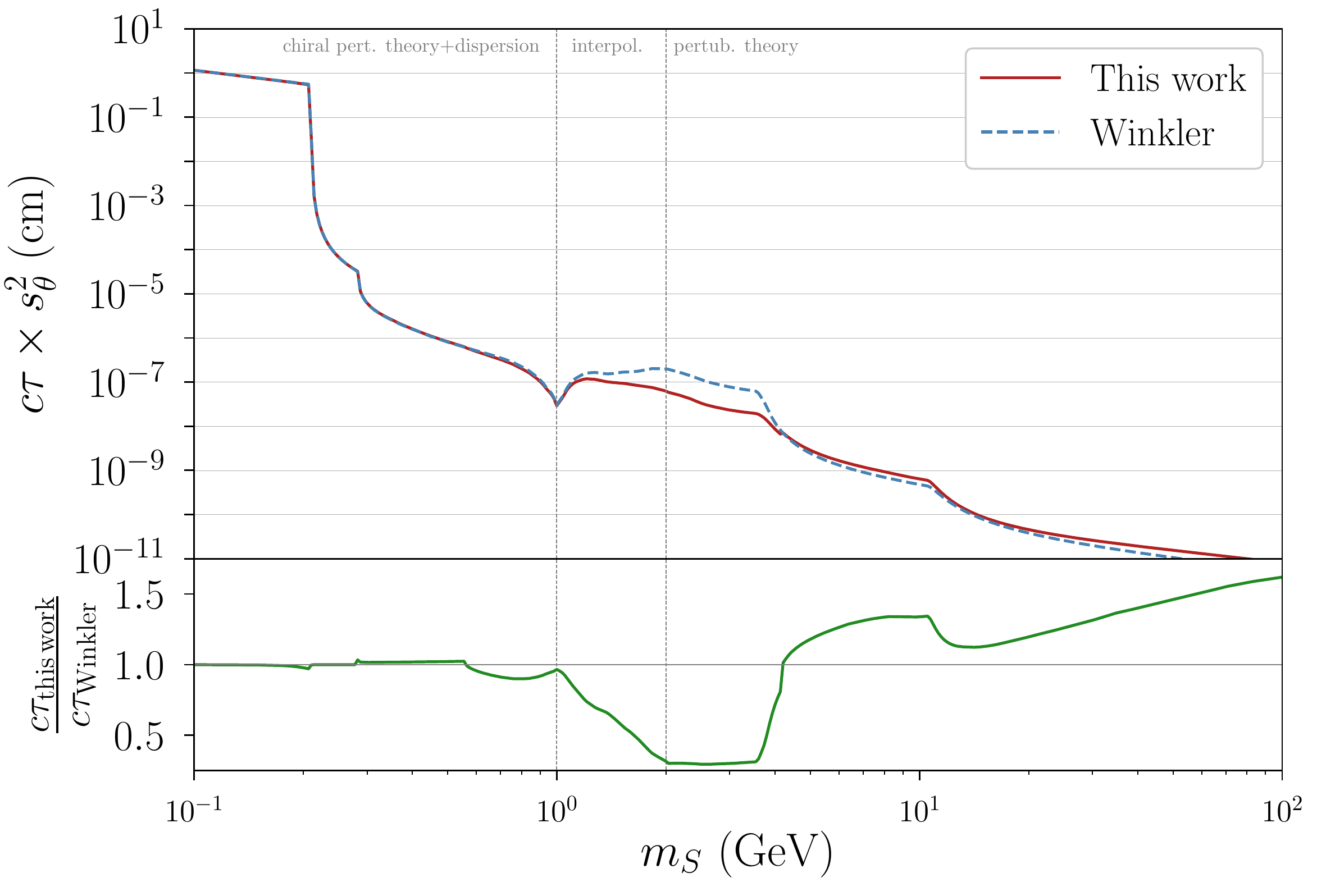}
\caption{{\bf Left:} Estimated branching ratios for the scalar benchmark, derived from perturbative calculations in \cite{Spira:1997dg} and dispersive calculations in \cite{Winkler:2018qyg}. {\bf Right:} Lifetime for the scalar benchmark, in comparison with the result in \cite{Winkler:2018qyg}.\label{fig:higgs}}
\end{figure*}
 
The remaining couplings of $S$ to the SM are induced by the off-diagonal terms in \eqref{eq:Smassmatrix}, which mix the singlet with the SM Higgs with a small angle $\theta$. The dependence of the mixing angle on the underlying theory parameters depends on whether the approximate $S\leftrightarrow -S$ parity is primarily broken by the $\mu$ parameter in \eqref{eq:benchmark}, or by the dynamics of $V_{\text{int}}(S)$. We will refer to these two cases as \emph{induced $\mathbb{Z}_2$ breaking} and \emph{direct $\mathbb{Z}_2$ breaking} respectively.  Using the common shorthand notation $s_\theta=\sin\theta$, we can write
\begin{equation}
s_\theta\simeq \frac{|\mathcal{M}_{12}|}{m_h^2}=\begin{cases}  \frac{\mu v}{m_h^2}\left|1-\frac{\lambda_{SH} v^2}{2m_S^2}\right|\ ,\quad s_0= -\frac{\mu v^2}{2m_S^2}\\
 \frac{\lambda_{SH}}{\sqrt{2\lambda_S}} \frac{v m_S}{m_h^2}\ ,\quad\quad\quad\ \ s_0= \frac{m_S}{\sqrt{2\lambda_S}}\ 
\end{cases},\label{eq:mixing}
\end{equation}
where the first and second line refer respectively to the induced and direct $\mathbb{Z}_2$ breaking. The second equation on the second line serves as the definition of the effective singlet quartic $\lambda_S$. By taking $\mu\ll v$, the induced $\mathbb{Z}_2$ breaking scenario allows for arbitrarily small values of the mixing angle $s_\theta$, without fine tuning. This is simply a manifestation of the fact that $\mu$ is a technically natural parameter. In the direct $\mathbb{Z}_2$ breaking scenario on the other hand, the singlet quartic $\lambda_S$ is bounded from above by perturbativity, which implies a lower bound\footnote{Two upper bounds on $s_\theta$ can also be derived from \emph{(i)} insisting that $\mathcal{M}$ in \eqref{eq:Smassmatrix} is a positive definite matrix and \emph{(ii)} from the naturalness of radiative correction to $\lambda_S$, induced by $\lambda_{SH}$. While neither is relevant for the parameter space considered here, they are important for very low mass scalars~\cite{Banerjee:2020kww}. } on $s_\theta$
\begin{equation}
 s_\theta\gtrsim 10^{-6}\times \left(\frac{m_S}{1\text{ GeV}}\right)\times \left(\frac{4\pi}{\sqrt{\lambda_S}}\right),\label{eq:lowermix}
\end{equation}
as indicated by the diagonal yellow line on Fig.~\ref{fig:higgsmoney}.

Assuming no other dark sector decay channels are open, the mixing angle $s_\theta$ fixes the singlet width as 
\begin{equation}
\Gamma_S(m_S)=s_\theta^2\times \Gamma_h(m_S)\ , \label{eq:singletwidth}
\end{equation}
with $\Gamma_h(m_S)$ the width of the SM Higgs evaluated at mass $m_S$. Especially for $m_S\sim$ GeV, the form of $\Gamma_h(m_S)$ is very complicated and subject to large theory uncertainties. For $m_S\lesssim 1.0$ GeV, the width can be estimated using chiral pertubation theory and dispersion methods, as most recently updated by Winkler \cite{Winkler:2018qyg}. For \mbox{$m_S\gtrsim$ 2 GeV}, perturbative calculations can be used, though still subject to sizable uncertainties. In the intermediate regime we must currently rely on an interpolation. In Fig.~\ref{fig:higgs} we provide a minor update to the analysis in \cite{Winkler:2018qyg} by including the NLO QCD correction to the width and the running of quark masses, as calculated in \cite{Spira:1997dg}. This has an $\mathcal{O}(1)$ effect on the upper boundary condition of the interpolation region, and thus also has some impact on the regime where interpolation is needed.

Direct singlet production in association with a jet does not lead to interesting displaced jet signatures: The signal strength is only large enough for $s_\theta\gtrsim10^{-4}$ and in most of this regime the singlet decays more or less promptly. We therefore find no substantial viable region associated with the $pp\to S j$ process, in sharp contrast to the ALP case. Concretely, from the left-hand panel of Fig.~\ref{fig:higgs} we see that the partial with to gluons is subdominant for $m_S\gtrsim 3$ GeV, which means that for a fixed gluon coupling the scalar will decay more promptly than the ALP. This results in the small but non-zero event yield shown in the right-hand panel of Fig.~\ref{fig:higgsmoney}.
We will return to this argument in Sec.~\ref{sec:ALP}.

Below the B meson threshold, the light singlet $S$ can be probed through exotic $B$ decays. One possible channel is $B\to S X_s$, generated through an electroweak penguin \cite{PhysRevD.26.3287,Chivukula:1988lo,Grinstein:1988yu} and with signal strength controlled by $s_\theta$. A second channel is  $B\to SS X_s$, also generated through an electroweak penguin \cite{Bird:2004ts} but with signal strength controlled by $\lambda_{SH}$. The branching ratios of the two processes are
\begin{align}
&\text{BR}(B\to S X_s)\simeq 3\times 10^{-8}\left(\frac{s_\theta}{10^{-4}}\right)^2\ ,\\
&\text{BR}(B\to S S X_s)\simeq 3\times 10^{-11} \left(\frac{\lambda_{\text{SH}}}{10^{-3}}\right)^2\ ,
\end{align}
where we neglected $\mathcal{O}(1)$ factors distinguishing between exclusive and inclusive channels~\cite{Boiarska:2019jym}. For the values of the quartic considered here, the dominant channel is $B\to S X_s$ as long as $s_\theta\gtrsim3\times 10^{-6}$. As shown in Fig.~\ref{fig:higgsmoney},  the exclusive searches for $B^+\to K^+\mu^+\mu^-$ \cite{Aaij:2016qsm} and $B^0\to K^*\mu^+\mu^-$ \cite{Aaij:2015tna} at LHCb already exclude the region where $s_\theta<10^{-3}$ and further improvements are expected from the full LHCb dataset and from Belle2~\cite{Kachanovich:2020yhi}. The implementation of a displaced dimuon vertex trigger at CMS would moreover allow to probe the region further down to $s_\theta\simeq10^{-4}$ \cite{Gershtein:2019dhy,Evans:2020aqs}. Probing lower mixing angles would require dedicated experiments such as FASER 2 \cite{Feng:2017vli}, MATHUSLA \cite{Alpigiani:2020tva}, CODEX-b \cite{Gligorov:2017nwh} or SHiP \cite{Alekhin:2015byh}. 

There also are a number of searches which can probe this model above the $B$ meson threshold, for our specific choice $\lambda_{SH}$. (We refer to \cite{Fuchs:2020cmm} for a more complete study of the full parameter space.) Concretely, there are two searches for displaced objects which do already have (limited) sensitivity: ATLAS performed a powerful search for anomalous activity in the muon chambers, using a dedicated trigger (``mu-ROI'' in Fig.~\ref{fig:higgsmoney}) \cite{Aaboud:2018aqj}. In addition, the CMS search relying on an $H_T$ cut plus two displaced vertices in the tracker (``$H_T$+2DV'') \cite{Sirunyan:2020cao} has some sensitivity for the higher end of the $m_S$ range. For phase II, it has been argued that the timing detector and high granularity calorimeters will add substantial additional sensitivity \cite{Liu:2018wte,Liu:2020vur}. A dedicated displaced vertex trigger would be complementary to all these strategies, as it can probe shorter lifetimes than the muon chambers and would have higher signal yield than triggers relying on MET, $H_T$ or a VBF tag.

Aside from the aforementioned displaced searches, one may wonder whether various prompt searches already have sensitivity in the high $s_\theta$ region. In this regime, low mass dimuon resonances \cite{CMS-PAS-EXO-19-018} can be sensitive to single $S$ production through gluon fusion. Similarly, there exist a number of relevant searches for prompt exotic higgs decays, most notably in the $h\to b\bar b \mu\mu$ channel \cite{Aaboud:2018esj}. In all such cases one however must either pay the branching ratio of $S\to \mu\mu$ (see Fig.~\ref{fig:higgs}) at least once in the signal rate or content with very large hadronic backgrounds. As a result, we find that these searches are not yet constraining the parameter space in Fig.~\ref{fig:higgsmoney}.

Before concluding this section it is important to note that the formulas derived in this section assume that the singlet couplings to the SM are exactly those of a lighter SM Higgs boson universally rescaled by the mixing angle $s_\theta$. Large deviations from this assumption occur in models where the coupling of the singlet to the SM is driven purely by higher dimensional operators. A prominent example would be the one of a light dilaton~\cite{Bellazzini:2013fga,Coradeschi:2013gda} where the coupling to fermions can be suppressed compared to the one into gluons by carefully engineering how the SM yukawas are generated by the conformal sector dynamics~\cite{Bellazzini:2012vz,Chacko:2012sy}.

\subsection{A light ALP}\label{sec:ALP} The second benchmark model, in \eqref{eq:ALPL}, describes the SM interactions of a pseudo-Nambu-Goldstone boson (pNGB) of a spontaneously broken Peccei-Quinn (PQ) symmetry which has a non-zero mixed anomaly with the QCD gauge group~\cite{Peccei:1977ur,Peccei:1977hh}. The spontaneous breaking of such a symmetry is a necessary requirement in axion solutions of the SM strong CP problem, and predicts an ALP with anomalous coupling to gluons. The anomalous ALP coupling to photons is controlled by the mixed anomaly with the $U(1)_{\text{em}}$ and will also be generically non-zero. As a consequence of the approximate shift symmetry acting on the ALP, the hierarchy $m_a\ll f_a$ is technically natural. We briefly discuss two different UV completions of the ALP lagrangian in \eqref{eq:ALPL}, which differ by the type of states we expect to be present in the UV. 

In the class of UV completions put forward by Kim, Shifman, Vainshtein and Zakharov (KSVZ)~\cite{Kim:1979if,Shifman:1979if}, the ALP is embedded in a complex scalar singlet $\Phi$ which couples to heavy, colored fermions. The latter are charged under the PQ symmetry such that the model is defined by
\begin{equation}
\mathcal{L}_{\text{KSVZ}}\supset g_*\Phi \tilde{\psi}\psi\quad , \quad \Phi=\frac{v_a+\varphi}{\sqrt{2}}e^{ia/v_a}\ ,\label{eq:KSVZ}
\end{equation}
where in this setup the singlet VEV ($v_a$) is related to the axion decay constant, defined in \eqref{eq:ALPL}, by $v_a=2 Nf_a $. The anomaly coefficients $N$ and $E$ can be related to the multiplicity and the representation of the fermions: $N=q_{\text{PQ}}\sum_{\psi} C_3(R_\psi)$ and $E=q_{\text{PQ}}\sum_{\psi} Q^2(R_\psi)$,  where $C_3$ is the color index of the fermion representation $\text{Tr} (R^a_\psi R^b_\psi)=C_3(R_\psi)\delta^{ab}$ and $Q(R_\psi)$ its electromagnetic charge. We can set the PQ charge $q_{\text{PQ}}=1$ without loss of generality. As a well-motivated example, we take $N_{\psi}$ flavors of heavy fermions in the $5+\bar{5}$ representation of the $SU(5)$ GUT group, which implies $N=N_\psi/2$ and $E=4/3 N_\psi$. The coupling $g_*$ sets the mass of the fermions $m_\psi=\sqrt{2}N g_*f_a $. 

At fixed $f_a$, a larger multiplicity $N_\psi$ or a strong coupling $g_*$ will make the colored fermions heavier, leaving the light ALP as the main phenomenological target of this setup. In the right-hand panel of Fig.~\ref{fig:higgsmoney} we indicate where the colored fermions would lie below 2 TeV for $g_\ast$ fixed at its unitary bound $g_*=4\pi/\sqrt{N_\psi}$, assuming $N_\psi=1$. As we see, ALP searches into diphotons will be able to probe a portion parameter space where the colored fermions are out of reach of the LHC.  

In models following the Dine-Fischler-Srednicki-Zhitnitsky (DFSZ) setup~\cite{Zhitnitsky:1980tq,Dine:1981rt} the SM Higgs sector is extended to a two Higgs doublet model ($H_u,H_d$), plus a complex scalar singlet ($\Phi$). The singlet couples to the two electroweak Higgs doublets through
\begin{equation}
\mathcal{L}_{\text{DFSZ}}\supset \lambda_* H_u H_d \Phi^2\ , \label{eq:DFSZ}
\end{equation}
fixing again the PQ charge of the singlet to be $q_{\text{PQ}}=1$ the PQ charges of the two Higgses are fixed in this model: $q_{H_u}=-2\cos^2\beta$, $q_{H_d}=-2\sin^2\beta$, where $\tan\beta\equiv v_u/v_d$ with $v_{u,d}$ the VEVs of $H_{u,d}$. The DFSZ ALP couples to the SM fermions through their couplings to the Higgs doublets. Here we take a type II two Higgs doublet as an example. The  SM fermions that are heavier than the ALP can be integrated out and contribute to the gluon and the photon anomaly, while the couplings to the light quarks and leptons are a new phenomenological feature of this setup. For $m_b<m_a<m_t$ we have $N=\cos\beta^2$ and $E=8/3\cos\beta^2$ from the top contribution. The axion decay constant $f_a$ is directly related to the mass of a heavy, CP-even Higgs bosons in this setup through $m_{H}\sim 2\sqrt{\lambda_*} f_a$. The LHC bounds on these extra Higgs doublets are mild relative to the bounds on new, colored fermions, which means that the DFSZ ALPs can generally have a lower decay constant than KSVZ ALPs. In Fig.~\ref{fig:higgsmoney} we indicate the regime where the CP Higgs bosons would roughly lay below 500 GeV, again saturating the unitarity bound $\lambda_*=16\pi^2$.

The ALP width into two gluons and two photons is given by
\begin{equation}
\Gamma_{gg}=K_{gg} \frac{\alpha_s^2}{32\pi^3}\frac{m_a^3}{f_a^2}\quad , \quad \Gamma_{\gamma\gamma}= \frac{E^2}{N^2}\frac{\alpha_{\text{em}}^2}{256\pi^3}\frac{m_a^3}{f_a^2}\ ,\label{eq:widthsALP}
\end{equation}
where $K_{gg}$ is the k-factor accounting for NNLO corrections to the the gluon width~\cite{Chetyrkin:1998mw}  
\begin{equation}
K_{gg}(m_a)=1 + \frac{\alpha_s}{\pi}E_A+\left(\frac{\alpha_s}{\pi}\right)^2E_A\left(\frac{3}{4}E_A+\frac{\beta_1}{\beta_0}\right)\ ,
\end{equation}
where $E_A=97/4-7N_f/6$, $\beta_0= 11/4 - N_f/6$ and $\beta_1= 51/8 - 19N_f/24$. $N_f$ is the number of quarks with mass below $m_a$. Numerically, the k-factor ranges from 3.5 at $m_a=5\text{ GeV}$ to $2.5$ for $m_a=20\text{ GeV}$.  From \eqref{eq:widthsALP} we see explicitly that the gluon width dominates the total width as long as $E/N\lesssim K_{gg}8(\alpha_s/\alpha_{\text{em}})^2\simeq 10^{4}$. The axion lifetime therefore scales as 
\begin{equation}
c\tau_a\simeq 0.2\text{ cm}\left(\frac{f_a}{10^6\text{ GeV}}\right)^2\left(\frac{10\text{ GeV}}{m_a}\right)^3\ ,\label{eq:ALPwidth}
\end{equation}
where in this simplified formula we neglected the k-factor dependence on the ALP mass. We see that for a decay constant as large as $10^6\text{ GeV}$ the ALP can produce a displaced vertex.

 \begin{figure}[t]\centering
\includegraphics[width=0.4\textwidth]{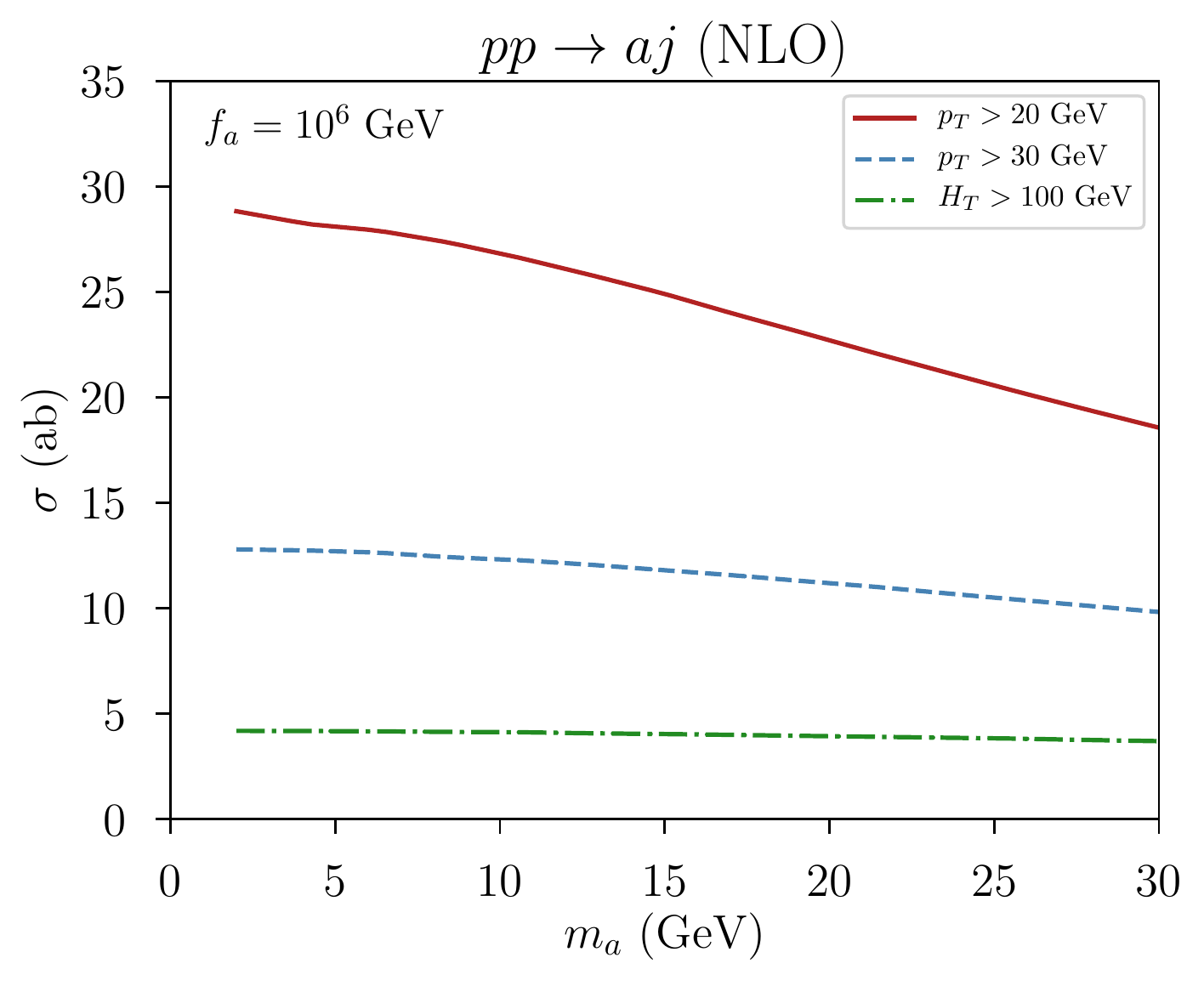}
\caption{Production cross section for an ALP in association with a jet, computed with Madgraph@NLO~\cite{Alwall:2014hca,Artoisenet:2013puc} for different fiducial cuts. The axion decay constant is fixed here to $f_a=10^6\text{ GeV}$ which is the typical size of the decay constant required for the ALP to decay displaced (see \eqref{eq:ALPwidth}). \label{fig:ALPxsec}}
\end{figure}

The signal events for the ALP model where generated at leading order with Madgraph~\cite{Alwall:2014hca} and decayed with Pythia 8 \cite{Sjostrand:2014zea}.  We normalized this sample to fixed order\footnote{We verified that matching up to two jets does not lead to relevant differences in efficiency and worked for simplicity with the fixed order cross section.} NLO cross section for $pp\to a+j$, as computed with Madgraph@NLO~\cite{Alwall:2014hca,Artoisenet:2013puc}. This is shown in Fig.~\ref{fig:ALPxsec}, where for reference we also show the cross sections for the fiducial cuts used in \cite{Hook:2019qoh}. The corresponding $k$-factor for this process is $\simeq 2$ in the full mass range under consideration.\footnote{Notice that our $k$-factor differs from the one of Ref.~\cite{Hook:2019qoh}, which was taken from the NLO corrections to $\sigma(pp\to a)$. The latter process has a very small efficiency and it is less well suited to assign a reliable $k$-factor.} Fig.~\ref{fig:ALPxsec} and \eqref{eq:ALPwidth} make it clear that at most $\mathcal{O}(100)$ long-lived ALPs could be produced at the HL-LHC, which explains the limited signal yield in Fig.~\ref{fig:higgsmoney}. 

We now comment on exotic Higgs decays into a pair of ALPs. These are induced by the dimension six operator 
\begin{equation}
\mathcal{L}_{\text{h-ALP}}=\frac{c_{\text{aH}}}{N^2}H^{\dagger}H\frac{(\partial_\mu a)^2}{f_a^2}\ ,\label{eq:mixingALP}
\end{equation}
which respects the ALP shift-symmetry. The branching ratio is 
\begin{equation}
\mathrm{Br}[h\to aa]\approx\frac{\Gamma_{h\to aa}}{\Gamma_{h\to b\bar b}}=\frac{2c_{\text{aH}}^2}{3N^4}\frac{ v^4}{f_a^4}\frac{\lambda_H}{y_b^2}\, ,
\end{equation}
which implies that $\mathrm{Br}[h\to aa]\simeq 0.01$ can be obtained for $f_a\simeq 1\text{ TeV}$ and $c_{aH}\simeq\mathcal{O}(1)$. For values of the ALP decay constant which leads displaced jet signatures (see \eqref{eq:ALPwidth}), the signal rate from Higgs decays is very suppressed unless the decay constant controlling the gluon width is taken to be very different from the one controlling the operator in \eqref{eq:mixingALP}. Achieving $c_{\text{aH}}\gg N^2$ is in principle possible since this hierarchy is protected by a symmetry, however it is fair to say that vanilla ALP scenarios would not lead to Higgs decay to long-lived ALPs.  It is worth noting that this is very different from the singlet case, because the ALP is a compact field (see \eqref{eq:KSVZ}), while the singlet is not. In practical terms, this means that a substantial model building effort is needed to enhance $c_{\text{aH}}/N^2$ in \eqref{eq:mixingALP}. In the singlet case on the other hand, we have shown in the previous section that a large hierarchy between $\lambda_{SH}$ and $s_\theta$ is completely natural.

Finally, the ALP could also be produced in $B$ decays, in particular in DFSZ style models, see e.g.~\cite{Batell:2009jf,Freytsis:2009ct}. (In KSVZ style models this process takes place either at two loops \cite{Aloni:2018vki} or through the ALP-$WW$ coupling \cite{Izaguirre:2016dfi}, and the reach is somewhat less promising.) Due to the $B$ meson's large cross section and exceptionally small width, it is possible to produce a sizable sample of long-lived ALPs at the LHC through an exotic $B$ meson decay. A displaced, hadronic vertex trigger would have the advantage that one does not need to pay the branching ratio of the ALP to muons, which tends to be very small in most models. We leave this case for a future study.

\section{Theory motivation\label{sec:theory}}
In this section we review the theory motivation for the simplified benchmark models in the previous section, and low mass, displaced vertices in general. Some of the models we review make a concrete prediction in the parameter space shown in Fig.~\ref{fig:higgsmoney}.  

\subsection{Hierarchy problem\label{sec:hierarchy}}
\paragraph*{Neutral Naturalness models:}

The non-observation of colored top partners at the LHC has revived the interest in a class of models where the top partners are charged under a dark $SU(3)$, rather than under the SM color group. The great ancestors of these models are the Twin Higgs \cite{Chacko:2005pe,Chacko:2005un}, folded supersymmetry \cite{Burdman:2006tz} and the quirky little higgs \cite{Cai:2008au}, which can all be seen as examples of a broader class of ``neutral naturalness'' models~\cite{Craig:2014roa,Craig:2014aea,Craig:2015pha,Craig:2016kue}. All these models relax the fine-tuning of the electroweak scale by introducing an extended Higgs sector together with non-trivial dynamics in a hidden sector. As a consequence, an experimental smoking gun of these scenarios are exotic higgs decays into a dark parton shower, followed by a dark hadronization process. The resulting spray of dark sector hadrons can decay to the standard model through displaced vertices. In this sense, neutral naturalness models are themselves examples of a broader class of a hidden valley models~\cite{Strassler:2006im,Han:2007ae}. (See \cite{Alimena:2019zri} for a recent review of dark shower phenomenology.) 

The phenonomenology of the fraternal Twin Higgs model \cite{Craig:2015pha} in particular was studied in great detail \cite{Craig:2015pha,Curtin:2015fna} and can be summarized as follows: The SM Higgs decays to dark sector bottom quarks ($b'$) with branching ratio $\mathrm{Br}[h\to b'b']\approx v^2/f^2$, where $f$ is the VEV of the twin Higgs.\footnote{The Twin Higgs itself could also be within the reach of the LHC, depending the value of the twin quartic~\cite{Alipour-fard:2018mre}.} The $b'$ subsequently shower and hadronize, and eventually produce a number of dark glueballs. The lightest glueball is known to be a CP-even scalar \cite{Morningstar:1999rf} and can be identified with $S$ in the simplified model in Sec.~\ref{sec:singlet}. Because most $h\to b'b'$ decays are expected to produce one or more $0^{++}$ glueballs, we  roughly identify $\mathrm{Br}[h\to b'b']\sim\mathrm{Br}[h\to SS]$. With this identification, the slice of the parameter space plotted in Fig.~\ref{fig:higgsmoney} corresponds to $v/f=1/10$, which is considered to be rather fine tuned region of the Twin Higgs construction already. In other words, the event rate in the natural parameter space of the fraternal Twin Higgs is likely higher than the rate we assumed in Fig.~\ref{fig:higgsmoney}.

 The glueball lifetime is controlled by its mixing with the SM Higgs, which is given by \cite{Craig:2015pha} 
\begin{equation}
s_\theta\simeq\frac{3.06 m_0^3 v}{24\pi^2 f^2m_h^2}\ ,\label{eq:twinglue}
\end{equation}
so that at fixed $f$, the mixing is predicted in terms of the mass of the glueball ($m_0$). This relation is shown by the yellow line in Fig.~\ref{fig:higgsmoney}.  Along the dotted line, $m_0$ is too low to be accommodated in the fraternal twin Higgs without substantial fine-tuning. It can be realized however by adding more matter to the dark sector, as in the vector-like twin Higgs~\cite{Craig:2016kue}. 

Crucially, the presence of a dark shower decouples the glueball's production and decay rates, such that its lifetime can be parametrically enhanced without suppressing its production rate. Our identification of this rather complicated scenario with the simple benchmark model in Sec.~\eqref{sec:singlet} necessarily neglects $\mathcal{O}(1)$ theory uncertainties in the event yield. However these uncertainties do not affect the qualitative result that a displaced vertex trigger would be a major asset in the search for neutral naturalness models. 

\paragraph*{Composite Higgs models:}
In composite Higgs models, the Higgs boson is a pseudo Nambu-Goldstone boson (pNGB) of a global symmetry, which is broken spontaneously by a strong sector in the UV. Depending on the size of the symmetry group in the model, additional light pNGB's may be present, which one can identify with the ALP benchmark. The ALP decay constant $f_a$ is of the same order as the compositeness scale of the model, which must be $\sim$ few TeV to preserve naturalness. As we have seen in Sec.~\ref{sec:benchmarks}, for $f_a\sim 1$ TeV, the ALP is copiously produced at the LHC, though it typically decays promptly. Searches for prompt ALPs decaying to $\gamma\gamma$ or $\tau\tau$ therefore tend to be the most promising, be it in direct production~\cite{Mariotti:2017vtv,CidVidal:2018blh,Cacciapaglia:2017iws} or in the exotic $h\to aa$ decay \cite{Bauer:2017ris}.

On the other hand, arguments have been presented in favor of pushing the compositeness scale to $\gtrsim 10$ TeV, which reduces tension with electroweak precision bounds, at the cost of moderate amount of fine tuning \cite{Barnard:2015rba}. In this case, any additional pNGBs aside the Higgs could have macroscopic lifetimes \cite{Curtin:2018mvb}.

\paragraph*{Supersymmetry:}
A generic pNGB arising in low energy SUSY-breaking is the R-axion related to the spontaneous breaking of the $U(1)$ R-symmetry~\cite{Goh:2008xz,Bellazzini:2017neg}. The R-symmetry is universally associated to the $N=1$ SUSY algebra as the only abelian symmetry that does not commute with the supercharges and has to be broken to realize a phenomenologically viable SUSY spectrum with (large) Majorana gaugino masses. The R-axion can behave like a DFSZ ALP or a KSVZ ALP, depending on the R-charge assignment of the MSSM fields \cite{Bellazzini:2017neg}. Moreover, the values of the R-axion decay constant can be related to the mass of the messengers, which can easily be of the order of $10^3\text{ TeV}$ in vanilla gauge mediation scenarios \cite{Komargodski:2009jf,Giudice:1998bp}. 

Some SUSY scenarios also predict exotic $Z^0$ decays. One of the most straightforward ones arises in R-parity violating  (RPV) supersymmetry with a light, bino-like lightest supersymmetric particle (LSP). If the colored superpartners are inaccessible at the LHC, the bounds on such a $\tilde \chi^0_1$ can be very mild, and even \mbox{$m_{\tilde \chi^0_1}\sim 1$ GeV} is still allowed. However, to prevent such a light $\tilde \chi^0_1$ from overclosing the universe, R-parity must be violated, rendering $\tilde \chi^0_1$ unstable~\cite{Bechtle:2015nua}. The decay of $\tilde \chi^0_1$ will go through a heavy, off-shell sfermion, and therefore often occurs displaced, even for relatively large RPV couplings. In addition, the current LHC bounds permit a higgsino as light as $\sim 150$ GeV ~\cite{Aaboud:2017leg}, as long as the wino is kinematically inaccessible. In this scenario, the mixing of the $ \tilde\chi^0_1$ with the higgsino can be large enough to induce an observable branching ratio for $Z^0\to  \tilde\chi^0_1 \tilde\chi^0_1$ \cite{BARTL1989233,Helo:2018qej}. Given that the final states could be fully hadronic and that the partonic center-of-mass energy is merely $\sim m_{Z^0}$, this scenario will benefit greatly from a dedicated displaced vertex trigger.

\paragraph*{Cosmological solutions:}
We first discuss the relaxion mechanism, which is the prototypical cosmological solution to the hierarchy problem~\cite{Graham:2015cka}. In this setup, the VEV of the SM Higgs is driven towards a small value by cosmological evolution of a light scalar field, the relaxion. Relaxion models behaves parametrically as the singlet models with induced $\mathbb{Z}_2$-symmetry breaking discussed in Sec.~\ref{sec:benchmarks}. Moreover, a single scale $\mu_b$ controls both the quartic coupling with the Higgs and the relaxion mass, such that $m_\phi^2\sim \lambda_{SH} v^2$ with $\lambda_{SH}=\mu_b^2/f^2$ where $f$ is the relaxion decay constant. For the value of $\lambda_{SH} $ chosen in Fig.~\ref{fig:higgsmoney}, the relaxion mass will be roughly around $\sim 10$ GeV.  Moreover, requiring the relaxion to address the (little) hierarchy problem one obtains 
\begin{equation}
\!\sin\theta=\!\frac{m_\phi^2}{m_h^2}\frac{f}{v}\gtrsim\! 2.6\times 10^{-1}\!\left(\frac{m_\phi}{10\text{ GeV}}\right)^2\!\left(\frac{f}{\text{10 TeV}}\right)\, ,
\end{equation}
where $f$ needs to be larger than the UV cutoff of the Higgs sector, which we took to be 10 TeV in this expression. The above equation shows that a successful relaxion mechanism is a fortiori associated to a promptly decaying relaxion for $m_\phi\gtrsim 1$ GeV. This feature is even more pronounced if one accounts for the suppression of the relaxion mass compare to its mixing with the Higgs~\cite{Banerjee:2020kww}. On the other hand, for $m_\phi\lesssim 1$ GeV CMS may have some sensitivity in the displaced dimuon channel, by leveraging the production through exotic $B$-meson decay.
   
A different cosmological solution of the hierarchy problem has recently been proposed in Ref.~\cite{Csaki:2020zqz}. Here, the idea is that the Higgs mixes with a dilaton of a conformal field theory (CFT) which has a large, negative vacuum energy. The flat dilaton potential, gets modified by the Higgs dynamics if the Higgs VEV is small enough and develops a new vacuum where the cosmological evolution is standard. Hubble patches with a large Higgs VEV on the other hand will crunch immediately under the influence of the CFT's large negative vacuum expectation value, and thus provide a form of anthropic selection. The existence of our Universe today can then be related to the smallness of the Higgs VEV. The light dilaton in \cite{Csaki:2020zqz} can be accessible at colliders, and its preferred mass range lies exactly where a displaced vertex trigger could probe interesting new parameter space. 

\subsection{Strong CP problem}

Here we comment on the possibility that the ALP phenomenology described in Sec.~\ref{sec:benchmarks} could arise in QCD axion models addressing the strong CP problem. While they are elegant and predictive solutions to the strong CP problem, axion models generally suffer from what is known as the \emph{axion quality problem}. The quality problem states that any sources of explicit breaking of the PQ symmetry must be extremely small compared to the contribution from the QCD sector. In particular, quantum gravity is expected to break the PQ symmetry at the Planck scale ($M_{\text{Pl}}$) with an order one coefficient. A successful solution of the strong CP problem thus implies an upper bound on the axion decay constant 
\begin{equation}
f_a\lesssim \left(10^{-10}m_a^2 M_{\text{Pl}}^{\Delta-4}\right)^{\frac{1}{\Delta-2}}\ ,\label{eq:qualityPQ}
\end{equation}
with $\Delta$ the dimension of the leading operator which breaks the PQ symmetry. Using the relation \mbox{$m_a\simeq \Lambda_{\text{QCD}}^2/f_a$} this implies 
 \begin{equation}
f_a\lesssim 10^{-\frac{10}{\Delta}}\left(\frac{\Lambda_{QCD}}{M_{\text{Pl}}}\right)^{\frac{4}{\Delta}} M_{\text{Pl}} \lesssim 10^{5}\,\mathrm{GeV} ,\label{eq:qualityPQvanilla}
\end{equation}
where in the last inequality we assumed $\Delta=6$, in analogy to baryon number violation in the SM. This regime however suffers from the rather unfortunate problem that it is already experimentally excluded by rare meson decays \cite{GEORGI198673} and the non-observation of anomalous stellar cooling \cite{Raffelt:1994ry,Viaux:2013lha,Bertolami:2014wua}. This leaves us with two possible avenues to salvage the axion solution to the strong CP problem: The most common approach is build more elaborate models in the UV which forbid or suppress all dangerous operators (see e.g.~\cite{Randall:1992ut,Redi:2016esr,DiLuzio:2017tjx,Duerr:2017amf}), effectively raising $\Delta$. This is the idea underlying all searches for the low mass QCD axions, e.g.~using high quality resonance cavities.

The second option is to attempt to raise $m_a$ by breaking the $m_a\simeq \Lambda_{\text{QCD}}^2/f_a$ relation. This can be accomplished by adding a heavy dark sector which also contributes to the axion mass. This is still a non-trivial model building exercise, as the phase of the new contribution to the axion potential must be perfectly aligned with that of the QCD contribution. Perhaps the simplest possibility is to introduce a mirror QCD sector~\cite{Rubakov:1997vp,Berezhiani:2000gh,Hook:2014cda,Fukuda:2015ana,Dimopoulos:2016lvn}. The possibility of probing this particular heavy axion model at the LHC was recently studied in Ref.~\cite{Hook:2019qoh}. Alternative models modifies the running of the QCD coupling constant at high energies~\cite{Holdom:1982ex,Choi:1988sy,Holdom:1985vx,Dine:1986bg,Flynn:1987rs,Choi:1998ep} or embed the color group in a larger product group~\cite{Agrawal:2017ksf}. Once accomplished however, the strong CP problem can be solved with \mbox{$m_a\gtrsim $ GeV} and an $f_a$ which could be within reach of the LHC. The green curves in Fig.~\ref{fig:higgsmoney} indicate the upper bounds on $f_a$ from \eqref{eq:qualityPQ}, assuming dimension $\Delta=5$ and $\Delta=6$. 

\subsection{Dark Matter\label{sec:darkmatter}}

The diversity of the SM forces can be taken as a motivation to go beyond the standard WIMP paradigm by introducing new forces that could control the Dark Matter freeze-out process. Both the ALP and singlet benchmark model are frequently used as new force mediators in this context, see e.g.~\cite{Krnjaic:2015mbs,Evans:2017kti,CidVidal:2018blh,Matsumoto:2018acr,DAgnolo:2018wcn,DAgnolo:2019zkf}. If the mediator can decay to the dark matter, the signal at the LHC will typically be MET, associated with jets, leptons, photons or electroweak gauge bosons, depending on the model. In this case, a robust prediction of the signal rate is possible. If the mediator is lighter than the DM, it does typically decay back to the SM but there is no longer a direct connection between the mediator's coupling to SM and the dark matter relic density. This is the scenario towards which a displaced vertex trigger can contribute. 

As long as the mediator is in thermal contact with the SM during the dark matter freeze-out epoch, it has the potential to play a role in setting the DM relic density. For the singlet benchmark, the most important process is $SS\leftrightarrow f\bar f$ scattering through an off-shell Higgs boson, with $f$ the SM fermions. Concretely, we find that $S$ always maintains thermal equilibrium with the SM as long as there is a fermion species $f$ for which
\begin{equation}
m_{\phi} \gtrsim \frac{4\pi m_h^4}{\lambda_{SH}^2 m_f^2 M_{\text{Pl}}}\qquad\mathrm{and}\qquad m_{\phi}> m_f\ ,
\end{equation} 
with $M_{\text{Pl}}$ the Planck mass. This is easily satisfied everywhere in the parameter space in Fig.~\ref{fig:higgsmoney}, regardless of the value of $s_\theta$. The ALP model is able to maintain thermal equilibrium with the SM down to $T\sim m_a$ as long as 
\begin{equation}
f_a \lesssim \frac{\alpha_s}{2\pi}\sqrt{m_a M_{\text{Pl}}}\ ,
\end{equation}
which is also satisfied everywhere in Fig.~\ref{fig:higgsmoney}. 

An exhaustive review of all relevant models that can explain the DM abundance is well beyond our scope, and we instead conclude this general discussion by mentioning a few selected examples. An elegant and minimal example was provided by Evans et.~al.~\cite{Evans:2017kti}, where they extend the scalar benchmark in \eqref{eq:benchmark} with a Majorana dark matter particle ($\chi$) which couple to the singlet $S$ as  
\begin{equation}
\mathcal{L}_{S\chi}\supset\frac{1}{2}y_{\chi} S \chi\chi\ ,\label{eq:DMmodelS}
\end{equation}
and obtains its mass from the vacuum expectation value of $S$. Their model has 4 independent parameters, $m_\chi$, $m_S$, $y_\chi$ and $s_\theta$, as they set $\mu=0$ in \eqref{eq:benchmark}. This corresponds to the direct $\mathbb{Z}_2$ breaking scenario in Sec.~\ref{sec:singlet}.
 Here we will slightly generalize their setup by allowing for $\mu\neq0$, and thus gain $\lambda_{SH}$ as a $5^{\text{th}}$ independent parameter. One can fix $y_\chi$ by requiring that the $\chi\chi\to SS$ process reproduces the correct dark matter relic abundance, using the calculations in \cite{Evans:2017kti}. In Fig.~\ref{fig:higgsmoney} we have also fixed $m_\chi/m_S=3$. $\lambda_{SH}$ was fixed by our choice of $\text{Br}[h\to SS]$, as explained in Sec.~\ref{sec:singlet}. The lower and upper orange curves saturate respectively the XENON1T limit \cite{Aprile:2018dbl} and the neutrino floor. The orange shaded region in Fig.~\ref{fig:higgsmoney} is therefore ruled out for this model with these particular choice for $m_S/m_\chi$. Interestingly, this also implies the region that will be probed by the next generation of large dark matter detectors coincides with the region that could be accessible to CMS. While the model does not predict that the dark matter must live in this specific region of parameter space, the prospect of a possible double discovery is nevertheless exciting.

In a similar spirit, the ALP model can be easily extended to a minimal model of DM freeze-out. For instance, we may add a Dirac DM candidate, charged under the $U(1)_{\text{PQ}}$ in the KSVZ model of \eqref{eq:KSVZ}~\cite{CidVidal:2018blh}. The interaction is then
\begin{equation}
\mathcal{L}_{a\chi}\supset y_{\chi} \Phi \tilde{\chi}\chi+\, \text{h.c.} \label{eq:DMmodelKSVZ}
\end{equation}
where the mass of the DM is $m_\psi=\sqrt{2} y_\chi N f_a$ with $N$ the anomaly coefficient\footnote{The appearance of $N$ in this equation is an artefact of the normalization we chose for $f_a$ in \eqref{eq:ALPL}.} as defined in Sec.~\ref{sec:ALP}. The dominant dark matter annihilation processes are $\chi\chi\to aa$ and $\chi\chi\to gg$. Whichever dominates, depends on the value of $N$. For $N=1/2$, corresponding to one heavy flavor, and fixing $y_\chi<4\pi$, we indicate in Fig.~\ref{fig:higgsmoney} the range of $f_a$ where the correct dark matter relic abundance can be obtained. 

Both benchmarks can also play a role in more exotic freeze-out mechanisms: For example, D'Angolo et.~al.~have shown that sub-GeV thermal dark matter candidates can exist through the coannihilation \cite{DAgnolo:2018wcn} or coscattering \cite{DAgnolo:2019zkf} of a compressed multiplet of dark sector particles, for which they used the singlet benchmark model as a mediator.

\subsection{Baryogenesis}
The presence of out-of-equilibrium dynamics in the early universe is necessary condition for generating the observed baryon asymmetry. It is therefore natural to consider a relatively small coupling in a hidden sector, which can be responsible for a particle species decaying out-of-equilibrium before the onset of Big Bang Nucleosynthesis (BBN). While a baryogenesis mechanisms at very high energy scales can be fairly easily constructed, there are a number of schemes which are explicitly tied to energy scales that are accessible to the LHC. In those models, displaced vertices are a generic expectation. 

For example, in WIMP baryogenesis \cite{Cui:2012jh,Cui:2014twa} an electroweak state freezes out in the early universe, much like a classic WIMP. Unlike the standard WIMP however, it is allowed to decay to a SM state carrying baryon number and another stable particle, which will be the dark matter. If this decay occurs sufficiently slowly, the baryon asymmetry can be generated. As it turns out, the necessary lifetimes correspond to displaced decays at the LHC. So far the main focus has been on high mass scenarios, partially inspired by supersymmetry, for which the MET and/or $H_T$ triggers are adequate. A displaced vertex trigger could however open a complementary, low mass parameter space.

In a more recent example, the baryon asymmetry is generated through the CP-violating oscillations of heavy flavor SM baryons~\cite{McKeen:2015cuz,Aitken:2017wie} or mesons~\cite{Elor:2018twp,Nelson:2019fln}. (See \cite{Alonso-Alvarez:2019fym} for a supersymmetric implementation.) Concretely, the first ingredient is a massive, non-relativistic particle, which decays out-of-equilibrium to the SM b-quarks at a low temperature. The $b$ quarks subsequently hadronize to heavy flavor mesons and hadrons, which start to oscillate due to the CP violating phase in the SM CKM matrix. Finally, a new, exotic decay mode of the $B$-mesons into SM baryons and a dark sector then generates the asymmetry. The dark sector states can be stable, or decay back to the SM, depending on the specific model. In the latter case, its decay must however occur through an operator with rather high dimension, which predicts a macroscopic lifetime. This scenario is likely difficult to detect, since the unstable hidden sector state must be lighter than $m_B$. On the other hand, the experimental constraints on such a decay are currently very limited and given the huge $b\bar b$ cross section, a very large number of events could be collected by a displaced vertex trigger, even if the trigger efficiency itself is relatively small.

\section{Simulation framework\label{sec:simulation}}
For this study we rely on a toy simulation of the CMS L1 track trigger, as developed and described in \cite{Gershtein:2017tsv,Gershtein:2019dhy}, with minor modifications, which we describe below.

\subsection{Track reconstruction}
Our toy tracking detector consists out of 6 cylindrical layers located at the radii of the CMS phase II outer tracker \cite{CERN-LHCC-2020-004}, for $|\eta|<2.4$. A track is propagated in the 3.8 T magnetic field, starting from the location of the decay vertex from which the track originated. In a realistic experimental setup, a charged particle would undergo multiple scattering in the detector material, slightly deflecting its trajectory and hereby degrading the vertex resolution. To model this, we deflect each track when passing through a layer with an angle drawn from a gaussian distribution centered around zero with width $4\cdot 10^{-4}/p_T$ \cite{Collaboration:2272264, Tanabashi:2018oca}. For this purpose, also the 4 layers of the inner tracker have been included, though they are otherwise ``dark'' from the point of view of the L1 track trigger. See \cite{Gershtein:2019dhy} for a quantitive estimate of the impact of multiple scattering on the vertex reconstruction in our toy simulation.

Once the track has been propagated, the intersections with all 6 layers of the outer tracker are found, and for each intersection the azimuthal offset is computed. Only ``stubs'' for which the offset is consistent with a prompt, $p_T>2$ GeV track are kept. We demand that at least 5 stubs per track pass this selection. All such surviving stubs are subsequently smeared with the expected resolution for the layer in question \cite{Collaboration:2272264}. Finally, the hits are fit to a 5 parameter helix, allowing the track to originate away from the beam line. The reconstruction efficiency is most impacted by the $p_T$ of the track, its transverse impact parameter ($d_0$) and the number of stubs required in the track. 

\begin{figure}[t]\centering
\includegraphics[width=0.4\textwidth]{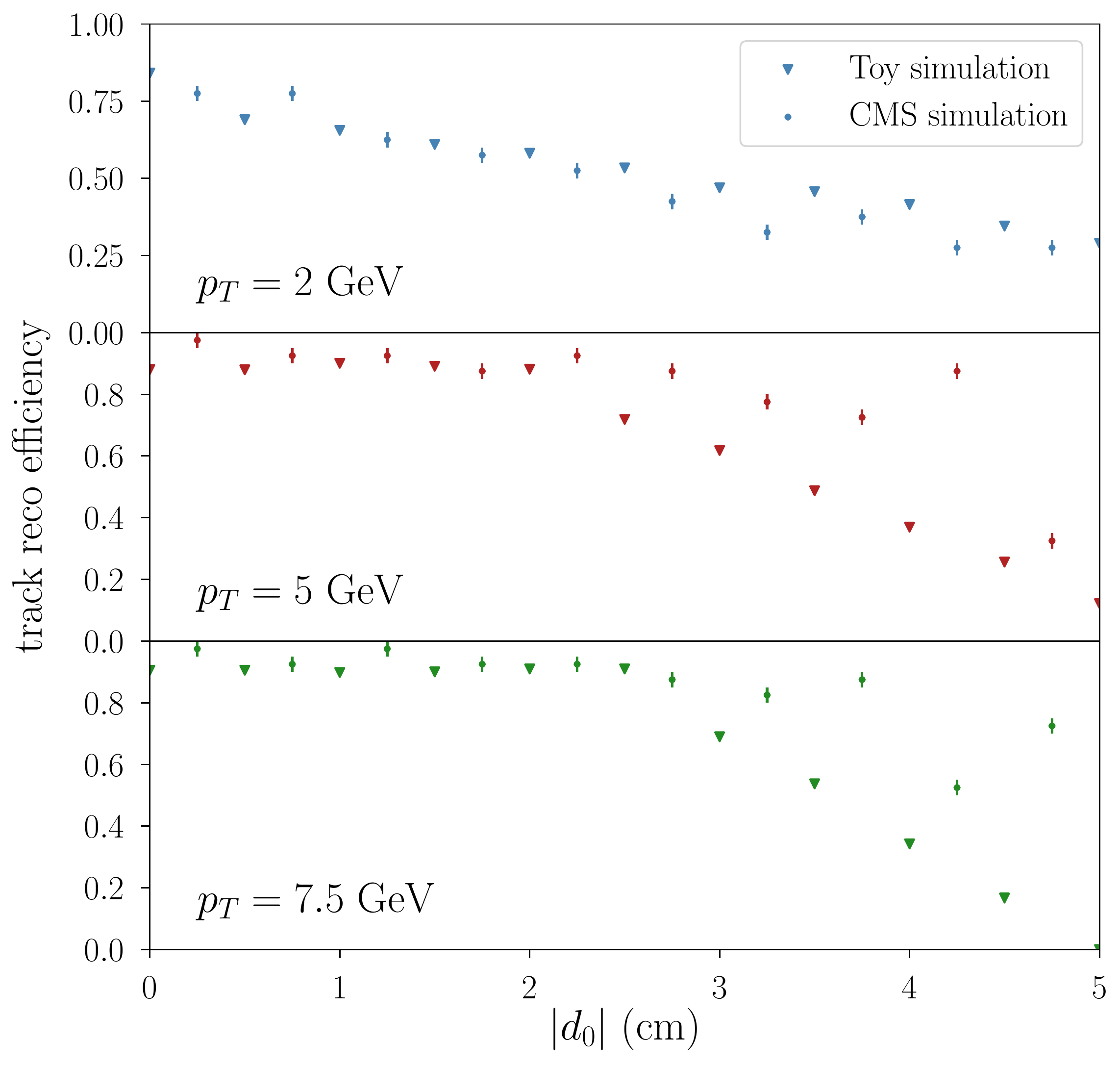}
\caption{The track reconstruction efficiency of our toy detector simulation as a function of the transverse impact parameter of the track ($|d_0|$), compared with the efficiency found by the CMS collaboration \cite{CMS-PAS-FTR-18-018}. Tracks were generated with a particle gun, uniform in $|\eta|<1$, and were required to have 5 reconstructed stubs. The error bars on the CMS result represent the bin size in which the efficiency was reported. \label{fig:particlegun} }
\end{figure}

In Fig.~\ref{fig:particlegun} we used a particle gun to compare the track reconstruction efficiency obtained with our toy simulation with the efficiency reported by the CMS collaboration \cite{CMS-PAS-FTR-18-018}. The agreement is excellent for low impact parameters, though the toy simulation is somewhat underperforming for high impact parameters. The signal trigger efficiencies we report in Sec.~\ref{sec:results} are in this sense conservative estimates. Moreover, the background which is most sensitive to the tracking efficiency is the $B$ meson background 
(Sec.~\ref{sec:Bmeson}). This background is however restricted to very low $d_0$, for which the toy simulation is performing well, as we will see in Fig.~\ref{fig:Brate}.

\subsection{Vertex reconstruction\label{sec:vertexreco}}
We employ a very simplified vertexing algorithm, deliberately avoiding any $\chi^2$-minimizations and/or numerical solutions for intersecting trajectories, as these operations are likely computationally prohibitive at the L1 track trigger. After the tracks have been reconstructed, our simplified vertex algorithm proceeds as follows:
\begin{enumerate}
\item The two hardest tracks of the candidate vertex are selected to seed the algorithm.
\item For these two tracks, their intersections in the transverse plane are found, if they exist. If the tracks do not intersect in the transverse plane, the transverse location of the candidate vertex is chosen to be the point on the line connecting the centers of the circles for which the distance to each circle is equal. The distance of a track to the candidate vertex in the transverse plane is recorded in the variable $\Delta_{xy}$, which is set to 0 if intersections are found. 

\item For each candidate vertex in the transverse plane, the distance along the $z$-direction between both tracks is computed. If two intersections were found in step (ii), the one with the smallest distance in the $z$-direction is chosen, with the z-coordinate of the candidate vertex being the midpoint between both tracks. The distance in the z-direction between a tracks and the vertex candidate is recorded in the variable $\Delta_z$.\footnote{Note that in \cite{Gershtein:2019dhy}, the definitions of $\Delta_{xy}$ and $\Delta_z$ differ with a factor of 2 from the ones we use here, as both variables defined as the distances between the two tracks defining the vertex, rather than the distance of the track to the vertex. }

\item Having defined the location of the candidate vertex by intersecting the two hardest tracks, we compute subsequently $\Delta_{xy}$ and $\Delta_z$ for the remaining tracks, without updating the vertex location. Only tracks satisfying $\Delta_{xy}<0.1$ cm and $\Delta_z<0.5$ cm are assigned to the vertex. We further demand that all tracks satisfy $|d_0|>0.1$ cm, to remove tracks originating from the beam line.

\item Only vertices with 4 or more tracks are being kept.

\end{enumerate}
 Better performance can likely be obtained by numerically finding the most optimal vertex location using the information of all tracks associated with the vertex candidate, rather than just the two hardest tracks. While speed is of course not an issue for our simple toy simulation, it would however likely be the most important bottleneck in a real life experimental implementation. We therefore deliberately refrain from using numerical optimization routines, and for all steps laid out above, closed form analytic expressions exist as a function of the track parameters, substantially speeding up the algorithm.

With this toy vertexing algorithm we have however left one important question unanswered: How can one efficiently identify the set of tracks corresponding to a candidate vertex? This is important given that a sizable number of fake, displaced tracks is expected in each event, and the rapid combinatorical growth of the possible combinations could be a major obstruction to implementing even a primitive vertex algorithm on the trigger. A possible approach could be to select the hardest displaced track(s) in the event and define a corresponding region of interest in $\phi$ and $\eta$ around the hardest track, excluding all other tracks in the event. 
In the necessary detail, this question is highly non-trivial, and can only be addressed adequately through a more detailed study within the CMS collaboration. It is not a priori obvious that a workable solution exists, subject to the latency constraints of the L1 track trigger. The purpose of our toy simulation is therefore to provide the motivation needed to justify allocating CMS resources to perform such a study.

\section{Background rate\label{sec:backgrounds}}
We consider three backgrounds which we expect to drive most of the rate for a displaced vertex trigger:
\begin{itemize}
\item Secondary vertices from long-lived SM hadrons interacting with the detector material,
\item true displaced vertices from SM hadron decays and
\item fake vertices from randomly crossing fake tracks.
\end{itemize}
Each background requires dedicated modeling, as we describe below. Given the total L1 bandwidth of $\sim$750 kHz after the phase II upgrade, we consider rates $\lesssim 1$ kHz as ``acceptable'' in this work.

\subsection{Material interactions\label{sec:matinteraction}}

In the current configuration of CMS, about 5\% of all pions with $p_T\gtrsim 5$ GeV were found to create a secondary vertex in the tracker \cite{CMS-PAS-TRK-10-003}. Extrapolating to HL-LHC conditions, this corresponds to a daunting rate of roughly 30 MHz. The selection criteria for the tracks in this measurement where however less stringent than those which will be imposed by the $p_T$-modules in the L1 track trigger: In  \cite{CMS-PAS-TRK-10-003} a vertex was required to have two (off-line) reconstructed tracks with $p_T>0.5$ GeV each, while our selections demand at least \emph{four} tracks with $p_T>2$ GeV each. A reliable extrapolation from this data is therefore not feasible, and we must perform a simplified GEANT4 \cite{AGOSTINELLI2003250} simulation instead. 

\begin{figure*}[t!]\centering
\includegraphics[height=6.cm]{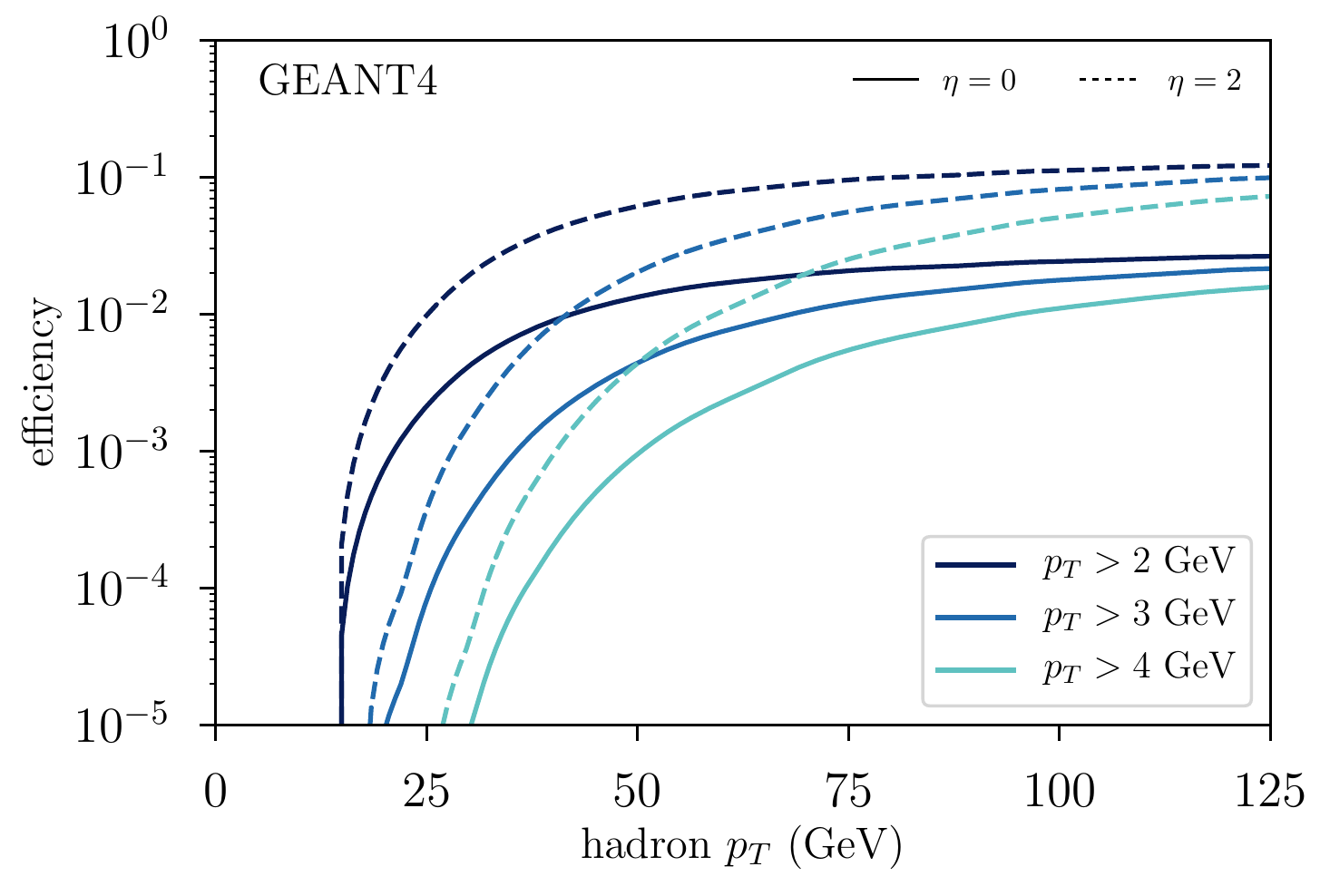}\hfill
\includegraphics[height=6.cm]{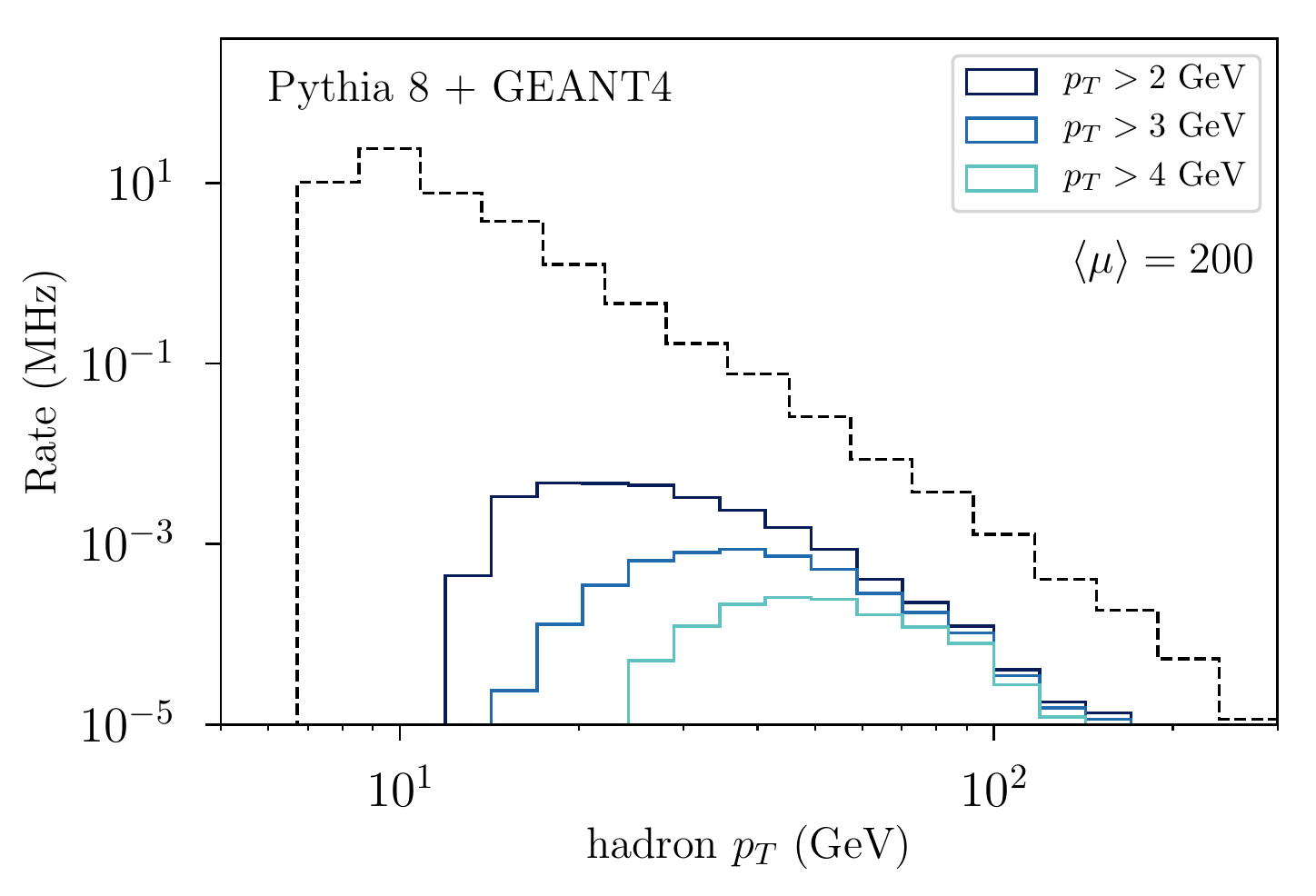}
\caption{\emph{Left:} Efficiency for a hadron to create a secondary vertex with at least 4 tracks subject to the listed $p_T$ cuts, as estimated with a GEANT4 particle gun. The dashed and solid lines represent different example values for the pseudo-rapidity of the incoming hadron, to illustrate the effect of the $\eta$-dependence of the material budget of the detector. (See text for details.) \emph{Right:} Inclusive rate for long-lived SM hadrons with $|\eta|<2.4$ (black dashed) and estimated rate for secondary vertices from a hadron interacting with detector material (solid) for different $p_T$ cuts on the outgoing tracks. The integrated rates for $p_T>2$ GeV, 3 GeV and 4 GeV are respectively 25 kHz, 5 kHz and 1 kHz. \label{fig:secondaries}} 
\end{figure*}

Concretely,  we set up a particle gun where we fired $\pi^+$ particles at a slab of silicon with thickness of 1 cm, using the standard \verb+FTFP BERT+ physics list, based on the Fritiof~\cite{ANDERSSON1987289,Andersson1996,NilssonAlmqvist:1986rx,Ganhuyag:1997gz} and Bertini intra-nuclear cascade~\cite{Guthrie:1968ue,Bertini:1971xb,Karmanov:1979if} models. The particle gun was then used to compute the probability for the resulting secondary vertex to have at least four hard tracks, as a function of the incident pion energy. For a given pseudo-rapidity ($\eta$), this can then be converted in a $p_T$-dependent efficiency by rotating the system and rescaling the GEANT4 output to account for the true, $\eta$-dependent material budget of the detector \cite{Collaboration:2272264}. The resulting efficiency curve is shown in the left-hand panel of Fig.~\ref{fig:secondaries} for two values of $\eta$ and three $p_T$ cuts on the outgoing tracks.  For the loosest $p_T$ requirement, the efficiency sharply turns on around 15 GeV, and plateaus around 30 GeV, where the differences between the \mbox{$\eta$-ranges} can be attributed to the differences in material budget. The plateau at high $p_T$ arises because the relevant process is deep inelastic scattering on the nuclei. The cross section is therefore essentially geometric in this regime and thus independent of the momentum of the incoming pion.

To estimate the total trigger rate, we simulate the rate for stable SM hadrons with Pythia 8~\cite{Sjostrand:2006za,Sjostrand:2014zea}, in the form of a weighted dijet sample with a minimal $p_T>5$ GeV cut. The resulting spectrum was conservatively normalized to a cross section of 68 mbn, the total inelastic cross section at $\sqrt{s}=$13 TeV \cite{Aaboud:2016mmw}. The combined rate for all stable or long-lived SM hadrons is shown by the dashed histogram in the right-hand panel of Fig.~\ref{fig:secondaries}. 

To estimate rate at which SM hadrons produce secondary vertices which pass our selection cuts, we fold the efficiency obtained by the GEANT4 particle gun against the hadron spectrum. The resulting rates are shown in the solid histograms in the right-hand panel of Fig.~\ref{fig:secondaries}. For a minimal $p_T$ cut of 2 GeV per track, the integrated rate for this background is roughly 25 kHz. While this is likely somewhat higher than what one may be able to accommodate comfortably, we stress that this estimate is very conservative and should be understood as an estimated upper bound on the rate. In particular, \emph{(i)} for this background, no track reconstruction efficiencies were accounted for, as this require the full CMS detector simulation. Instead, we assumed that every track above the threshold was reconstructed with 100\% efficiency, \emph{regardless of its impact parameters}. We moreover assume that the vertex quality requirements in Sec.~\ref{sec:vertexreco} are always satisfied for these background events. \emph{(ii)} No material veto\footnote{Given that the pixel layers will not be available to the L1 trigger, the spacial resolution on the vertex location is expected to be relatively poor \cite{Gershtein:2019dhy}. A material veto would therefore be less effective than in a full, offline analysis and we therefore chose to not rely on it in our study. } was attempted and \emph{(iii)} from Fig.~\ref{fig:secondaries} one can see that the rate is primarily driven by hadrons with $p_T\gtrsim 15$ GeV. Such hard hadrons will typically not be isolated, providing an additional handle to reduce this background if needed.  Finally, the rate drops substantially when the threshold per track is raised slightly, to \mbox{5 kHz} \mbox{(1 kHz)} for \mbox{$p_T>3$ GeV} \mbox{($p_T>4$ GeV).} We therefore present our subsequent results for those 3 different threshold choices.

\subsection{SM meson decays\label{sec:Bmeson}}
First, we simulate an inclusive sample of $K_S$ and $K_L$ samples with Pythia 8 by generating a $p_T$-weighted dijet sample, where we subject the kaons to the fiducial cuts of $p_T>8$ GeV and $|\eta|<2.4$. Conservatively normalizing this sample to the total measured inelastic cross section of 68 mbn \cite{Aaboud:2016mmw}, this yields a rate of roughly 21 MHz for each species, which corresponds roughly one Kaon per event. The dominant decay modes which feature 4 charged final states are \mbox{$K_S\to \pi^+\pi^-e^+e^-$}  with branching ratio $4.79\times 10^{-5}$and \mbox{$K_L\to \pi^{\pm}e^{\mp} \nu e^+e^-$} with branching ratio  $1.26\times 10^{-5}$ \cite{10.1093/ptep/ptaa104}. Both modes therefore yield rates $<$ kHz, even before other fiducial and reconstruction efficiencies are factored in. We therefore focus on $B$-meson decays as the most dangerous hadron background.

\begin{figure}[t]\centering
\includegraphics[height=6.cm]{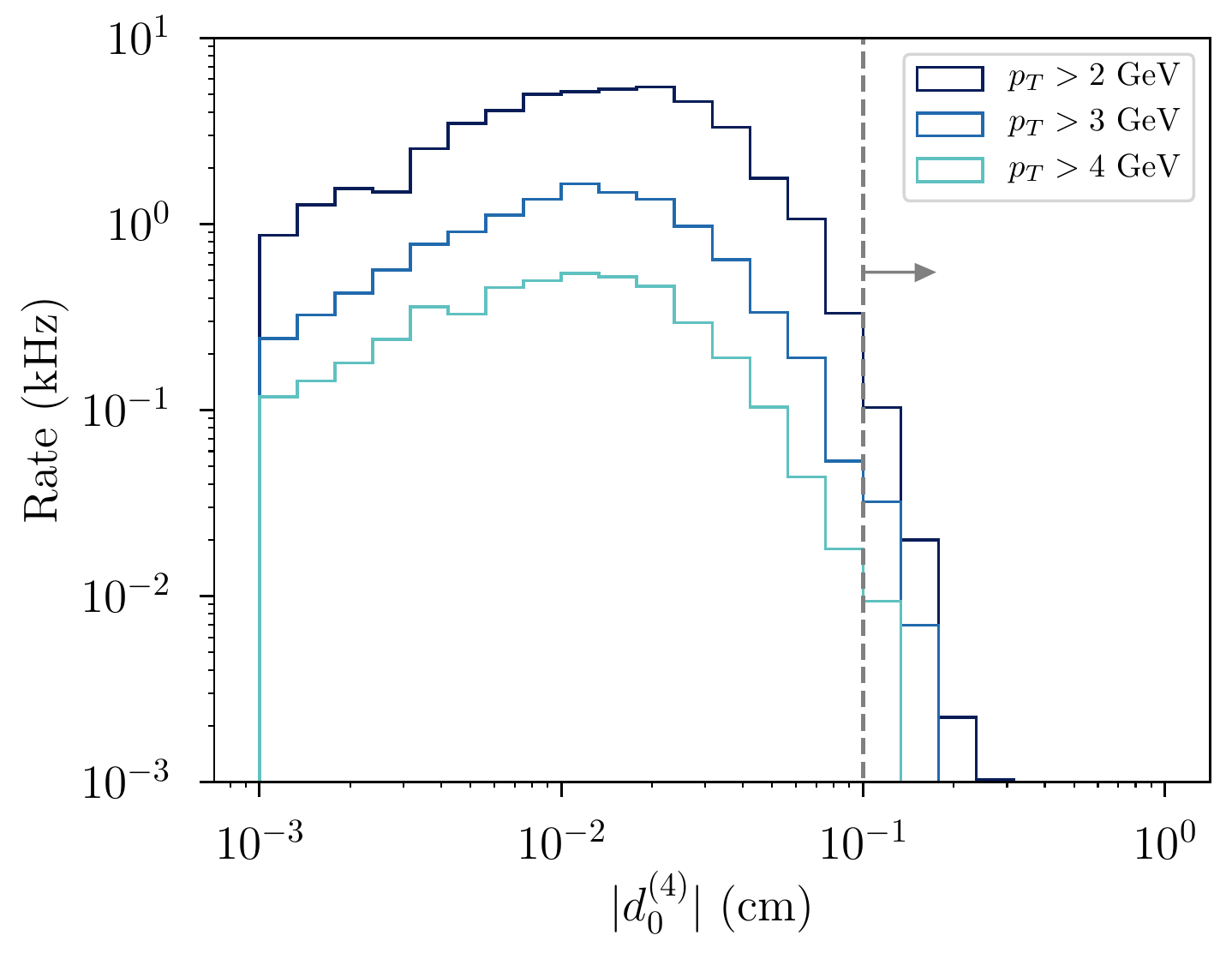}
\caption{Estimated L1 trigger rate from $B$-meson backgrounds, as a function of the transverse impact parameter of the $4^{\text{th}}$ reconstructed track ($|d^{(4)}_0|$), where the tracks were ordered according to decreasing $|d_0|$. We require $|d^{(4)}_0|>0.1$ cm, as indicated by the dashed vertical line.\label{fig:Brate}}
\end{figure}

We use Pythia 8 to simulate an inclusive sample of $B$-mesons, which are subsequently processed through the toy detector simulation described in Sec.~\ref{sec:simulation}. The sample is normalized to the inclusive $b\bar b$ cross section, as calculated with FONLL \cite{Cacciari:1998it,Cacciari:2001td,Cacciari:2012ny,Cacciari:2015fta}. Demanding at least 4 tracks forming a vertex, the resulting rate is approximately 50 kHz. Most tracks originating from B decays however have small impact parameters, and a cut of $|d_0|>1$ mm for at least 4 tracks in the vertex further drops the rate to roughly 130 Hz for the most loose $p_T$ cut (see Fig.~\ref{fig:Brate}). The rate moreover drops substantially if the $p_T$ cuts are tightened, as shown in Tab.~\ref{tab:backgrounds}. If needed, an additional cut on the transverse radius of the vertex ($L_{xy}$) is extremely effective at reducing this background (see e.g.~\cite{Evans:2020aqs}), at a minimal cost to the signal.

\begin{table}[b]\centering
\begin{tabular}{p{3cm}>{\centering\arraybackslash}p{1.5cm}>{\centering\arraybackslash}p{1.5cm}>{\centering\arraybackslash}p{1.5cm}}
min track $p_T$& 2 GeV & 3 GeV & 4 GeV\\\hline\hline
secondaries (kHz)&25&5&1\\
B-mesons (kHz)&0.13&0.04&0.01\\
fake vertices (kHz)&0.04&0.01&0.004
\end{tabular}
\caption{Estimated rates for various backgrounds, for different choices of the minimum track $p_T$. The rate for the secondaries is to be understood as a conservative upper bound. See text for details.\label{tab:backgrounds}}
\end{table}

\subsection{Fake vertices}
Finally, to estimate the probability of 4 fake tracks forming a vertex, we generated $10^7$ pairs of fake tracks, uniformly distributed in the track parameters. (See \cite{Gershtein:2019dhy} for details.) The pairs of fake tracks where processed through the toy vertex reconstruction algorithm and discarded if the $\Delta_{xy}$ and $\Delta_z$ parameters of the candidate vertex failed our vertex quality requirements in Sec.~\ref{sec:vertexreco}. The remaining two-track vertices are then pairwise compared with one another, where we select only those pairs for which the distance between both vertices in the transverse and longitudinal directions is compatible with the respective vertex resolution cuts. Finally, all remaining candidate four-track vertices are again processed through the algorithm in Sec.~\ref{sec:vertexreco}. With this method, the probability of 4 random fake tracks forming a vertex is found to be approximately $3\times 10^{-11}$.

Based on the expected occupancy of the trigger system \cite{CMSCollaboration:2015zni}, we assume on average 30 fake tracks per event. We thus compute the average combinatorical enhancement factor per event to be 
\begin{equation}
\left\langle \frac{n_{\text{fakes}}! }{(n_{\text{fakes}}-4)! 4!}\right\rangle \approx 3.4\times 10^4\ , 
\end{equation}
where the brackets indicates the Poisson weighted average, assuming $\langle n_{\text{fakes}}\rangle=30$. Combining this with the fake vertex probability, we find the rate to about 40 Hz or lower (See Tab.~\ref{tab:backgrounds}), giving a substantial buffer in case our estimate of 30 tracks per event proves to be insufficiently conservative.

Finally, one may be concerned that a significant contribution from the rate may come from real, 3-track vertices to which a fourth, fake track is erroneously asigned. With the assumptions laid out in Sec.~\ref{sec:matinteraction} we find that the rate for a 3-track vertex from hadronic interactions with the detector material is roughly 50 kHz for a $p_T$ cut on the tracks of 2 GeV. Following Sec.~\ref{sec:Bmeson}, the rate for a 3-track vertex from the $K^\pm\to \pi^\pm \pi^+ \pi^-$ process is bounded from above by roughly 30 kHz. We hereby imposed a truth-level $p_T>6$ GeV cut for the $K^\pm$ and required it to decay within 30 cm of the beamline, but no requirements were made on the kinematics of the individual pions.  With $\mathcal{O}(30)$ fake tracks per event, the rate for both cases therefore remains below 1 kHz if the probability for a fake track to be associated with a real displaced vertex is below $\mathcal{O}(10^{-3})$. This appears plausible, but in particular for the vertices from material interactions, we cannot accurately estimate this probability without the full CMS detector simulation. Our estimates of the rate for real, 3-track vertices are moreover conservative, and in practice we expect that a $\mathcal{O}(10^{-2})$ suppression could be sufficient. Should this process nevertheless prove to be problematic, it can be suppressed further by tightening the $p_T$ cuts on the tracks: In particular, by increasing the cut to 3 GeV (4 GeV), the rate for 3-track material interactions drops to 13 kHz (5 kHz). Similarly, increasing the $p_T$ cut on the $K^{\pm}$ to 9 GeV (12 GeV) reduces the rate to 5 kHz (0.8 kHz).

\section{Discussion and outlook\label{sec:results}}
As we showed in the previous section, if a displaced vertex could be reconstructed at the trigger level, the $H_T>100$ GeV requirement used in \cite{Gershtein:2017tsv,CMS-PAS-FTR-18-018,Bhattacherjee:2020nno,Hook:2019qoh} may no longer be needed to bring the background rate down to a manageable level. This in turn substantially increases the signal efficiency for scenarios where the final states are rather soft.

\begin{figure}[t]\centering
\includegraphics[width=0.48\textwidth]{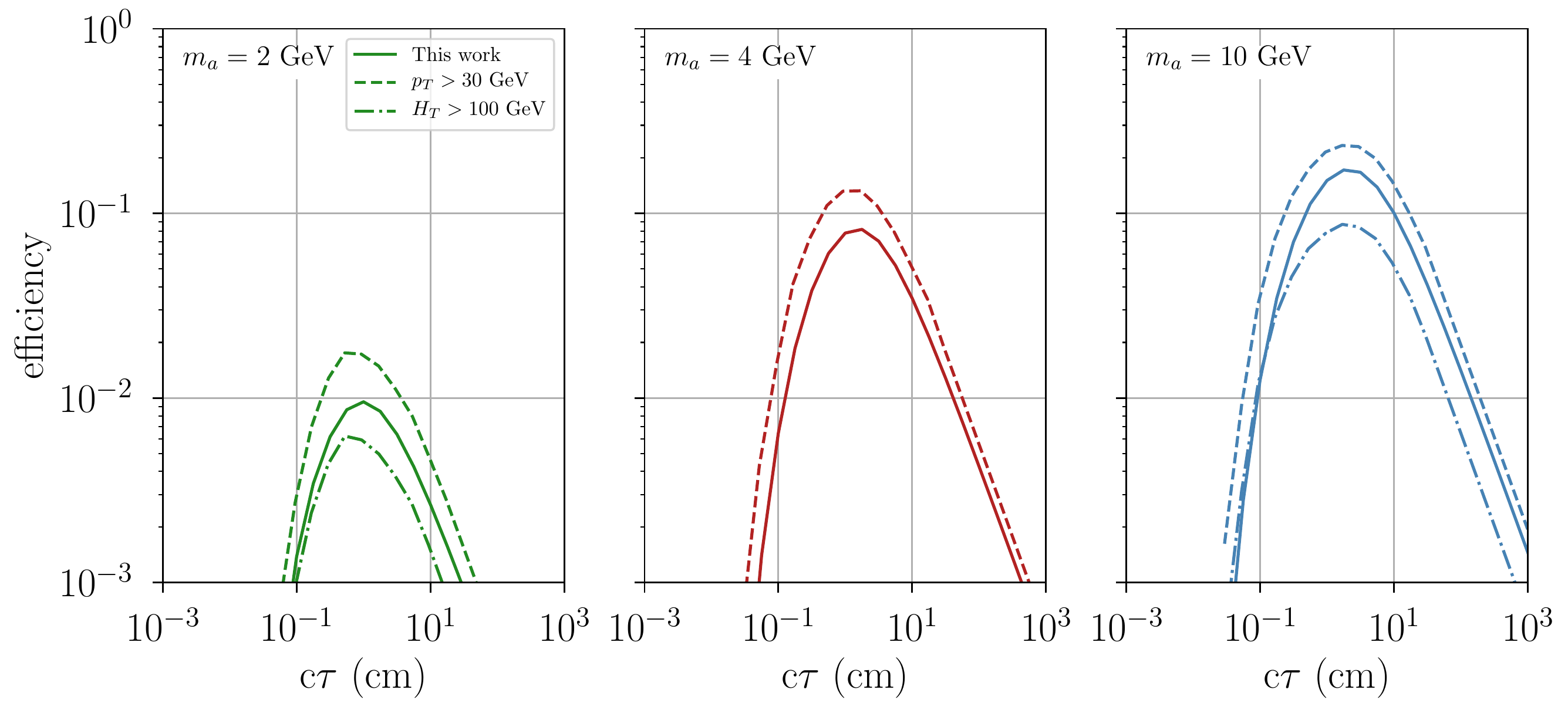}
\caption{Estimated L1 trigger efficiency for a CMS displaced vertex trigger for the ALP model (solid). Also shown are the estimates based on the $H_T>100$ GeV and $p_T>30$ GeV selections from Hook et.~al.~\cite{Hook:2019qoh} All efficiencies were normalized to a truth-level cross section with $p_T>20$ GeV cut on the ALP. (See text for details) \label{fig:alpeff}}
\end{figure}

Fig.~\ref{fig:alpeff} illustrates this point for the ALP benchmark, where we normalized all efficiencies to a sample with a $20$ GeV cut on the truth-level $p_T$ of the ALP. The trigger efficiency is $\mathcal{O}(10\%)$, which we consider fairly decent for a signal this soft. We find that a displaced vertex trigger would roughly be a factor of 3 improvement over the displaced jet trigger with $H_T>100$ GeV, as studied Hook et.~al.~\cite{Hook:2019qoh}.\footnote{We thank Soubhik Kumar for providing us with this unpublished data.} Hook et.~al.~also provided a second, more speculative selection consisting out of a single, displaced jet with $p_T>30$ GeV. This selection unsurprisingly outperforms a displaced vertex trigger, as the former only requires 3 displaced tracks vs 4 displaced tracks for the latter. Given the high rate of $p_T>30$ GeV jets, secondary vertices from material interactions and fake tracks, we however suspect that the background rate for this proposal could be prohibitive.
%
%
 %
In sum, Fig.~\ref{fig:alpeff} makes it clear that the reason for the rather low signal yield in the right-hand panel of Fig.~\ref{fig:higgsmoney} is not related to the low reconstruction efficiency per se. Rather, both the lifetime and the production cross section of the ALP are controlled by $f_a$ and the lifetime is simply very short for values of $f_a$ which yield interesting cross sections (see \eqref{eq:ALPwidth} and Fig.~\ref{fig:ALPxsec}). 

This is in sharp contrast with the scalar model, where the lifetime and production rate are controlled by \emph{independent} paramaters, $s_\theta$ and $\lambda_{SH}$ respectively. As a result, a displaced vertex trigger could open a large portion of the parameter space where the mass of the singlet is natural with respect to its quartic coupling with the SM Higgs. We showed that the values of the mixing angles that could be explored are further motivated by simple models of DM freeze-out and neutral naturalness scenarios.

\begin{figure}[t!]\centering
\includegraphics[width=0.48\textwidth]{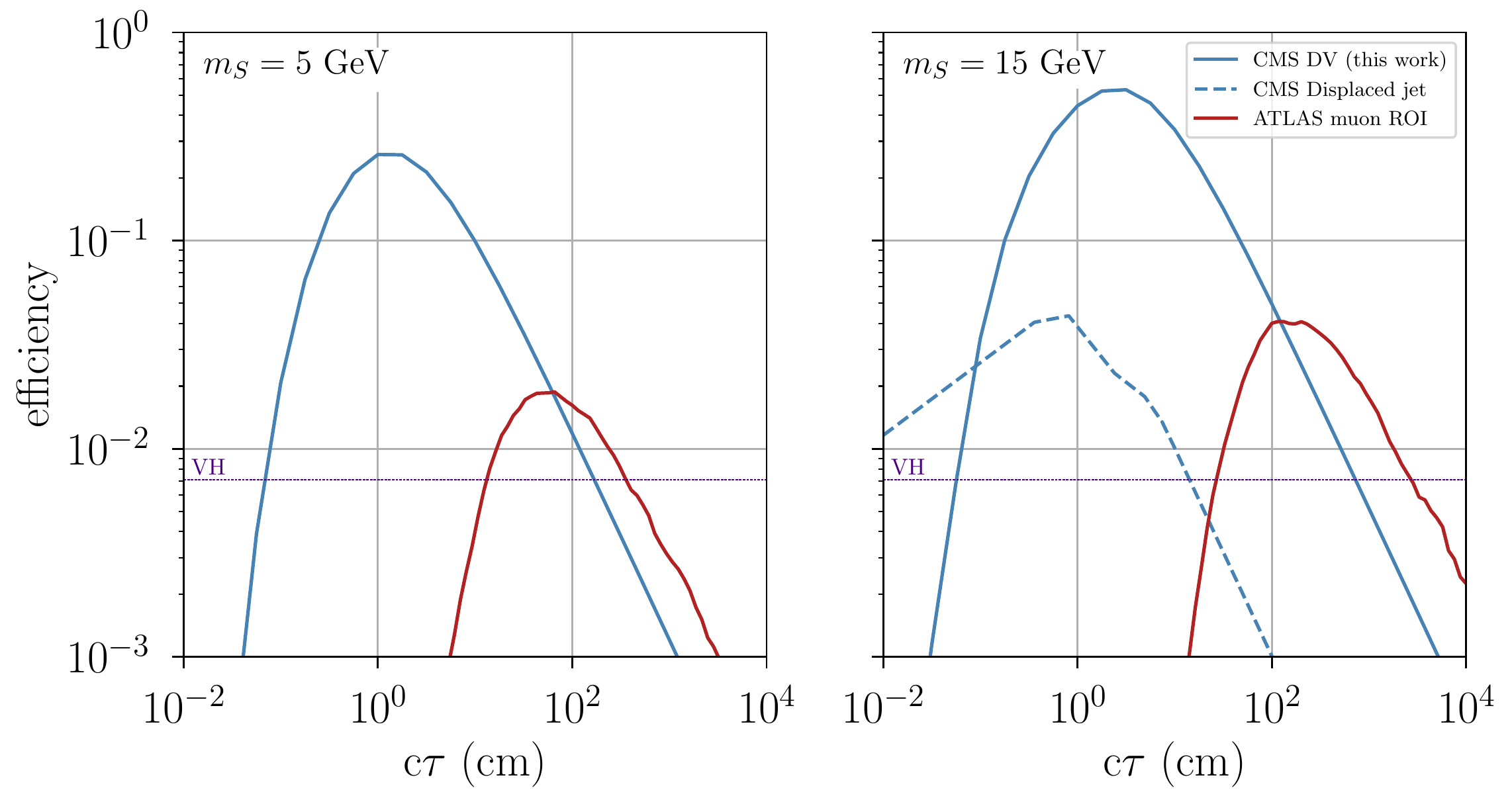}
\caption{Estimated L1 trigger efficiency for a CMS displaced vertex trigger (solid blue). Also shown is the projected efficiency for CMS L1 track jet trigger for displaced jets \cite{Gershtein:2017tsv,CMS-PAS-FTR-18-018}, assuming 5 kHz rate (dashed blue) and the efficiency of the existing ATLAS muon region of interest trigger (solid red) \cite{Aaboud:2018aqj}. The dashed purple line indicates the efficiency to trigger on an associated lepton in VH production, normalized against the gluon fusion cross section.\label{fig:higgseff}}
\end{figure}

The model independent trigger efficiency for the $h\to SS$ topology is shown Fig.~\ref{fig:higgseff}, where a displaced vertex trigger would improve on the displaced jet + $H_T$ trigger by more than an order of magnitude. For reference, we also include the efficiency for the existing ATLAS trigger which relies on anomalous activity in a region of interest in the muon chamber (``muon ROI'') \cite{Aaboud:2018aqj}. As expected, both approaches are highly complimentary, as a trigger on a displaced vertex in the tracker could cover shorter lifetimes.

We conclude by commenting on another opportunity for the track trigger, which we intend explore in upcoming work \cite{futurework}. Traditionally, dark matter searches rely heavily on the MET trigger, which has low efficiency for low mass DM or soft production modes. This is particularly so for inelastic DM models, where the DM resides in a narrowly split multiplet. Inelastic DM models predict a soft displaced vertex with a moderate amount of MET, typically well below the current trigger thresholds. Models of heavy neutral leptons produce a similar signature. By identifying the displaced tracks with the track trigger, CMS may be able to substantially lower the MET requirement, while maintaining a manageable background rate. The implementation of such a trigger may allow CMS to probe a large portion of the parameter space of these models, complementing existing proposals~\cite{Izaguirre:2015zva,Berlin:2018jbm}.

\FloatBarrier

 \section*{Acknowledgments}
We are grateful to Brando Bellazzini, Matthew Citron, Soubhik Kumar, Jared Evans, Sam Junius,  Zhen Liu, Gaia Lianfranchi, Alberto Mariotti, Matthew McCullough, Toby Opferkuch, Simone Pagan Griso, Michele Papucci, Dean Robinson, Filippo Sala, Ennio Salvioni, Jessie Shelton and Scott Thomas for useful discussions. We thank especially Gauthier Durieux, Davide Pagani, Pier Monni and Marco Zaro for important clarifications on the QCD corrections to the cross section and for help with $MadGraph5@NLO$.

\bibliography{DV_trigger}

\providecommand{\noopsort}[1]{}\providecommand{\singleletter}[1]{#1}%
\begin{thebibliography}{142}%
\makeatletter
\providecommand \@ifxundefined [1]{%
 \@ifx{#1\undefined}
}%
\providecommand \@ifnum [1]{%
 \ifnum #1\expandafter \@firstoftwo
 \else \expandafter \@secondoftwo
 \fi
}%
\providecommand \@ifx [1]{%
 \ifx #1\expandafter \@firstoftwo
 \else \expandafter \@secondoftwo
 \fi
}%
\providecommand \natexlab [1]{#1}%
\providecommand \enquote  [1]{``#1''}%
\providecommand \bibnamefont  [1]{#1}%
\providecommand \bibfnamefont [1]{#1}%
\providecommand \citenamefont [1]{#1}%
\providecommand \href@noop [0]{\@secondoftwo}%
\providecommand \href [0]{\begingroup \@sanitize@url \@href}%
\providecommand \@href[1]{\@@startlink{#1}\@@href}%
\providecommand \@@href[1]{\endgroup#1\@@endlink}%
\providecommand \@sanitize@url [0]{\catcode `\\12\catcode `\$12\catcode
  `\&12\catcode `\#12\catcode `\^12\catcode `\_12\catcode `\%12\relax}%
\providecommand \@@startlink[1]{}%
\providecommand \@@endlink[0]{}%
\providecommand \url  [0]{\begingroup\@sanitize@url \@url }%
\providecommand \@url [1]{\endgroup\@href {#1}{\urlprefix }}%
\providecommand \urlprefix  [0]{URL }%
\providecommand \Eprint [0]{\href }%
\providecommand \doibase [0]{https://doi.org/}%
\providecommand \selectlanguage [0]{\@gobble}%
\providecommand \bibinfo  [0]{\@secondoftwo}%
\providecommand \bibfield  [0]{\@secondoftwo}%
\providecommand \translation [1]{[#1]}%
\providecommand \BibitemOpen [0]{}%
\providecommand \bibitemStop [0]{}%
\providecommand \bibitemNoStop [0]{.\EOS\space}%
\providecommand \EOS [0]{\spacefactor3000\relax}%
\providecommand \BibitemShut  [1]{\csname bibitem#1\endcsname}%
\let\auto@bib@innerbib\@empty
\bibitem [{\citenamefont {Einsweiler}\ \emph {et~al.}(2017)\citenamefont
  {Einsweiler} \emph {et~al.}}]{Collaboration:2285585}%
  \BibitemOpen
  \bibfield  {author} {\bibinfo {author} {\bibfnamefont {K.}~\bibnamefont
  {Einsweiler}} \emph {et~al.} (\bibinfo {collaboration} {ATLAS
  collaboration}),\ }\href {https://cds.cern.ch/record/2285585} {\emph
  {\bibinfo {title} {{Technical Design Report for the ATLAS Inner Tracker Pixel
  Detector}}}},\ \bibinfo {type} {Tech. Rep.}\ \bibinfo {number}
  {CERN-LHCC-2017-021. ATLAS-TDR-030}\ (\bibinfo  {institution} {CERN},\
  \bibinfo {address} {Geneva},\ \bibinfo {year} {2017})\BibitemShut {NoStop}%
\bibitem [{\citenamefont {{CMS Collaboration}}(2020)}]{collaboration:2714892}%
  \BibitemOpen
  \bibfield  {author} {\bibinfo {author} {\bibnamefont {{CMS Collaboration}}},\
  }\href {https://cds.cern.ch/record/2714892} {\emph {\bibinfo {title} {{The
  Phase-2 Upgrade of the CMS Level-1 Trigger}}}},\ \bibinfo {type} {Tech.
  Rep.}\ \bibinfo {number} {CERN-LHCC-2020-004. CMS-TDR-021}\ (\bibinfo
  {institution} {CERN},\ \bibinfo {address} {Geneva},\ \bibinfo {year} {2020})\
  \bibinfo {note} {final version}\BibitemShut {NoStop}%
\bibitem [{\citenamefont {Gershtein}(2017)}]{Gershtein:2017tsv}%
  \BibitemOpen
  \bibfield  {author} {\bibinfo {author} {\bibfnamefont {Y.}~\bibnamefont
  {Gershtein}},\ }\bibfield  {title} {\bibinfo {title} {{CMS Hardware Track
  Trigger: New Opportunities for Long-Lived Particle Searches at the HL-LHC}},\
  }\href {https://doi.org/10.1103/PhysRevD.96.035027} {\bibfield  {journal}
  {\bibinfo  {journal} {Phys. Rev.}\ }\textbf {\bibinfo {volume} {D96}},\
  \bibinfo {pages} {035027} (\bibinfo {year} {2017})},\ \Eprint
  {https://arxiv.org/abs/1705.04321} {arXiv:1705.04321 [hep-ph]} \BibitemShut
  {NoStop}%
\bibitem [{\citenamefont {{CMS Collaboration}}(2018)}]{CMS-PAS-FTR-18-018}%
  \BibitemOpen
  \bibfield  {author} {\bibinfo {author} {\bibnamefont {{CMS Collaboration}}},\
  }\href {https://cds.cern.ch/record/2647987} {\emph {\bibinfo {title} {{First
  Level Track Jet Trigger for Displaced Jets at High Luminosity LHC}}}},\
  \bibinfo {type} {Tech. Rep.}\ \bibinfo {number} {CMS-PAS-FTR-18-018}\
  (\bibinfo {address} {Geneva},\ \bibinfo {year} {2018})\BibitemShut {NoStop}%
\bibitem [{\citenamefont {Bhattacherjee}\ \emph {et~al.}(2020)\citenamefont
  {Bhattacherjee}, \citenamefont {Mukherjee}, \citenamefont {Sengupta},\ and\
  \citenamefont {Solanki}}]{Bhattacherjee:2020nno}%
  \BibitemOpen
  \bibfield  {author} {\bibinfo {author} {\bibfnamefont {B.}~\bibnamefont
  {Bhattacherjee}}, \bibinfo {author} {\bibfnamefont {S.}~\bibnamefont
  {Mukherjee}}, \bibinfo {author} {\bibfnamefont {R.}~\bibnamefont {Sengupta}},
  and\ \bibinfo {author} {\bibfnamefont {P.}~\bibnamefont {Solanki}},\
  }\bibfield  {title} {\bibinfo {title} {{Triggering long-lived particles in
  HL-LHC and the challenges in the first stage of the trigger system}},\ }\href
  {https://doi.org/10.1007/JHEP08(2020)141} {\bibfield  {journal} {\bibinfo
  {journal} {JHEP}\ }\textbf {\bibinfo {volume} {08}},\ \bibinfo {pages}
  {141}},\ \Eprint {https://arxiv.org/abs/2003.03943} {arXiv:2003.03943
  [hep-ph]} \BibitemShut {NoStop}%
\bibitem [{\citenamefont {{Y. Gershtein and S.
  Knapen}}(2020)}]{Gershtein:2019dhy}%
  \BibitemOpen
  \bibfield  {author} {\bibinfo {author} {\bibnamefont {{Y. Gershtein and S.
  Knapen}}},\ }\bibfield  {title} {\bibinfo {title} {{Trigger strategy for
  displaced muon pairs following the CMS phase II upgrades}},\ }\href
  {https://doi.org/10.1103/PhysRevD.101.032003} {\bibfield  {journal} {\bibinfo
   {journal} {Phys. Rev. D}\ }\textbf {\bibinfo {volume} {101}},\ \bibinfo
  {pages} {032003} (\bibinfo {year} {2020})},\ \Eprint
  {https://arxiv.org/abs/1907.00007} {arXiv:1907.00007 [hep-ex]} \BibitemShut
  {NoStop}%
\bibitem [{\citenamefont {Hook}\ \emph {et~al.}(2020)\citenamefont {Hook},
  \citenamefont {Kumar}, \citenamefont {Liu},\ and\ \citenamefont
  {Sundrum}}]{Hook:2019qoh}%
  \BibitemOpen
  \bibfield  {author} {\bibinfo {author} {\bibfnamefont {A.}~\bibnamefont
  {Hook}}, \bibinfo {author} {\bibfnamefont {S.}~\bibnamefont {Kumar}},
  \bibinfo {author} {\bibfnamefont {Z.}~\bibnamefont {Liu}}, and\ \bibinfo
  {author} {\bibfnamefont {R.}~\bibnamefont {Sundrum}},\ }\bibfield  {title}
  {\bibinfo {title} {{High Quality QCD Axion and the LHC}},\ }\href
  {https://doi.org/10.1103/PhysRevLett.124.221801} {\bibfield  {journal}
  {\bibinfo  {journal} {Phys. Rev. Lett.}\ }\textbf {\bibinfo {volume} {124}},\
  \bibinfo {pages} {221801} (\bibinfo {year} {2020})},\ \Eprint
  {https://arxiv.org/abs/1911.12364} {arXiv:1911.12364 [hep-ph]} \BibitemShut
  {NoStop}%
\bibitem [{\citenamefont {Evans}\ \emph {et~al.}(2020)\citenamefont {Evans},
  \citenamefont {Gandrakota}, \citenamefont {Knapen},\ and\ \citenamefont
  {Routray}}]{Evans:2020aqs}%
  \BibitemOpen
  \bibfield  {author} {\bibinfo {author} {\bibfnamefont {J.~A.}\ \bibnamefont
  {Evans}}, \bibinfo {author} {\bibfnamefont {A.}~\bibnamefont {Gandrakota}},
  \bibinfo {author} {\bibfnamefont {S.}~\bibnamefont {Knapen}}, and\ \bibinfo
  {author} {\bibfnamefont {H.}~\bibnamefont {Routray}},\ }\bibfield  {title}
  {\bibinfo {title} {{Searching for exotic B meson decays with the CMS L1 track
  trigger}},\ }\href@noop {} {\  (\bibinfo {year} {2020})},\ \Eprint
  {https://arxiv.org/abs/2008.06918} {arXiv:2008.06918 [hep-ph]} \BibitemShut
  {NoStop}%
\bibitem [{\citenamefont {{S. Foroughi-Abari and A.
  Ritz}}(2020)}]{Foroughi-Abari:2020gju}%
  \BibitemOpen
  \bibfield  {author} {\bibinfo {author} {\bibnamefont {{S. Foroughi-Abari and
  A. Ritz}}},\ }\bibfield  {title} {\bibinfo {title} {{LSND Constraints on the
  Higgs Portal}},\ }\href {https://doi.org/10.1103/PhysRevD.102.035015}
  {\bibfield  {journal} {\bibinfo  {journal} {Phys. Rev. D}\ }\textbf {\bibinfo
  {volume} {102}},\ \bibinfo {pages} {035015} (\bibinfo {year} {2020})},\
  \Eprint {https://arxiv.org/abs/2004.14515} {arXiv:2004.14515 [hep-ph]}
  \BibitemShut {NoStop}%
\bibitem [{\citenamefont {Aaij}\ \emph {et~al.}(2017)\citenamefont {Aaij} \emph
  {et~al.}}]{Aaij:2016qsm}%
  \BibitemOpen
  \bibfield  {author} {\bibinfo {author} {\bibfnamefont {R.}~\bibnamefont
  {Aaij}} \emph {et~al.} (\bibinfo {collaboration} {LHCb collaboration}),\
  }\bibfield  {title} {\bibinfo {title} {{Search for long-lived scalar
  particles in $B^+ \to K^+ \chi (\mu^+\mu^-)$ decays}},\ }\href
  {https://doi.org/10.1103/PhysRevD.95.071101} {\bibfield  {journal} {\bibinfo
  {journal} {Phys. Rev.}\ }\textbf {\bibinfo {volume} {D95}},\ \bibinfo {pages}
  {071101} (\bibinfo {year} {2017})},\ \Eprint
  {https://arxiv.org/abs/1612.07818} {arXiv:1612.07818 [hep-ex]} \BibitemShut
  {NoStop}%
\bibitem [{\citenamefont {Aaboud}\ \emph
  {et~al.}(2019{\natexlab{a}})\citenamefont {Aaboud} \emph
  {et~al.}}]{Aaboud:2018aqj}%
  \BibitemOpen
  \bibfield  {author} {\bibinfo {author} {\bibfnamefont {M.}~\bibnamefont
  {Aaboud}} \emph {et~al.} (\bibinfo {collaboration} {ATLAS}),\ }\bibfield
  {title} {\bibinfo {title} {{Search for long-lived particles produced in $pp$
  collisions at $\sqrt{s}=13$ TeV that decay into displaced hadronic jets in
  the ATLAS muon spectrometer}},\ }\href
  {https://doi.org/10.1103/PhysRevD.99.052005} {\bibfield  {journal} {\bibinfo
  {journal} {Phys. Rev. D}\ }\textbf {\bibinfo {volume} {99}},\ \bibinfo
  {pages} {052005} (\bibinfo {year} {2019}{\natexlab{a}})},\ \Eprint
  {https://arxiv.org/abs/1811.07370} {arXiv:1811.07370 [hep-ex]} \BibitemShut
  {NoStop}%
\bibitem [{\citenamefont {Sirunyan}\ \emph {et~al.}(2020)\citenamefont
  {Sirunyan} \emph {et~al.}}]{Sirunyan:2020cao}%
  \BibitemOpen
  \bibfield  {author} {\bibinfo {author} {\bibfnamefont {A.~M.}\ \bibnamefont
  {Sirunyan}} \emph {et~al.} (\bibinfo {collaboration} {CMS}),\ }\bibfield
  {title} {\bibinfo {title} {{Search for long-lived particles using displaced
  jets in proton-proton collisions at $\sqrt{s} = $ 13 TeV}},\ }\href@noop {}
  {\  (\bibinfo {year} {2020})},\ \Eprint {https://arxiv.org/abs/2012.01581}
  {arXiv:2012.01581 [hep-ex]} \BibitemShut {NoStop}%
\bibitem [{\citenamefont {Kachanovich}\ \emph {et~al.}(2020)\citenamefont
  {Kachanovich}, \citenamefont {Nierste},\ and\ \citenamefont {Ni\v~sand\v
  zi\'c}}]{Kachanovich:2020yhi}%
  \BibitemOpen
  \bibfield  {author} {\bibinfo {author} {\bibfnamefont {A.}~\bibnamefont
  {Kachanovich}}, \bibinfo {author} {\bibfnamefont {U.}~\bibnamefont
  {Nierste}}, and\ \bibinfo {author} {\bibfnamefont {I.}~\bibnamefont
  {Ni\v~sand\v zi\'c}},\ }\bibfield  {title} {\bibinfo {title} {{Higgs portal
  to dark matter and $B\to K^{(*)}$ decays}},\ }\href
  {https://doi.org/10.1140/epjc/s10052-020-8240-z} {\bibfield  {journal}
  {\bibinfo  {journal} {Eur. Phys. J. C}\ }\textbf {\bibinfo {volume} {80}},\
  \bibinfo {pages} {669} (\bibinfo {year} {2020})},\ \Eprint
  {https://arxiv.org/abs/2003.01788} {arXiv:2003.01788 [hep-ph]} \BibitemShut
  {NoStop}%
\bibitem [{\citenamefont {Feng}\ \emph {et~al.}(2018)\citenamefont {Feng},
  \citenamefont {Galon}, \citenamefont {Kling},\ and\ \citenamefont
  {Trojanowski}}]{Feng:2017vli}%
  \BibitemOpen
  \bibfield  {author} {\bibinfo {author} {\bibfnamefont {J.~L.}\ \bibnamefont
  {Feng}}, \bibinfo {author} {\bibfnamefont {I.}~\bibnamefont {Galon}},
  \bibinfo {author} {\bibfnamefont {F.}~\bibnamefont {Kling}}, and\ \bibinfo
  {author} {\bibfnamefont {S.}~\bibnamefont {Trojanowski}},\ }\bibfield
  {title} {\bibinfo {title} {{Dark Higgs bosons at the ForwArd Search
  ExpeRiment}},\ }\href {https://doi.org/10.1103/PhysRevD.97.055034} {\bibfield
   {journal} {\bibinfo  {journal} {Phys. Rev. D}\ }\textbf {\bibinfo {volume}
  {97}},\ \bibinfo {pages} {055034} (\bibinfo {year} {2018})},\ \Eprint
  {https://arxiv.org/abs/1710.09387} {arXiv:1710.09387 [hep-ph]} \BibitemShut
  {NoStop}%
\bibitem [{\citenamefont {Alpigiani}\ \emph {et~al.}(2020)\citenamefont
  {Alpigiani} \emph {et~al.}}]{Alpigiani:2020tva}%
  \BibitemOpen
  \bibfield  {author} {\bibinfo {author} {\bibfnamefont {C.}~\bibnamefont
  {Alpigiani}} \emph {et~al.} (\bibinfo {collaboration} {MATHUSLA}),\
  }\bibfield  {title} {\bibinfo {title} {{An Update to the Letter of Intent for
  MATHUSLA: Search for Long-Lived Particles at the HL-LHC}},\ }\href@noop {} {\
   (\bibinfo {year} {2020})},\ \Eprint {https://arxiv.org/abs/2009.01693}
  {arXiv:2009.01693 [physics.ins-det]} \BibitemShut {NoStop}%
\bibitem [{\citenamefont {Aielli}\ \emph {et~al.}(2019)\citenamefont {Aielli}
  \emph {et~al.}}]{Aielli:2019ivi}%
  \BibitemOpen
  \bibfield  {author} {\bibinfo {author} {\bibfnamefont {G.}~\bibnamefont
  {Aielli}} \emph {et~al.},\ }\bibfield  {title} {\bibinfo {title} {{Expression
  of Interest for the CODEX-b Detector}},\ }\href@noop {} {\  (\bibinfo {year}
  {2019})},\ \Eprint {https://arxiv.org/abs/1911.00481} {arXiv:1911.00481
  [hep-ex]} \BibitemShut {NoStop}%
\bibitem [{\citenamefont {Aprile}\ \emph {et~al.}(2018)\citenamefont {Aprile}
  \emph {et~al.}}]{Aprile:2018dbl}%
  \BibitemOpen
  \bibfield  {author} {\bibinfo {author} {\bibfnamefont {E.}~\bibnamefont
  {Aprile}} \emph {et~al.} (\bibinfo {collaboration} {XENON}),\ }\bibfield
  {title} {\bibinfo {title} {{Dark Matter Search Results from a One Ton-Year
  Exposure of XENON1T}},\ }\href
  {https://doi.org/10.1103/PhysRevLett.121.111302} {\bibfield  {journal}
  {\bibinfo  {journal} {Phys. Rev. Lett.}\ }\textbf {\bibinfo {volume} {121}},\
  \bibinfo {pages} {111302} (\bibinfo {year} {2018})},\ \Eprint
  {https://arxiv.org/abs/1805.12562} {arXiv:1805.12562 [astro-ph.CO]}
  \BibitemShut {NoStop}%
\bibitem [{\citenamefont {Evans}\ \emph {et~al.}(2018)\citenamefont {Evans},
  \citenamefont {Gori},\ and\ \citenamefont {Shelton}}]{Evans:2017kti}%
  \BibitemOpen
  \bibfield  {author} {\bibinfo {author} {\bibfnamefont {J.~A.}\ \bibnamefont
  {Evans}}, \bibinfo {author} {\bibfnamefont {S.}~\bibnamefont {Gori}}, and\
  \bibinfo {author} {\bibfnamefont {J.}~\bibnamefont {Shelton}},\ }\bibfield
  {title} {\bibinfo {title} {{Looking for the WIMP Next Door}},\ }\href
  {https://doi.org/10.1007/JHEP02(2018)100} {\bibfield  {journal} {\bibinfo
  {journal} {JHEP}\ }\textbf {\bibinfo {volume} {02}},\ \bibinfo {pages}
  {100}},\ \Eprint {https://arxiv.org/abs/1712.03974} {arXiv:1712.03974
  [hep-ph]} \BibitemShut {NoStop}%
\bibitem [{\citenamefont {Aad}\ \emph {et~al.}(2014)\citenamefont {Aad} \emph
  {et~al.}}]{Aad:2014ioa}%
  \BibitemOpen
  \bibfield  {author} {\bibinfo {author} {\bibfnamefont {G.}~\bibnamefont
  {Aad}} \emph {et~al.} (\bibinfo {collaboration} {ATLAS}),\ }\bibfield
  {title} {\bibinfo {title} {{Search for Scalar Diphoton Resonances in the Mass
  Range $65-600$ GeV with the ATLAS Detector in $pp$ Collision Data at
  $\sqrt{s}$ = 8 $TeV$}},\ }\href
  {https://doi.org/10.1103/PhysRevLett.113.171801} {\bibfield  {journal}
  {\bibinfo  {journal} {Phys. Rev. Lett.}\ }\textbf {\bibinfo {volume} {113}},\
  \bibinfo {pages} {171801} (\bibinfo {year} {2014})},\ \Eprint
  {https://arxiv.org/abs/1407.6583} {arXiv:1407.6583 [hep-ex]} \BibitemShut
  {NoStop}%
\bibitem [{\citenamefont {Mariotti}\ \emph {et~al.}(2018)\citenamefont
  {Mariotti}, \citenamefont {Redigolo}, \citenamefont {Sala},\ and\
  \citenamefont {Tobioka}}]{Mariotti:2017vtv}%
  \BibitemOpen
  \bibfield  {author} {\bibinfo {author} {\bibfnamefont {A.}~\bibnamefont
  {Mariotti}}, \bibinfo {author} {\bibfnamefont {D.}~\bibnamefont {Redigolo}},
  \bibinfo {author} {\bibfnamefont {F.}~\bibnamefont {Sala}}, and\ \bibinfo
  {author} {\bibfnamefont {K.}~\bibnamefont {Tobioka}},\ }\bibfield  {title}
  {\bibinfo {title} {{New LHC bound on low-mass diphoton resonances}},\ }\href
  {https://doi.org/10.1016/j.physletb.2018.06.039} {\bibfield  {journal}
  {\bibinfo  {journal} {Phys. Lett. B}\ }\textbf {\bibinfo {volume} {783}},\
  \bibinfo {pages} {13} (\bibinfo {year} {2018})},\ \Eprint
  {https://arxiv.org/abs/1710.01743} {arXiv:1710.01743 [hep-ph]} \BibitemShut
  {NoStop}%
\bibitem [{\citenamefont {Cid~Vidal}\ \emph {et~al.}(2019)\citenamefont
  {Cid~Vidal}, \citenamefont {Mariotti}, \citenamefont {Redigolo},
  \citenamefont {Sala},\ and\ \citenamefont {Tobioka}}]{CidVidal:2018blh}%
  \BibitemOpen
  \bibfield  {author} {\bibinfo {author} {\bibfnamefont {X.}~\bibnamefont
  {Cid~Vidal}}, \bibinfo {author} {\bibfnamefont {A.}~\bibnamefont {Mariotti}},
  \bibinfo {author} {\bibfnamefont {D.}~\bibnamefont {Redigolo}}, \bibinfo
  {author} {\bibfnamefont {F.}~\bibnamefont {Sala}}, and\ \bibinfo {author}
  {\bibfnamefont {K.}~\bibnamefont {Tobioka}},\ }\bibfield  {title} {\bibinfo
  {title} {{New Axion Searches at Flavor Factories}},\ }\href
  {https://doi.org/10.1007/JHEP01(2019)113} {\bibfield  {journal} {\bibinfo
  {journal} {JHEP}\ }\textbf {\bibinfo {volume} {01}},\ \bibinfo {pages}
  {113}},\ \bibinfo {note} {[Erratum: JHEP 06, 141 (2020)]},\ \Eprint
  {https://arxiv.org/abs/1810.09452} {arXiv:1810.09452 [hep-ph]} \BibitemShut
  {NoStop}%
\bibitem [{\citenamefont {Sirunyan}\ \emph {et~al.}(2018)\citenamefont
  {Sirunyan} \emph {et~al.}}]{Sirunyan:2017nvi}%
  \BibitemOpen
  \bibfield  {author} {\bibinfo {author} {\bibfnamefont {A.~M.}\ \bibnamefont
  {Sirunyan}} \emph {et~al.} (\bibinfo {collaboration} {CMS}),\ }\bibfield
  {title} {\bibinfo {title} {{Search for low mass vector resonances decaying
  into quark-antiquark pairs in proton-proton collisions at $ \sqrt{s}=13 $
  TeV}},\ }\href {https://doi.org/10.1007/JHEP01(2018)097} {\bibfield
  {journal} {\bibinfo  {journal} {JHEP}\ }\textbf {\bibinfo {volume} {01}},\
  \bibinfo {pages} {097}},\ \Eprint {https://arxiv.org/abs/1710.00159}
  {arXiv:1710.00159 [hep-ex]} \BibitemShut {NoStop}%
\bibitem [{\citenamefont {Lees}\ \emph {et~al.}(2011)\citenamefont {Lees} \emph
  {et~al.}}]{Lees:2011wb}%
  \BibitemOpen
  \bibfield  {author} {\bibinfo {author} {\bibfnamefont {J.}~\bibnamefont
  {Lees}} \emph {et~al.} (\bibinfo {collaboration} {BaBar}),\ }\bibfield
  {title} {\bibinfo {title} {{Search for hadronic decays of a light Higgs boson
  in the radiative decay $\Upsilon \to \gamma A^0$}},\ }\href
  {https://doi.org/10.1103/PhysRevLett.107.221803} {\bibfield  {journal}
  {\bibinfo  {journal} {Phys. Rev. Lett.}\ }\textbf {\bibinfo {volume} {107}},\
  \bibinfo {pages} {221803} (\bibinfo {year} {2011})},\ \Eprint
  {https://arxiv.org/abs/1108.3549} {arXiv:1108.3549 [hep-ex]} \BibitemShut
  {NoStop}%
\bibitem [{\citenamefont {Bechtle}\ \emph {et~al.}(2014)\citenamefont
  {Bechtle}, \citenamefont {Heinemeyer}, \citenamefont {St\r{a}l},
  \citenamefont {Stefaniak},\ and\ \citenamefont {Weiglein}}]{Bechtle:2014ewa}%
  \BibitemOpen
  \bibfield  {author} {\bibinfo {author} {\bibfnamefont {P.}~\bibnamefont
  {Bechtle}}, \bibinfo {author} {\bibfnamefont {S.}~\bibnamefont {Heinemeyer}},
  \bibinfo {author} {\bibfnamefont {O.}~\bibnamefont {St\r{a}l}}, \bibinfo
  {author} {\bibfnamefont {T.}~\bibnamefont {Stefaniak}}, and\ \bibinfo
  {author} {\bibfnamefont {G.}~\bibnamefont {Weiglein}},\ }\bibfield  {title}
  {\bibinfo {title} {{Probing the Standard Model with Higgs signal rates from
  the Tevatron, the LHC and a future ILC}},\ }\href
  {https://doi.org/10.1007/JHEP11(2014)039} {\bibfield  {journal} {\bibinfo
  {journal} {JHEP}\ }\textbf {\bibinfo {volume} {11}},\ \bibinfo {pages}
  {039}},\ \Eprint {https://arxiv.org/abs/1403.1582} {arXiv:1403.1582 [hep-ph]}
  \BibitemShut {NoStop}%
\bibitem [{\citenamefont {Belanger}\ \emph {et~al.}(2013)\citenamefont
  {Belanger}, \citenamefont {Dumont}, \citenamefont {Ellwanger}, \citenamefont
  {Gunion},\ and\ \citenamefont {Kraml}}]{Belanger:2013xza}%
  \BibitemOpen
  \bibfield  {author} {\bibinfo {author} {\bibfnamefont {G.}~\bibnamefont
  {Belanger}}, \bibinfo {author} {\bibfnamefont {B.}~\bibnamefont {Dumont}},
  \bibinfo {author} {\bibfnamefont {U.}~\bibnamefont {Ellwanger}}, \bibinfo
  {author} {\bibfnamefont {J.}~\bibnamefont {Gunion}}, and\ \bibinfo {author}
  {\bibfnamefont {S.}~\bibnamefont {Kraml}},\ }\bibfield  {title} {\bibinfo
  {title} {{Global fit to Higgs signal strengths and couplings and implications
  for extended Higgs sectors}},\ }\href
  {https://doi.org/10.1103/PhysRevD.88.075008} {\bibfield  {journal} {\bibinfo
  {journal} {Phys. Rev. D}\ }\textbf {\bibinfo {volume} {88}},\ \bibinfo
  {pages} {075008} (\bibinfo {year} {2013})},\ \Eprint
  {https://arxiv.org/abs/1306.2941} {arXiv:1306.2941 [hep-ph]} \BibitemShut
  {NoStop}%
\bibitem [{\citenamefont {Frugiuele}\ \emph {et~al.}(2018)\citenamefont
  {Frugiuele}, \citenamefont {Fuchs}, \citenamefont {Perez},\ and\
  \citenamefont {Schlaffer}}]{Frugiuele:2018coc}%
  \BibitemOpen
  \bibfield  {author} {\bibinfo {author} {\bibfnamefont {C.}~\bibnamefont
  {Frugiuele}}, \bibinfo {author} {\bibfnamefont {E.}~\bibnamefont {Fuchs}},
  \bibinfo {author} {\bibfnamefont {G.}~\bibnamefont {Perez}}, and\ \bibinfo
  {author} {\bibfnamefont {M.}~\bibnamefont {Schlaffer}},\ }\bibfield  {title}
  {\bibinfo {title} {{Relaxion and light (pseudo)scalars at the HL-LHC and
  lepton colliders}},\ }\href {https://doi.org/10.1007/JHEP10(2018)151}
  {\bibfield  {journal} {\bibinfo  {journal} {JHEP}\ }\textbf {\bibinfo
  {volume} {10}},\ \bibinfo {pages} {151}},\ \Eprint
  {https://arxiv.org/abs/1807.10842} {arXiv:1807.10842 [hep-ph]} \BibitemShut
  {NoStop}%
\bibitem [{\citenamefont {Spira}(1998)}]{Spira:1997dg}%
  \BibitemOpen
  \bibfield  {author} {\bibinfo {author} {\bibfnamefont {M.}~\bibnamefont
  {Spira}},\ }\bibfield  {title} {\bibinfo {title} {{QCD effects in Higgs
  physics}},\ }\href
  {https://doi.org/10.1002/(SICI)1521-3978(199804)46:3<203::AID-PROP203>3.0.CO;2-4}
  {\bibfield  {journal} {\bibinfo  {journal} {Fortsch. Phys.}\ }\textbf
  {\bibinfo {volume} {46}},\ \bibinfo {pages} {203} (\bibinfo {year} {1998})},\
  \Eprint {https://arxiv.org/abs/hep-ph/9705337} {arXiv:hep-ph/9705337}
  \BibitemShut {NoStop}%
\bibitem [{\citenamefont {Winkler}(2019)}]{Winkler:2018qyg}%
  \BibitemOpen
  \bibfield  {author} {\bibinfo {author} {\bibfnamefont {M.~W.}\ \bibnamefont
  {Winkler}},\ }\bibfield  {title} {\bibinfo {title} {{Decay and detection of a
  light scalar boson mixing with the Higgs boson}},\ }\href
  {https://doi.org/10.1103/PhysRevD.99.015018} {\bibfield  {journal} {\bibinfo
  {journal} {Phys. Rev.}\ }\textbf {\bibinfo {volume} {D99}},\ \bibinfo {pages}
  {015018} (\bibinfo {year} {2019})},\ \Eprint
  {https://arxiv.org/abs/1809.01876} {arXiv:1809.01876 [hep-ph]} \BibitemShut
  {NoStop}%
\bibitem [{\citenamefont {Banerjee}\ \emph {et~al.}(2020)\citenamefont
  {Banerjee}, \citenamefont {Kim}, \citenamefont {Matsedonskyi}, \citenamefont
  {Perez},\ and\ \citenamefont {Safronova}}]{Banerjee:2020kww}%
  \BibitemOpen
  \bibfield  {author} {\bibinfo {author} {\bibfnamefont {A.}~\bibnamefont
  {Banerjee}}, \bibinfo {author} {\bibfnamefont {H.}~\bibnamefont {Kim}},
  \bibinfo {author} {\bibfnamefont {O.}~\bibnamefont {Matsedonskyi}}, \bibinfo
  {author} {\bibfnamefont {G.}~\bibnamefont {Perez}}, and\ \bibinfo {author}
  {\bibfnamefont {M.~S.}\ \bibnamefont {Safronova}},\ }\bibfield  {title}
  {\bibinfo {title} {{Probing the Relaxed Relaxion at the Luminosity and
  Precision Frontiers}},\ }\href {https://doi.org/10.1007/JHEP07(2020)153}
  {\bibfield  {journal} {\bibinfo  {journal} {JHEP}\ }\textbf {\bibinfo
  {volume} {07}},\ \bibinfo {pages} {153}},\ \Eprint
  {https://arxiv.org/abs/2004.02899} {arXiv:2004.02899 [hep-ph]} \BibitemShut
  {NoStop}%
\bibitem [{\citenamefont {{R. S. Willey and H. L.
  Yu}}(1982)}]{PhysRevD.26.3287}%
  \BibitemOpen
  \bibfield  {author} {\bibinfo {author} {\bibnamefont {{R. S. Willey and H. L.
  Yu}}},\ }\bibfield  {title} {\bibinfo {title} {Decays
  ${K}^{\ifmmode\pm\else\textpm\fi{}}\ensuremath{\rightarrow}{\ensuremath{\pi}}^{\ifmmode\pm\else\textpm\fi{}}{l}^{+}{l}^{\ensuremath{-}}$
  and limits on the mass of the neutral higgs boson},\ }\href
  {https://doi.org/10.1103/PhysRevD.26.3287} {\bibfield  {journal} {\bibinfo
  {journal} {Phys. Rev. D}\ }\textbf {\bibinfo {volume} {26}},\ \bibinfo
  {pages} {3287} (\bibinfo {year} {1982})}\BibitemShut {NoStop}%
\bibitem [{\citenamefont {{R.S. Chivukula and A. V.
  Manohar}}(1988)}]{Chivukula:1988lo}%
  \BibitemOpen
  \bibfield  {author} {\bibinfo {author} {\bibnamefont {{R.S. Chivukula and A.
  V. Manohar}}},\ }\bibfield  {title} {\bibinfo {title} {Limits on a light
  higgs boson},\ }\href
  {https://doi.org/http://dx.doi.org/10.1016/0370-2693(88)90891-X} {\bibfield
  {journal} {\bibinfo  {journal} {Physics Letters B}\ }\textbf {\bibinfo
  {volume} {207}},\ \bibinfo {pages} {86 } (\bibinfo {year}
  {1988})}\BibitemShut {NoStop}%
\bibitem [{\citenamefont {Grinstein}\ \emph {et~al.}(1988)\citenamefont
  {Grinstein}, \citenamefont {Hall},\ and\ \citenamefont
  {Randall}}]{Grinstein:1988yu}%
  \BibitemOpen
  \bibfield  {author} {\bibinfo {author} {\bibfnamefont {B.}~\bibnamefont
  {Grinstein}}, \bibinfo {author} {\bibfnamefont {L.~J.}\ \bibnamefont {Hall}},
  and\ \bibinfo {author} {\bibfnamefont {L.}~\bibnamefont {Randall}},\
  }\bibfield  {title} {\bibinfo {title} {{Do B meson decays exclude a light
  Higgs?}},\ }\href {https://doi.org/10.1016/0370-2693(88)90916-1} {\bibfield
  {journal} {\bibinfo  {journal} {Phys. Lett.}\ }\textbf {\bibinfo {volume}
  {B211}},\ \bibinfo {pages} {363} (\bibinfo {year} {1988})}\BibitemShut
  {NoStop}%
\bibitem [{\citenamefont {Bird}\ \emph {et~al.}(2004)\citenamefont {Bird},
  \citenamefont {Jackson}, \citenamefont {Kowalewski},\ and\ \citenamefont
  {Pospelov}}]{Bird:2004ts}%
  \BibitemOpen
  \bibfield  {author} {\bibinfo {author} {\bibfnamefont {C.}~\bibnamefont
  {Bird}}, \bibinfo {author} {\bibfnamefont {P.}~\bibnamefont {Jackson}},
  \bibinfo {author} {\bibfnamefont {R.~V.}\ \bibnamefont {Kowalewski}}, and\
  \bibinfo {author} {\bibfnamefont {M.}~\bibnamefont {Pospelov}},\ }\bibfield
  {title} {\bibinfo {title} {{Search for dark matter in b ---\ensuremath{>} s
  transitions with missing energy}},\ }\href
  {https://doi.org/10.1103/PhysRevLett.93.201803} {\bibfield  {journal}
  {\bibinfo  {journal} {Phys. Rev. Lett.}\ }\textbf {\bibinfo {volume} {93}},\
  \bibinfo {pages} {201803} (\bibinfo {year} {2004})},\ \Eprint
  {https://arxiv.org/abs/hep-ph/0401195} {arXiv:hep-ph/0401195} \BibitemShut
  {NoStop}%
\bibitem [{\citenamefont {Boiarska}\ \emph {et~al.}(2019)\citenamefont
  {Boiarska}, \citenamefont {Bondarenko}, \citenamefont {Boyarsky},
  \citenamefont {Gorkavenko}, \citenamefont {Ovchynnikov},\ and\ \citenamefont
  {Sokolenko}}]{Boiarska:2019jym}%
  \BibitemOpen
  \bibfield  {author} {\bibinfo {author} {\bibfnamefont {I.}~\bibnamefont
  {Boiarska}}, \bibinfo {author} {\bibfnamefont {K.}~\bibnamefont
  {Bondarenko}}, \bibinfo {author} {\bibfnamefont {A.}~\bibnamefont
  {Boyarsky}}, \bibinfo {author} {\bibfnamefont {V.}~\bibnamefont
  {Gorkavenko}}, \bibinfo {author} {\bibfnamefont {M.}~\bibnamefont
  {Ovchynnikov}}, and\ \bibinfo {author} {\bibfnamefont {A.}~\bibnamefont
  {Sokolenko}},\ }\bibfield  {title} {\bibinfo {title} {{Phenomenology of
  GeV-scale scalar portal}},\ }\href {https://doi.org/10.1007/JHEP11(2019)162}
  {\bibfield  {journal} {\bibinfo  {journal} {JHEP}\ }\textbf {\bibinfo
  {volume} {11}},\ \bibinfo {pages} {162}},\ \Eprint
  {https://arxiv.org/abs/1904.10447} {arXiv:1904.10447 [hep-ph]} \BibitemShut
  {NoStop}%
\bibitem [{\citenamefont {Aaij}\ \emph {et~al.}(2015)\citenamefont {Aaij} \emph
  {et~al.}}]{Aaij:2015tna}%
  \BibitemOpen
  \bibfield  {author} {\bibinfo {author} {\bibfnamefont {R.}~\bibnamefont
  {Aaij}} \emph {et~al.} (\bibinfo {collaboration} {LHCb}),\ }\bibfield
  {title} {\bibinfo {title} {{Search for hidden-sector bosons in $B^0 \!\to
  K^{*0}\mu^+\mu^-$ decays}},\ }\href
  {https://doi.org/10.1103/PhysRevLett.115.161802} {\bibfield  {journal}
  {\bibinfo  {journal} {Phys. Rev. Lett.}\ }\textbf {\bibinfo {volume} {115}},\
  \bibinfo {pages} {161802} (\bibinfo {year} {2015})},\ \Eprint
  {https://arxiv.org/abs/1508.04094} {arXiv:1508.04094 [hep-ex]} \BibitemShut
  {NoStop}%
\bibitem [{\citenamefont {Gligorov}\ \emph {et~al.}(2018)\citenamefont
  {Gligorov}, \citenamefont {Knapen}, \citenamefont {Papucci},\ and\
  \citenamefont {Robinson}}]{Gligorov:2017nwh}%
  \BibitemOpen
  \bibfield  {author} {\bibinfo {author} {\bibfnamefont {V.~V.}\ \bibnamefont
  {Gligorov}}, \bibinfo {author} {\bibfnamefont {S.}~\bibnamefont {Knapen}},
  \bibinfo {author} {\bibfnamefont {M.}~\bibnamefont {Papucci}}, and\ \bibinfo
  {author} {\bibfnamefont {D.~J.}\ \bibnamefont {Robinson}},\ }\bibfield
  {title} {\bibinfo {title} {{Searching for Long-lived Particles: A Compact
  Detector for Exotics at LHCb}},\ }\href
  {https://doi.org/10.1103/PhysRevD.97.015023} {\bibfield  {journal} {\bibinfo
  {journal} {Phys. Rev. D}\ }\textbf {\bibinfo {volume} {97}},\ \bibinfo
  {pages} {015023} (\bibinfo {year} {2018})},\ \Eprint
  {https://arxiv.org/abs/1708.09395} {arXiv:1708.09395 [hep-ph]} \BibitemShut
  {NoStop}%
\bibitem [{\citenamefont {Alekhin}\ \emph {et~al.}(2016)\citenamefont {Alekhin}
  \emph {et~al.}}]{Alekhin:2015byh}%
  \BibitemOpen
  \bibfield  {author} {\bibinfo {author} {\bibfnamefont {S.}~\bibnamefont
  {Alekhin}} \emph {et~al.},\ }\bibfield  {title} {\bibinfo {title} {{A
  facility to Search for Hidden Particles at the CERN SPS: the SHiP physics
  case}},\ }\href {https://doi.org/10.1088/0034-4885/79/12/124201} {\bibfield
  {journal} {\bibinfo  {journal} {Rept. Prog. Phys.}\ }\textbf {\bibinfo
  {volume} {79}},\ \bibinfo {pages} {124201} (\bibinfo {year} {2016})},\
  \Eprint {https://arxiv.org/abs/1504.04855} {arXiv:1504.04855 [hep-ph]}
  \BibitemShut {NoStop}%
\bibitem [{\citenamefont {Fuchs}\ \emph {et~al.}(2020)\citenamefont {Fuchs},
  \citenamefont {Matsedonskyi}, \citenamefont {Savoray},\ and\ \citenamefont
  {Schlaffer}}]{Fuchs:2020cmm}%
  \BibitemOpen
  \bibfield  {author} {\bibinfo {author} {\bibfnamefont {E.}~\bibnamefont
  {Fuchs}}, \bibinfo {author} {\bibfnamefont {O.}~\bibnamefont {Matsedonskyi}},
  \bibinfo {author} {\bibfnamefont {I.}~\bibnamefont {Savoray}}, and\ \bibinfo
  {author} {\bibfnamefont {M.}~\bibnamefont {Schlaffer}},\ }\bibfield  {title}
  {\bibinfo {title} {{Collider searches of scalar singlets across lifetimes}},\
  }\href@noop {} {\  (\bibinfo {year} {2020})},\ \Eprint
  {https://arxiv.org/abs/2008.12773} {arXiv:2008.12773 [hep-ph]} \BibitemShut
  {NoStop}%
\bibitem [{\citenamefont {Liu}\ \emph {et~al.}(2019)\citenamefont {Liu},
  \citenamefont {Liu},\ and\ \citenamefont {Wang}}]{Liu:2018wte}%
  \BibitemOpen
  \bibfield  {author} {\bibinfo {author} {\bibfnamefont {J.}~\bibnamefont
  {Liu}}, \bibinfo {author} {\bibfnamefont {Z.}~\bibnamefont {Liu}}, and\
  \bibinfo {author} {\bibfnamefont {L.-T.}\ \bibnamefont {Wang}},\ }\bibfield
  {title} {\bibinfo {title} {{Enhancing Long-Lived Particles Searches at the
  LHC with Precision Timing Information}},\ }\href
  {https://doi.org/10.1103/PhysRevLett.122.131801} {\bibfield  {journal}
  {\bibinfo  {journal} {Phys. Rev. Lett.}\ }\textbf {\bibinfo {volume} {122}},\
  \bibinfo {pages} {131801} (\bibinfo {year} {2019})},\ \Eprint
  {https://arxiv.org/abs/1805.05957} {arXiv:1805.05957 [hep-ph]} \BibitemShut
  {NoStop}%
\bibitem [{\citenamefont {Liu}\ \emph {et~al.}(2020)\citenamefont {Liu},
  \citenamefont {Liu}, \citenamefont {Wang},\ and\ \citenamefont
  {Wang}}]{Liu:2020vur}%
  \BibitemOpen
  \bibfield  {author} {\bibinfo {author} {\bibfnamefont {J.}~\bibnamefont
  {Liu}}, \bibinfo {author} {\bibfnamefont {Z.}~\bibnamefont {Liu}}, \bibinfo
  {author} {\bibfnamefont {L.-T.}\ \bibnamefont {Wang}}, and\ \bibinfo {author}
  {\bibfnamefont {X.-P.}\ \bibnamefont {Wang}},\ }\bibfield  {title} {\bibinfo
  {title} {{Enhancing Sensitivities to Long-lived Particles with High
  Granularity Calorimeters at the LHC}},\ }\href
  {https://doi.org/10.1007/JHEP11(2020)066} {\bibfield  {journal} {\bibinfo
  {journal} {JHEP}\ }\textbf {\bibinfo {volume} {11}},\ \bibinfo {pages}
  {066}},\ \Eprint {https://arxiv.org/abs/2005.10836} {arXiv:2005.10836
  [hep-ph]} \BibitemShut {NoStop}%
\bibitem [{CMS(2019)}]{CMS-PAS-EXO-19-018}%
  \BibitemOpen
  \href {https://cds.cern.ch/record/2684861} {\emph {\bibinfo {title} {{Search
  for a narrow resonance decaying to a pair of muons in proton-proton
  collisions at 13 TeV}}}},\ \bibinfo {type} {Tech. Rep.}\ \bibinfo {number}
  {CMS-PAS-EXO-19-018}\ (\bibinfo  {institution} {CERN},\ \bibinfo {address}
  {Geneva},\ \bibinfo {year} {2019})\BibitemShut {NoStop}%
\bibitem [{\citenamefont {Aaboud}\ \emph
  {et~al.}(2019{\natexlab{b}})\citenamefont {Aaboud} \emph
  {et~al.}}]{Aaboud:2018esj}%
  \BibitemOpen
  \bibfield  {author} {\bibinfo {author} {\bibfnamefont {M.}~\bibnamefont
  {Aaboud}} \emph {et~al.} (\bibinfo {collaboration} {ATLAS}),\ }\bibfield
  {title} {\bibinfo {title} {{Search for Higgs boson decays into a pair of
  light bosons in the $bb\mu\mu$ final state in $pp$ collision at $\sqrt{s} =
  $13 TeV with the ATLAS detector}},\ }\href
  {https://doi.org/10.1016/j.physletb.2018.10.073} {\bibfield  {journal}
  {\bibinfo  {journal} {Phys. Lett. B}\ }\textbf {\bibinfo {volume} {790}},\
  \bibinfo {pages} {1} (\bibinfo {year} {2019}{\natexlab{b}})},\ \Eprint
  {https://arxiv.org/abs/1807.00539} {arXiv:1807.00539 [hep-ex]} \BibitemShut
  {NoStop}%
\bibitem [{\citenamefont {Bellazzini}\ \emph {et~al.}(2014)\citenamefont
  {Bellazzini}, \citenamefont {Csaki}, \citenamefont {Hubisz}, \citenamefont
  {Serra},\ and\ \citenamefont {Terning}}]{Bellazzini:2013fga}%
  \BibitemOpen
  \bibfield  {author} {\bibinfo {author} {\bibfnamefont {B.}~\bibnamefont
  {Bellazzini}}, \bibinfo {author} {\bibfnamefont {C.}~\bibnamefont {Csaki}},
  \bibinfo {author} {\bibfnamefont {J.}~\bibnamefont {Hubisz}}, \bibinfo
  {author} {\bibfnamefont {J.}~\bibnamefont {Serra}}, and\ \bibinfo {author}
  {\bibfnamefont {J.}~\bibnamefont {Terning}},\ }\bibfield  {title} {\bibinfo
  {title} {{A Naturally Light Dilaton and a Small Cosmological Constant}},\
  }\href {https://doi.org/10.1140/epjc/s10052-014-2790-x} {\bibfield  {journal}
  {\bibinfo  {journal} {Eur. Phys. J. C}\ }\textbf {\bibinfo {volume} {74}},\
  \bibinfo {pages} {2790} (\bibinfo {year} {2014})},\ \Eprint
  {https://arxiv.org/abs/1305.3919} {arXiv:1305.3919 [hep-th]} \BibitemShut
  {NoStop}%
\bibitem [{\citenamefont {Coradeschi}\ \emph {et~al.}(2013)\citenamefont
  {Coradeschi}, \citenamefont {Lodone}, \citenamefont {Pappadopulo},
  \citenamefont {Rattazzi},\ and\ \citenamefont {Vitale}}]{Coradeschi:2013gda}%
  \BibitemOpen
  \bibfield  {author} {\bibinfo {author} {\bibfnamefont {F.}~\bibnamefont
  {Coradeschi}}, \bibinfo {author} {\bibfnamefont {P.}~\bibnamefont {Lodone}},
  \bibinfo {author} {\bibfnamefont {D.}~\bibnamefont {Pappadopulo}}, \bibinfo
  {author} {\bibfnamefont {R.}~\bibnamefont {Rattazzi}}, and\ \bibinfo {author}
  {\bibfnamefont {L.}~\bibnamefont {Vitale}},\ }\bibfield  {title} {\bibinfo
  {title} {{A naturally light dilaton}},\ }\href
  {https://doi.org/10.1007/JHEP11(2013)057} {\bibfield  {journal} {\bibinfo
  {journal} {JHEP}\ }\textbf {\bibinfo {volume} {11}},\ \bibinfo {pages}
  {057}},\ \Eprint {https://arxiv.org/abs/1306.4601} {arXiv:1306.4601 [hep-th]}
  \BibitemShut {NoStop}%
\bibitem [{\citenamefont {Bellazzini}\ \emph {et~al.}(2013)\citenamefont
  {Bellazzini}, \citenamefont {Csaki}, \citenamefont {Hubisz}, \citenamefont
  {Serra},\ and\ \citenamefont {Terning}}]{Bellazzini:2012vz}%
  \BibitemOpen
  \bibfield  {author} {\bibinfo {author} {\bibfnamefont {B.}~\bibnamefont
  {Bellazzini}}, \bibinfo {author} {\bibfnamefont {C.}~\bibnamefont {Csaki}},
  \bibinfo {author} {\bibfnamefont {J.}~\bibnamefont {Hubisz}}, \bibinfo
  {author} {\bibfnamefont {J.}~\bibnamefont {Serra}}, and\ \bibinfo {author}
  {\bibfnamefont {J.}~\bibnamefont {Terning}},\ }\bibfield  {title} {\bibinfo
  {title} {{A Higgslike Dilaton}},\ }\href
  {https://doi.org/10.1140/epjc/s10052-013-2333-x} {\bibfield  {journal}
  {\bibinfo  {journal} {Eur. Phys. J. C}\ }\textbf {\bibinfo {volume} {73}},\
  \bibinfo {pages} {2333} (\bibinfo {year} {2013})},\ \Eprint
  {https://arxiv.org/abs/1209.3299} {arXiv:1209.3299 [hep-ph]} \BibitemShut
  {NoStop}%
\bibitem [{\citenamefont {{Z. Chacko and R. K. Mishra}}(2013)}]{Chacko:2012sy}%
  \BibitemOpen
  \bibfield  {author} {\bibinfo {author} {\bibnamefont {{Z. Chacko and R. K.
  Mishra}}},\ }\bibfield  {title} {\bibinfo {title} {{Effective Theory of a
  Light Dilaton}},\ }\href {https://doi.org/10.1103/PhysRevD.87.115006}
  {\bibfield  {journal} {\bibinfo  {journal} {Phys. Rev. D}\ }\textbf {\bibinfo
  {volume} {87}},\ \bibinfo {pages} {115006} (\bibinfo {year} {2013})},\
  \Eprint {https://arxiv.org/abs/1209.3022} {arXiv:1209.3022 [hep-ph]}
  \BibitemShut {NoStop}%
\bibitem [{\citenamefont {{R.D. Peccei and H. R.
  Quinn}}(1977{\natexlab{a}})}]{Peccei:1977ur}%
  \BibitemOpen
  \bibfield  {author} {\bibinfo {author} {\bibnamefont {{R.D. Peccei and H. R.
  Quinn}}},\ }\bibfield  {title} {\bibinfo {title} {{Constraints Imposed by CP
  Conservation in the Presence of Instantons}},\ }\href
  {https://doi.org/10.1103/PhysRevD.16.1791} {\bibfield  {journal} {\bibinfo
  {journal} {Phys. Rev. D}\ }\textbf {\bibinfo {volume} {16}},\ \bibinfo
  {pages} {1791} (\bibinfo {year} {1977}{\natexlab{a}})}\BibitemShut {NoStop}%
\bibitem [{\citenamefont {{R.D. Peccei and H. R.
  Quinn}}(1977{\natexlab{b}})}]{Peccei:1977hh}%
  \BibitemOpen
  \bibfield  {author} {\bibinfo {author} {\bibnamefont {{R.D. Peccei and H. R.
  Quinn}}},\ }\bibfield  {title} {\bibinfo {title} {{CP Conservation in the
  Presence of Instantons}},\ }\href
  {https://doi.org/10.1103/PhysRevLett.38.1440} {\bibfield  {journal} {\bibinfo
   {journal} {Phys. Rev. Lett.}\ }\textbf {\bibinfo {volume} {38}},\ \bibinfo
  {pages} {1440} (\bibinfo {year} {1977}{\natexlab{b}})}\BibitemShut {NoStop}%
\bibitem [{\citenamefont {Kim}(1979)}]{Kim:1979if}%
  \BibitemOpen
  \bibfield  {author} {\bibinfo {author} {\bibfnamefont {J.~E.}\ \bibnamefont
  {Kim}},\ }\bibfield  {title} {\bibinfo {title} {{Weak Interaction Singlet and
  Strong CP Invariance}},\ }\href {https://doi.org/10.1103/PhysRevLett.43.103}
  {\bibfield  {journal} {\bibinfo  {journal} {Phys. Rev. Lett.}\ }\textbf
  {\bibinfo {volume} {43}},\ \bibinfo {pages} {103} (\bibinfo {year}
  {1979})}\BibitemShut {NoStop}%
\bibitem [{\citenamefont {Shifman}\ \emph {et~al.}(1980)\citenamefont
  {Shifman}, \citenamefont {Vainshtein},\ and\ \citenamefont
  {Zakharov}}]{Shifman:1979if}%
  \BibitemOpen
  \bibfield  {author} {\bibinfo {author} {\bibfnamefont {M.~A.}\ \bibnamefont
  {Shifman}}, \bibinfo {author} {\bibfnamefont {A.}~\bibnamefont {Vainshtein}},
  and\ \bibinfo {author} {\bibfnamefont {V.~I.}\ \bibnamefont {Zakharov}},\
  }\bibfield  {title} {\bibinfo {title} {{Can Confinement Ensure Natural CP
  Invariance of Strong Interactions?}},\ }\href
  {https://doi.org/10.1016/0550-3213(80)90209-6} {\bibfield  {journal}
  {\bibinfo  {journal} {Nucl. Phys. B}\ }\textbf {\bibinfo {volume} {166}},\
  \bibinfo {pages} {493} (\bibinfo {year} {1980})}\BibitemShut {NoStop}%
\bibitem [{\citenamefont {Zhitnitsky}(1980)}]{Zhitnitsky:1980tq}%
  \BibitemOpen
  \bibfield  {author} {\bibinfo {author} {\bibfnamefont {A.}~\bibnamefont
  {Zhitnitsky}},\ }\bibfield  {title} {\bibinfo {title} {{On Possible
  Suppression of the Axion Hadron Interactions. (In Russian)}},\ }\href@noop {}
  {\bibfield  {journal} {\bibinfo  {journal} {Sov. J. Nucl. Phys.}\ }\textbf
  {\bibinfo {volume} {31}},\ \bibinfo {pages} {260} (\bibinfo {year}
  {1980})}\BibitemShut {NoStop}%
\bibitem [{\citenamefont {Dine}\ \emph {et~al.}(1981)\citenamefont {Dine},
  \citenamefont {Fischler},\ and\ \citenamefont {Srednicki}}]{Dine:1981rt}%
  \BibitemOpen
  \bibfield  {author} {\bibinfo {author} {\bibfnamefont {M.}~\bibnamefont
  {Dine}}, \bibinfo {author} {\bibfnamefont {W.}~\bibnamefont {Fischler}}, and\
  \bibinfo {author} {\bibfnamefont {M.}~\bibnamefont {Srednicki}},\ }\bibfield
  {title} {\bibinfo {title} {{A Simple Solution to the Strong CP Problem with a
  Harmless Axion}},\ }\href {https://doi.org/10.1016/0370-2693(81)90590-6}
  {\bibfield  {journal} {\bibinfo  {journal} {Phys. Lett. B}\ }\textbf
  {\bibinfo {volume} {104}},\ \bibinfo {pages} {199} (\bibinfo {year}
  {1981})}\BibitemShut {NoStop}%
\bibitem [{\citenamefont {Chetyrkin}\ \emph {et~al.}(1998)\citenamefont
  {Chetyrkin}, \citenamefont {Kniehl}, \citenamefont {Steinhauser},\ and\
  \citenamefont {Bardeen}}]{Chetyrkin:1998mw}%
  \BibitemOpen
  \bibfield  {author} {\bibinfo {author} {\bibfnamefont {K.}~\bibnamefont
  {Chetyrkin}}, \bibinfo {author} {\bibfnamefont {B.~A.}\ \bibnamefont
  {Kniehl}}, \bibinfo {author} {\bibfnamefont {M.}~\bibnamefont {Steinhauser}},
  and\ \bibinfo {author} {\bibfnamefont {W.~A.}\ \bibnamefont {Bardeen}},\
  }\bibfield  {title} {\bibinfo {title} {{Effective QCD interactions of CP odd
  Higgs bosons at three loops}},\ }\href
  {https://doi.org/10.1016/S0550-3213(98)00594-X} {\bibfield  {journal}
  {\bibinfo  {journal} {Nucl. Phys. B}\ }\textbf {\bibinfo {volume} {535}},\
  \bibinfo {pages} {3} (\bibinfo {year} {1998})},\ \Eprint
  {https://arxiv.org/abs/hep-ph/9807241} {arXiv:hep-ph/9807241} \BibitemShut
  {NoStop}%
\bibitem [{\citenamefont {Alwall}\ \emph {et~al.}(2014)\citenamefont {Alwall},
  \citenamefont {Frederix}, \citenamefont {Frixione}, \citenamefont {Hirschi},
  \citenamefont {Maltoni}, \citenamefont {Mattelaer}, \citenamefont {Shao},
  \citenamefont {Stelzer}, \citenamefont {Torrielli},\ and\ \citenamefont
  {Zaro}}]{Alwall:2014hca}%
  \BibitemOpen
  \bibfield  {author} {\bibinfo {author} {\bibfnamefont {J.}~\bibnamefont
  {Alwall}}, \bibinfo {author} {\bibfnamefont {R.}~\bibnamefont {Frederix}},
  \bibinfo {author} {\bibfnamefont {S.}~\bibnamefont {Frixione}}, \bibinfo
  {author} {\bibfnamefont {V.}~\bibnamefont {Hirschi}}, \bibinfo {author}
  {\bibfnamefont {F.}~\bibnamefont {Maltoni}}, \bibinfo {author} {\bibfnamefont
  {O.}~\bibnamefont {Mattelaer}}, \bibinfo {author} {\bibfnamefont {H.~S.}\
  \bibnamefont {Shao}}, \bibinfo {author} {\bibfnamefont {T.}~\bibnamefont
  {Stelzer}}, \bibinfo {author} {\bibfnamefont {P.}~\bibnamefont {Torrielli}},
  and\ \bibinfo {author} {\bibfnamefont {M.}~\bibnamefont {Zaro}},\ }\bibfield
  {title} {\bibinfo {title} {{The automated computation of tree-level and
  next-to-leading order differential cross sections, and their matching to
  parton shower simulations}},\ }\href
  {https://doi.org/10.1007/JHEP07(2014)079} {\bibfield  {journal} {\bibinfo
  {journal} {JHEP}\ }\textbf {\bibinfo {volume} {07}},\ \bibinfo {pages}
  {079}},\ \Eprint {https://arxiv.org/abs/1405.0301} {arXiv:1405.0301 [hep-ph]}
  \BibitemShut {NoStop}%
\bibitem [{\citenamefont {Artoisenet}\ \emph {et~al.}(2013)\citenamefont
  {Artoisenet} \emph {et~al.}}]{Artoisenet:2013puc}%
  \BibitemOpen
  \bibfield  {author} {\bibinfo {author} {\bibfnamefont {P.}~\bibnamefont
  {Artoisenet}} \emph {et~al.},\ }\bibfield  {title} {\bibinfo {title} {{A
  framework for Higgs characterisation}},\ }\href
  {https://doi.org/10.1007/JHEP11(2013)043} {\bibfield  {journal} {\bibinfo
  {journal} {JHEP}\ }\textbf {\bibinfo {volume} {11}},\ \bibinfo {pages}
  {043}},\ \Eprint {https://arxiv.org/abs/1306.6464} {arXiv:1306.6464 [hep-ph]}
  \BibitemShut {NoStop}%
\bibitem [{\citenamefont {Sjostrand}\ \emph {et~al.}(2015)\citenamefont
  {Sjostrand} \emph {et~al.}}]{Sjostrand:2014zea}%
  \BibitemOpen
  \bibfield  {author} {\bibinfo {author} {\bibfnamefont {T.}~\bibnamefont
  {Sjostrand}} \emph {et~al.},\ }\bibfield  {title} {\bibinfo {title} {{An
  Introduction to PYTHIA 8.2}},\ }\href
  {https://doi.org/10.1016/j.cpc.2015.01.024} {\bibfield  {journal} {\bibinfo
  {journal} {Comput. Phys. Commun.}\ }\textbf {\bibinfo {volume} {191}},\
  \bibinfo {pages} {159} (\bibinfo {year} {2015})},\ \Eprint
  {https://arxiv.org/abs/1410.3012} {arXiv:1410.3012 [hep-ph]} \BibitemShut
  {NoStop}%
\bibitem [{\citenamefont {Batell}\ \emph {et~al.}(2011)\citenamefont {Batell},
  \citenamefont {Pospelov},\ and\ \citenamefont {Ritz}}]{Batell:2009jf}%
  \BibitemOpen
  \bibfield  {author} {\bibinfo {author} {\bibfnamefont {B.}~\bibnamefont
  {Batell}}, \bibinfo {author} {\bibfnamefont {M.}~\bibnamefont {Pospelov}},
  and\ \bibinfo {author} {\bibfnamefont {A.}~\bibnamefont {Ritz}},\ }\bibfield
  {title} {\bibinfo {title} {{Multi-lepton Signatures of a Hidden Sector in
  Rare B Decays}},\ }\href {https://doi.org/10.1103/PhysRevD.83.054005}
  {\bibfield  {journal} {\bibinfo  {journal} {Phys. Rev. D}\ }\textbf {\bibinfo
  {volume} {83}},\ \bibinfo {pages} {054005} (\bibinfo {year} {2011})},\
  \Eprint {https://arxiv.org/abs/0911.4938} {arXiv:0911.4938 [hep-ph]}
  \BibitemShut {NoStop}%
\bibitem [{\citenamefont {Freytsis}\ \emph {et~al.}(2010)\citenamefont
  {Freytsis}, \citenamefont {Ligeti},\ and\ \citenamefont
  {Thaler}}]{Freytsis:2009ct}%
  \BibitemOpen
  \bibfield  {author} {\bibinfo {author} {\bibfnamefont {M.}~\bibnamefont
  {Freytsis}}, \bibinfo {author} {\bibfnamefont {Z.}~\bibnamefont {Ligeti}},
  and\ \bibinfo {author} {\bibfnamefont {J.}~\bibnamefont {Thaler}},\
  }\bibfield  {title} {\bibinfo {title} {{Constraining the Axion Portal with $B
  \to K l^+ l^-$}},\ }\href {https://doi.org/10.1103/PhysRevD.81.034001}
  {\bibfield  {journal} {\bibinfo  {journal} {Phys. Rev. D}\ }\textbf {\bibinfo
  {volume} {81}},\ \bibinfo {pages} {034001} (\bibinfo {year} {2010})},\
  \Eprint {https://arxiv.org/abs/0911.5355} {arXiv:0911.5355 [hep-ph]}
  \BibitemShut {NoStop}%
\bibitem [{\citenamefont {Aloni}\ \emph {et~al.}(2019)\citenamefont {Aloni},
  \citenamefont {Soreq},\ and\ \citenamefont {Williams}}]{Aloni:2018vki}%
  \BibitemOpen
  \bibfield  {author} {\bibinfo {author} {\bibfnamefont {D.}~\bibnamefont
  {Aloni}}, \bibinfo {author} {\bibfnamefont {Y.}~\bibnamefont {Soreq}}, and\
  \bibinfo {author} {\bibfnamefont {M.}~\bibnamefont {Williams}},\ }\bibfield
  {title} {\bibinfo {title} {{Coupling QCD-Scale Axionlike Particles to
  Gluons}},\ }\href {https://doi.org/10.1103/PhysRevLett.123.031803} {\bibfield
   {journal} {\bibinfo  {journal} {Phys. Rev. Lett.}\ }\textbf {\bibinfo
  {volume} {123}},\ \bibinfo {pages} {031803} (\bibinfo {year} {2019})},\
  \Eprint {https://arxiv.org/abs/1811.03474} {arXiv:1811.03474 [hep-ph]}
  \BibitemShut {NoStop}%
\bibitem [{\citenamefont {Izaguirre}\ \emph {et~al.}(2017)\citenamefont
  {Izaguirre}, \citenamefont {Lin},\ and\ \citenamefont
  {Shuve}}]{Izaguirre:2016dfi}%
  \BibitemOpen
  \bibfield  {author} {\bibinfo {author} {\bibfnamefont {E.}~\bibnamefont
  {Izaguirre}}, \bibinfo {author} {\bibfnamefont {T.}~\bibnamefont {Lin}}, and\
  \bibinfo {author} {\bibfnamefont {B.}~\bibnamefont {Shuve}},\ }\bibfield
  {title} {\bibinfo {title} {{Searching for Axionlike Particles in
  Flavor-Changing Neutral Current Processes}},\ }\href
  {https://doi.org/10.1103/PhysRevLett.118.111802} {\bibfield  {journal}
  {\bibinfo  {journal} {Phys. Rev. Lett.}\ }\textbf {\bibinfo {volume} {118}},\
  \bibinfo {pages} {111802} (\bibinfo {year} {2017})},\ \Eprint
  {https://arxiv.org/abs/1611.09355} {arXiv:1611.09355 [hep-ph]} \BibitemShut
  {NoStop}%
\bibitem [{\citenamefont {Chacko}\ \emph
  {et~al.}(2006{\natexlab{a}})\citenamefont {Chacko}, \citenamefont {Goh},\
  and\ \citenamefont {Harnik}}]{Chacko:2005pe}%
  \BibitemOpen
  \bibfield  {author} {\bibinfo {author} {\bibfnamefont {Z.}~\bibnamefont
  {Chacko}}, \bibinfo {author} {\bibfnamefont {H.-S.}\ \bibnamefont {Goh}},
  and\ \bibinfo {author} {\bibfnamefont {R.}~\bibnamefont {Harnik}},\
  }\bibfield  {title} {\bibinfo {title} {{The Twin Higgs: Natural electroweak
  breaking from mirror symmetry}},\ }\href
  {https://doi.org/10.1103/PhysRevLett.96.231802} {\bibfield  {journal}
  {\bibinfo  {journal} {Phys. Rev. Lett.}\ }\textbf {\bibinfo {volume} {96}},\
  \bibinfo {pages} {231802} (\bibinfo {year} {2006}{\natexlab{a}})},\ \Eprint
  {https://arxiv.org/abs/hep-ph/0506256} {arXiv:hep-ph/0506256 [hep-ph]}
  \BibitemShut {NoStop}%
\bibitem [{\citenamefont {Chacko}\ \emph
  {et~al.}(2006{\natexlab{b}})\citenamefont {Chacko}, \citenamefont {Goh},\
  and\ \citenamefont {Harnik}}]{Chacko:2005un}%
  \BibitemOpen
  \bibfield  {author} {\bibinfo {author} {\bibfnamefont {Z.}~\bibnamefont
  {Chacko}}, \bibinfo {author} {\bibfnamefont {H.-S.}\ \bibnamefont {Goh}},
  and\ \bibinfo {author} {\bibfnamefont {R.}~\bibnamefont {Harnik}},\
  }\bibfield  {title} {\bibinfo {title} {{A Twin Higgs model from left-right
  symmetry}},\ }\href {https://doi.org/10.1088/1126-6708/2006/01/108}
  {\bibfield  {journal} {\bibinfo  {journal} {JHEP}\ }\textbf {\bibinfo
  {volume} {01}},\ \bibinfo {pages} {108}},\ \Eprint
  {https://arxiv.org/abs/hep-ph/0512088} {arXiv:hep-ph/0512088 [hep-ph]}
  \BibitemShut {NoStop}%
\bibitem [{\citenamefont {Burdman}\ \emph {et~al.}(2007)\citenamefont
  {Burdman}, \citenamefont {Chacko}, \citenamefont {Goh},\ and\ \citenamefont
  {Harnik}}]{Burdman:2006tz}%
  \BibitemOpen
  \bibfield  {author} {\bibinfo {author} {\bibfnamefont {G.}~\bibnamefont
  {Burdman}}, \bibinfo {author} {\bibfnamefont {Z.}~\bibnamefont {Chacko}},
  \bibinfo {author} {\bibfnamefont {H.-S.}\ \bibnamefont {Goh}}, and\ \bibinfo
  {author} {\bibfnamefont {R.}~\bibnamefont {Harnik}},\ }\bibfield  {title}
  {\bibinfo {title} {{Folded supersymmetry and the LEP paradox}},\ }\href
  {https://doi.org/10.1088/1126-6708/2007/02/009} {\bibfield  {journal}
  {\bibinfo  {journal} {JHEP}\ }\textbf {\bibinfo {volume} {02}},\ \bibinfo
  {pages} {009}},\ \Eprint {https://arxiv.org/abs/hep-ph/0609152}
  {arXiv:hep-ph/0609152 [hep-ph]} \BibitemShut {NoStop}%
\bibitem [{\citenamefont {Cai}\ \emph {et~al.}(2009)\citenamefont {Cai},
  \citenamefont {Cheng},\ and\ \citenamefont {Terning}}]{Cai:2008au}%
  \BibitemOpen
  \bibfield  {author} {\bibinfo {author} {\bibfnamefont {H.}~\bibnamefont
  {Cai}}, \bibinfo {author} {\bibfnamefont {H.-C.}\ \bibnamefont {Cheng}}, and\
  \bibinfo {author} {\bibfnamefont {J.}~\bibnamefont {Terning}},\ }\bibfield
  {title} {\bibinfo {title} {{A Quirky Little Higgs Model}},\ }\href
  {https://doi.org/10.1088/1126-6708/2009/05/045} {\bibfield  {journal}
  {\bibinfo  {journal} {JHEP}\ }\textbf {\bibinfo {volume} {05}},\ \bibinfo
  {pages} {045}},\ \Eprint {https://arxiv.org/abs/0812.0843} {arXiv:0812.0843
  [hep-ph]} \BibitemShut {NoStop}%
\bibitem [{\citenamefont {Craig}\ \emph
  {et~al.}(2015{\natexlab{a}})\citenamefont {Craig}, \citenamefont {Knapen},\
  and\ \citenamefont {Longhi}}]{Craig:2014roa}%
  \BibitemOpen
  \bibfield  {author} {\bibinfo {author} {\bibfnamefont {N.}~\bibnamefont
  {Craig}}, \bibinfo {author} {\bibfnamefont {S.}~\bibnamefont {Knapen}}, and\
  \bibinfo {author} {\bibfnamefont {P.}~\bibnamefont {Longhi}},\ }\bibfield
  {title} {\bibinfo {title} {{The Orbifold Higgs}},\ }\href
  {https://doi.org/10.1007/JHEP03(2015)106} {\bibfield  {journal} {\bibinfo
  {journal} {JHEP}\ }\textbf {\bibinfo {volume} {03}},\ \bibinfo {pages}
  {106}},\ \Eprint {https://arxiv.org/abs/1411.7393} {arXiv:1411.7393 [hep-ph]}
  \BibitemShut {NoStop}%
\bibitem [{\citenamefont {Craig}\ \emph
  {et~al.}(2015{\natexlab{b}})\citenamefont {Craig}, \citenamefont {Knapen},\
  and\ \citenamefont {Longhi}}]{Craig:2014aea}%
  \BibitemOpen
  \bibfield  {author} {\bibinfo {author} {\bibfnamefont {N.}~\bibnamefont
  {Craig}}, \bibinfo {author} {\bibfnamefont {S.}~\bibnamefont {Knapen}}, and\
  \bibinfo {author} {\bibfnamefont {P.}~\bibnamefont {Longhi}},\ }\bibfield
  {title} {\bibinfo {title} {{Neutral Naturalness from Orbifold Higgs
  Models}},\ }\href {https://doi.org/10.1103/PhysRevLett.114.061803} {\bibfield
   {journal} {\bibinfo  {journal} {Phys. Rev. Lett.}\ }\textbf {\bibinfo
  {volume} {114}},\ \bibinfo {pages} {061803} (\bibinfo {year}
  {2015}{\natexlab{b}})},\ \Eprint {https://arxiv.org/abs/1410.6808}
  {arXiv:1410.6808 [hep-ph]} \BibitemShut {NoStop}%
\bibitem [{\citenamefont {Craig}\ \emph
  {et~al.}(2015{\natexlab{c}})\citenamefont {Craig}, \citenamefont {Katz},
  \citenamefont {Strassler},\ and\ \citenamefont {Sundrum}}]{Craig:2015pha}%
  \BibitemOpen
  \bibfield  {author} {\bibinfo {author} {\bibfnamefont {N.}~\bibnamefont
  {Craig}}, \bibinfo {author} {\bibfnamefont {A.}~\bibnamefont {Katz}},
  \bibinfo {author} {\bibfnamefont {M.}~\bibnamefont {Strassler}}, and\
  \bibinfo {author} {\bibfnamefont {R.}~\bibnamefont {Sundrum}},\ }\bibfield
  {title} {\bibinfo {title} {{Naturalness in the Dark at the LHC}},\ }\href
  {https://doi.org/10.1007/JHEP07(2015)105} {\bibfield  {journal} {\bibinfo
  {journal} {JHEP}\ }\textbf {\bibinfo {volume} {07}},\ \bibinfo {pages}
  {105}},\ \Eprint {https://arxiv.org/abs/1501.05310} {arXiv:1501.05310
  [hep-ph]} \BibitemShut {NoStop}%
\bibitem [{\citenamefont {Craig}\ \emph {et~al.}(2016)\citenamefont {Craig},
  \citenamefont {Knapen}, \citenamefont {Longhi},\ and\ \citenamefont
  {Strassler}}]{Craig:2016kue}%
  \BibitemOpen
  \bibfield  {author} {\bibinfo {author} {\bibfnamefont {N.}~\bibnamefont
  {Craig}}, \bibinfo {author} {\bibfnamefont {S.}~\bibnamefont {Knapen}},
  \bibinfo {author} {\bibfnamefont {P.}~\bibnamefont {Longhi}}, and\ \bibinfo
  {author} {\bibfnamefont {M.}~\bibnamefont {Strassler}},\ }\bibfield  {title}
  {\bibinfo {title} {{The Vector-like Twin Higgs}},\ }\href
  {https://doi.org/10.1007/JHEP07(2016)002} {\bibfield  {journal} {\bibinfo
  {journal} {JHEP}\ }\textbf {\bibinfo {volume} {07}},\ \bibinfo {pages}
  {002}},\ \Eprint {https://arxiv.org/abs/1601.07181} {arXiv:1601.07181
  [hep-ph]} \BibitemShut {NoStop}%
\bibitem [{\citenamefont {{M. J. Strassler and K. M.
  Zurek}}(2007)}]{Strassler:2006im}%
  \BibitemOpen
  \bibfield  {author} {\bibinfo {author} {\bibnamefont {{M. J. Strassler and K.
  M. Zurek}}},\ }\bibfield  {title} {\bibinfo {title} {{Echoes of a hidden
  valley at hadron colliders}},\ }\href
  {https://doi.org/10.1016/j.physletb.2007.06.055} {\bibfield  {journal}
  {\bibinfo  {journal} {Phys. Lett.}\ }\textbf {\bibinfo {volume} {B651}},\
  \bibinfo {pages} {374} (\bibinfo {year} {2007})},\ \Eprint
  {https://arxiv.org/abs/hep-ph/0604261} {arXiv:hep-ph/0604261 [hep-ph]}
  \BibitemShut {NoStop}%
\bibitem [{\citenamefont {Han}\ \emph {et~al.}(2008)\citenamefont {Han},
  \citenamefont {Si}, \citenamefont {Zurek},\ and\ \citenamefont
  {Strassler}}]{Han:2007ae}%
  \BibitemOpen
  \bibfield  {author} {\bibinfo {author} {\bibfnamefont {T.}~\bibnamefont
  {Han}}, \bibinfo {author} {\bibfnamefont {Z.}~\bibnamefont {Si}}, \bibinfo
  {author} {\bibfnamefont {K.~M.}\ \bibnamefont {Zurek}}, and\ \bibinfo
  {author} {\bibfnamefont {M.~J.}\ \bibnamefont {Strassler}},\ }\bibfield
  {title} {\bibinfo {title} {{Phenomenology of hidden valleys at hadron
  colliders}},\ }\href {https://doi.org/10.1088/1126-6708/2008/07/008}
  {\bibfield  {journal} {\bibinfo  {journal} {JHEP}\ }\textbf {\bibinfo
  {volume} {07}},\ \bibinfo {pages} {008}},\ \Eprint
  {https://arxiv.org/abs/0712.2041} {arXiv:0712.2041 [hep-ph]} \BibitemShut
  {NoStop}%
\bibitem [{\citenamefont {Alimena}\ \emph {et~al.}(2020)\citenamefont {Alimena}
  \emph {et~al.}}]{Alimena:2019zri}%
  \BibitemOpen
  \bibfield  {author} {\bibinfo {author} {\bibfnamefont {J.}~\bibnamefont
  {Alimena}} \emph {et~al.},\ }\bibfield  {title} {\bibinfo {title} {{Searching
  for Long-Lived Particles beyond the Standard Model at the Large Hadron
  Collider}},\ }\href {https://doi.org/10.1088/1361-6471/ab4574} {\bibfield
  {journal} {\bibinfo  {journal} {J. Phys. G}\ }\textbf {\bibinfo {volume}
  {47}},\ \bibinfo {pages} {090501} (\bibinfo {year} {2020})},\ \Eprint
  {https://arxiv.org/abs/1903.04497} {arXiv:1903.04497 [hep-ex]} \BibitemShut
  {NoStop}%
\bibitem [{\citenamefont {{D. Curtin and C. B.
  Verhaaren}}(2015)}]{Curtin:2015fna}%
  \BibitemOpen
  \bibfield  {author} {\bibinfo {author} {\bibnamefont {{D. Curtin and C. B.
  Verhaaren}}},\ }\bibfield  {title} {\bibinfo {title} {{Discovering Uncolored
  Naturalness in Exotic Higgs Decays}},\ }\href
  {https://doi.org/10.1007/JHEP12(2015)072} {\bibfield  {journal} {\bibinfo
  {journal} {JHEP}\ }\textbf {\bibinfo {volume} {12}},\ \bibinfo {pages}
  {072}},\ \Eprint {https://arxiv.org/abs/1506.06141} {arXiv:1506.06141
  [hep-ph]} \BibitemShut {NoStop}%
\bibitem [{\citenamefont {Alipour-Fard}\ \emph {et~al.}(2020)\citenamefont
  {Alipour-Fard}, \citenamefont {Craig}, \citenamefont {Gori}, \citenamefont
  {Koren},\ and\ \citenamefont {Redigolo}}]{Alipour-fard:2018mre}%
  \BibitemOpen
  \bibfield  {author} {\bibinfo {author} {\bibfnamefont {S.}~\bibnamefont
  {Alipour-Fard}}, \bibinfo {author} {\bibfnamefont {N.}~\bibnamefont {Craig}},
  \bibinfo {author} {\bibfnamefont {S.}~\bibnamefont {Gori}}, \bibinfo {author}
  {\bibfnamefont {S.}~\bibnamefont {Koren}}, and\ \bibinfo {author}
  {\bibfnamefont {D.}~\bibnamefont {Redigolo}},\ }\bibfield  {title} {\bibinfo
  {title} {{The second Higgs at the lifetime frontier}},\ }\href
  {https://doi.org/10.1007/JHEP07(2020)029} {\bibfield  {journal} {\bibinfo
  {journal} {JHEP}\ }\textbf {\bibinfo {volume} {07}},\ \bibinfo {pages}
  {029}},\ \Eprint {https://arxiv.org/abs/1812.09315} {arXiv:1812.09315
  [hep-ph]} \BibitemShut {NoStop}%
\bibitem [{\citenamefont {{C. J. Morningstar and M. J.
  Peardon}}(1999)}]{Morningstar:1999rf}%
  \BibitemOpen
  \bibfield  {author} {\bibinfo {author} {\bibnamefont {{C. J. Morningstar and
  M. J. Peardon}}},\ }\bibfield  {title} {\bibinfo {title} {{The Glueball
  spectrum from an anisotropic lattice study}},\ }\href
  {https://doi.org/10.1103/PhysRevD.60.034509} {\bibfield  {journal} {\bibinfo
  {journal} {Phys. Rev. D}\ }\textbf {\bibinfo {volume} {60}},\ \bibinfo
  {pages} {034509} (\bibinfo {year} {1999})},\ \Eprint
  {https://arxiv.org/abs/hep-lat/9901004} {arXiv:hep-lat/9901004} \BibitemShut
  {NoStop}%
\bibitem [{\citenamefont {Cacciapaglia}\ \emph {et~al.}(2018)\citenamefont
  {Cacciapaglia}, \citenamefont {Ferretti}, \citenamefont {Flacke},\ and\
  \citenamefont {Serodio}}]{Cacciapaglia:2017iws}%
  \BibitemOpen
  \bibfield  {author} {\bibinfo {author} {\bibfnamefont {G.}~\bibnamefont
  {Cacciapaglia}}, \bibinfo {author} {\bibfnamefont {G.}~\bibnamefont
  {Ferretti}}, \bibinfo {author} {\bibfnamefont {T.}~\bibnamefont {Flacke}},
  and\ \bibinfo {author} {\bibfnamefont {H.}~\bibnamefont {Serodio}},\
  }\bibfield  {title} {\bibinfo {title} {{Revealing timid pseudo-scalars with
  taus at the LHC}},\ }\href {https://doi.org/10.1140/epjc/s10052-018-6183-4}
  {\bibfield  {journal} {\bibinfo  {journal} {Eur. Phys. J. C}\ }\textbf
  {\bibinfo {volume} {78}},\ \bibinfo {pages} {724} (\bibinfo {year} {2018})},\
  \Eprint {https://arxiv.org/abs/1710.11142} {arXiv:1710.11142 [hep-ph]}
  \BibitemShut {NoStop}%
\bibitem [{\citenamefont {Bauer}\ \emph {et~al.}(2017)\citenamefont {Bauer},
  \citenamefont {Neubert},\ and\ \citenamefont {Thamm}}]{Bauer:2017ris}%
  \BibitemOpen
  \bibfield  {author} {\bibinfo {author} {\bibfnamefont {M.}~\bibnamefont
  {Bauer}}, \bibinfo {author} {\bibfnamefont {M.}~\bibnamefont {Neubert}}, and\
  \bibinfo {author} {\bibfnamefont {A.}~\bibnamefont {Thamm}},\ }\bibfield
  {title} {\bibinfo {title} {{Collider Probes of Axion-Like Particles}},\
  }\href {https://doi.org/10.1007/JHEP12(2017)044} {\bibfield  {journal}
  {\bibinfo  {journal} {JHEP}\ }\textbf {\bibinfo {volume} {12}},\ \bibinfo
  {pages} {044}},\ \Eprint {https://arxiv.org/abs/1708.00443} {arXiv:1708.00443
  [hep-ph]} \BibitemShut {NoStop}%
\bibitem [{\citenamefont {Barnard}\ \emph {et~al.}(2016)\citenamefont
  {Barnard}, \citenamefont {Cox}, \citenamefont {Gherghetta},\ and\
  \citenamefont {Spray}}]{Barnard:2015rba}%
  \BibitemOpen
  \bibfield  {author} {\bibinfo {author} {\bibfnamefont {J.}~\bibnamefont
  {Barnard}}, \bibinfo {author} {\bibfnamefont {P.}~\bibnamefont {Cox}},
  \bibinfo {author} {\bibfnamefont {T.}~\bibnamefont {Gherghetta}}, and\
  \bibinfo {author} {\bibfnamefont {A.}~\bibnamefont {Spray}},\ }\bibfield
  {title} {\bibinfo {title} {{Long-Lived, Colour-Triplet Scalars from
  Unnaturalness}},\ }\href {https://doi.org/10.1007/JHEP03(2016)003} {\bibfield
   {journal} {\bibinfo  {journal} {JHEP}\ }\textbf {\bibinfo {volume} {03}},\
  \bibinfo {pages} {003}},\ \Eprint {https://arxiv.org/abs/1510.06405}
  {arXiv:1510.06405 [hep-ph]} \BibitemShut {NoStop}%
\bibitem [{\citenamefont {Curtin}\ \emph {et~al.}(2019)\citenamefont {Curtin}
  \emph {et~al.}}]{Curtin:2018mvb}%
  \BibitemOpen
  \bibfield  {author} {\bibinfo {author} {\bibfnamefont {D.}~\bibnamefont
  {Curtin}} \emph {et~al.},\ }\bibfield  {title} {\bibinfo {title} {{Long-Lived
  Particles at the Energy Frontier: The MATHUSLA Physics Case}},\ }\href
  {https://doi.org/10.1088/1361-6633/ab28d6} {\bibfield  {journal} {\bibinfo
  {journal} {Rept. Prog. Phys.}\ }\textbf {\bibinfo {volume} {82}},\ \bibinfo
  {pages} {116201} (\bibinfo {year} {2019})},\ \Eprint
  {https://arxiv.org/abs/1806.07396} {arXiv:1806.07396 [hep-ph]} \BibitemShut
  {NoStop}%
\bibitem [{\citenamefont {{H-S Goh and M. Ibe}}(2009)}]{Goh:2008xz}%
  \BibitemOpen
  \bibfield  {author} {\bibinfo {author} {\bibnamefont {{H-S Goh and M.
  Ibe}}},\ }\bibfield  {title} {\bibinfo {title} {{R-axion detection at LHC}},\
  }\href {https://doi.org/10.1088/1126-6708/2009/03/049} {\bibfield  {journal}
  {\bibinfo  {journal} {JHEP}\ }\textbf {\bibinfo {volume} {03}},\ \bibinfo
  {pages} {049}},\ \Eprint {https://arxiv.org/abs/0810.5773} {arXiv:0810.5773
  [hep-ph]} \BibitemShut {NoStop}%
\bibitem [{\citenamefont {Bellazzini}\ \emph {et~al.}(2017)\citenamefont
  {Bellazzini}, \citenamefont {Mariotti}, \citenamefont {Redigolo},
  \citenamefont {Sala},\ and\ \citenamefont {Serra}}]{Bellazzini:2017neg}%
  \BibitemOpen
  \bibfield  {author} {\bibinfo {author} {\bibfnamefont {B.}~\bibnamefont
  {Bellazzini}}, \bibinfo {author} {\bibfnamefont {A.}~\bibnamefont
  {Mariotti}}, \bibinfo {author} {\bibfnamefont {D.}~\bibnamefont {Redigolo}},
  \bibinfo {author} {\bibfnamefont {F.}~\bibnamefont {Sala}}, and\ \bibinfo
  {author} {\bibfnamefont {J.}~\bibnamefont {Serra}},\ }\bibfield  {title}
  {\bibinfo {title} {{$R$-axion at colliders}},\ }\href
  {https://doi.org/10.1103/PhysRevLett.119.141804} {\bibfield  {journal}
  {\bibinfo  {journal} {Phys. Rev. Lett.}\ }\textbf {\bibinfo {volume} {119}},\
  \bibinfo {pages} {141804} (\bibinfo {year} {2017})},\ \Eprint
  {https://arxiv.org/abs/1702.02152} {arXiv:1702.02152 [hep-ph]} \BibitemShut
  {NoStop}%
\bibitem [{\citenamefont {{Z. Komargodski and D.
  Shih}}(2009)}]{Komargodski:2009jf}%
  \BibitemOpen
  \bibfield  {author} {\bibinfo {author} {\bibnamefont {{Z. Komargodski and D.
  Shih}}},\ }\bibfield  {title} {\bibinfo {title} {{Notes on SUSY and
  R-Symmetry Breaking in Wess-Zumino Models}},\ }\href
  {https://doi.org/10.1088/1126-6708/2009/04/093} {\bibfield  {journal}
  {\bibinfo  {journal} {JHEP}\ }\textbf {\bibinfo {volume} {04}},\ \bibinfo
  {pages} {093}},\ \Eprint {https://arxiv.org/abs/0902.0030} {arXiv:0902.0030
  [hep-th]} \BibitemShut {NoStop}%
\bibitem [{\citenamefont {{G.F. Giudice and R.
  Rattazzi}}(1999)}]{Giudice:1998bp}%
  \BibitemOpen
  \bibfield  {author} {\bibinfo {author} {\bibnamefont {{G.F. Giudice and R.
  Rattazzi}}},\ }\bibfield  {title} {\bibinfo {title} {{Theories with gauge
  mediated supersymmetry breaking}},\ }\href
  {https://doi.org/10.1016/S0370-1573(99)00042-3} {\bibfield  {journal}
  {\bibinfo  {journal} {Phys. Rept.}\ }\textbf {\bibinfo {volume} {322}},\
  \bibinfo {pages} {419} (\bibinfo {year} {1999})},\ \Eprint
  {https://arxiv.org/abs/hep-ph/9801271} {arXiv:hep-ph/9801271} \BibitemShut
  {NoStop}%
\bibitem [{\citenamefont {Bechtle}\ \emph {et~al.}(2016)\citenamefont {Bechtle}
  \emph {et~al.}}]{Bechtle:2015nua}%
  \BibitemOpen
  \bibfield  {author} {\bibinfo {author} {\bibfnamefont {P.}~\bibnamefont
  {Bechtle}} \emph {et~al.},\ }\bibfield  {title} {\bibinfo {title} {{Killing
  the cMSSM softly}},\ }\href {https://doi.org/10.1140/epjc/s10052-015-3864-0}
  {\bibfield  {journal} {\bibinfo  {journal} {Eur. Phys. J.}\ }\textbf
  {\bibinfo {volume} {C76}},\ \bibinfo {pages} {96} (\bibinfo {year} {2016})},\
  \Eprint {https://arxiv.org/abs/1508.05951} {arXiv:1508.05951 [hep-ph]}
  \BibitemShut {NoStop}%
\bibitem [{\citenamefont {Aaboud}\ \emph {et~al.}(2018)\citenamefont {Aaboud}
  \emph {et~al.}}]{Aaboud:2017leg}%
  \BibitemOpen
  \bibfield  {author} {\bibinfo {author} {\bibfnamefont {M.}~\bibnamefont
  {Aaboud}} \emph {et~al.} (\bibinfo {collaboration} {ATLAS}),\ }\bibfield
  {title} {\bibinfo {title} {{Search for electroweak production of
  supersymmetric states in scenarios with compressed mass spectra at
  $\sqrt{s}=13$ TeV with the ATLAS detector}},\ }\href
  {https://doi.org/10.1103/PhysRevD.97.052010} {\bibfield  {journal} {\bibinfo
  {journal} {Phys. Rev.}\ }\textbf {\bibinfo {volume} {D97}},\ \bibinfo {pages}
  {052010} (\bibinfo {year} {2018})},\ \Eprint
  {https://arxiv.org/abs/1712.08119} {arXiv:1712.08119 [hep-ex]} \BibitemShut
  {NoStop}%
\bibitem [{\citenamefont {Bartl}\ \emph {et~al.}(1989)\citenamefont {Bartl},
  \citenamefont {Majerotto},\ and\ \citenamefont {Oshimo}}]{BARTL1989233}%
  \BibitemOpen
  \bibfield  {author} {\bibinfo {author} {\bibfnamefont {A.}~\bibnamefont
  {Bartl}}, \bibinfo {author} {\bibfnamefont {W.}~\bibnamefont {Majerotto}},
  and\ \bibinfo {author} {\bibfnamefont {N.}~\bibnamefont {Oshimo}},\
  }\bibfield  {title} {\bibinfo {title} {On the production of neutralinos at
  the z and w and their decay into higgs bosons},\ }\href
  {https://doi.org/https://doi.org/10.1016/0370-2693(89)91401-9} {\bibfield
  {journal} {\bibinfo  {journal} {Physics Letters B}\ }\textbf {\bibinfo
  {volume} {216}},\ \bibinfo {pages} {233 } (\bibinfo {year}
  {1989})}\BibitemShut {NoStop}%
\bibitem [{\citenamefont {Helo}\ \emph {et~al.}(2018)\citenamefont {Helo},
  \citenamefont {Hirsch},\ and\ \citenamefont {Wang}}]{Helo:2018qej}%
  \BibitemOpen
  \bibfield  {author} {\bibinfo {author} {\bibfnamefont {J.~C.}\ \bibnamefont
  {Helo}}, \bibinfo {author} {\bibfnamefont {M.}~\bibnamefont {Hirsch}}, and\
  \bibinfo {author} {\bibfnamefont {Z.~S.}\ \bibnamefont {Wang}},\ }\bibfield
  {title} {\bibinfo {title} {{Heavy neutral fermions at the high-luminosity
  LHC}},\ }\href {https://doi.org/10.1007/JHEP07(2018)056} {\bibfield
  {journal} {\bibinfo  {journal} {JHEP}\ }\textbf {\bibinfo {volume} {07}},\
  \bibinfo {pages} {056}},\ \Eprint {https://arxiv.org/abs/1803.02212}
  {arXiv:1803.02212 [hep-ph]} \BibitemShut {NoStop}%
\bibitem [{\citenamefont {Graham}\ \emph {et~al.}(2015)\citenamefont {Graham},
  \citenamefont {Kaplan},\ and\ \citenamefont {Rajendran}}]{Graham:2015cka}%
  \BibitemOpen
  \bibfield  {author} {\bibinfo {author} {\bibfnamefont {P.~W.}\ \bibnamefont
  {Graham}}, \bibinfo {author} {\bibfnamefont {D.~E.}\ \bibnamefont {Kaplan}},
  and\ \bibinfo {author} {\bibfnamefont {S.}~\bibnamefont {Rajendran}},\
  }\bibfield  {title} {\bibinfo {title} {{Cosmological Relaxation of the
  Electroweak Scale}},\ }\href {https://doi.org/10.1103/PhysRevLett.115.221801}
  {\bibfield  {journal} {\bibinfo  {journal} {Phys. Rev. Lett.}\ }\textbf
  {\bibinfo {volume} {115}},\ \bibinfo {pages} {221801} (\bibinfo {year}
  {2015})},\ \Eprint {https://arxiv.org/abs/1504.07551} {arXiv:1504.07551
  [hep-ph]} \BibitemShut {NoStop}%
\bibitem [{\citenamefont {Cs\'aki}\ \emph {et~al.}(2020)\citenamefont
  {Cs\'aki}, \citenamefont {D'Agnolo}, \citenamefont {Geller},\ and\
  \citenamefont {Ismail}}]{Csaki:2020zqz}%
  \BibitemOpen
  \bibfield  {author} {\bibinfo {author} {\bibfnamefont {C.}~\bibnamefont
  {Cs\'aki}}, \bibinfo {author} {\bibfnamefont {R.~T.}\ \bibnamefont
  {D'Agnolo}}, \bibinfo {author} {\bibfnamefont {M.}~\bibnamefont {Geller}},
  and\ \bibinfo {author} {\bibfnamefont {A.}~\bibnamefont {Ismail}},\
  }\bibfield  {title} {\bibinfo {title} {{Crunching Dilaton, Hidden
  Naturalness}},\ }\href@noop {} {\  (\bibinfo {year} {2020})},\ \Eprint
  {https://arxiv.org/abs/2007.14396} {arXiv:2007.14396 [hep-ph]} \BibitemShut
  {NoStop}%
\bibitem [{\citenamefont {Georgi}\ \emph {et~al.}(1986)\citenamefont {Georgi},
  \citenamefont {{B. Kaplan}},\ and\ \citenamefont {Randall}}]{GEORGI198673}%
  \BibitemOpen
  \bibfield  {author} {\bibinfo {author} {\bibfnamefont {H.}~\bibnamefont
  {Georgi}}, \bibinfo {author} {\bibfnamefont {D.}~\bibnamefont {{B. Kaplan}}},
  and\ \bibinfo {author} {\bibfnamefont {L.}~\bibnamefont {Randall}},\
  }\bibfield  {title} {\bibinfo {title} {Manifesting the invisible axion at low
  energies},\ }\href
  {https://doi.org/https://doi.org/10.1016/0370-2693(86)90688-X} {\bibfield
  {journal} {\bibinfo  {journal} {Physics Letters B}\ }\textbf {\bibinfo
  {volume} {169}},\ \bibinfo {pages} {73 } (\bibinfo {year}
  {1986})}\BibitemShut {NoStop}%
\bibitem [{\citenamefont {{G. Raffelt and A. Weiss}}(1995)}]{Raffelt:1994ry}%
  \BibitemOpen
  \bibfield  {author} {\bibinfo {author} {\bibnamefont {{G. Raffelt and A.
  Weiss}}},\ }\bibfield  {title} {\bibinfo {title} {{Red giant bound on the
  axion - electron coupling revisited}},\ }\href
  {https://doi.org/10.1103/PhysRevD.51.1495} {\bibfield  {journal} {\bibinfo
  {journal} {Phys. Rev. D}\ }\textbf {\bibinfo {volume} {51}},\ \bibinfo
  {pages} {1495} (\bibinfo {year} {1995})},\ \Eprint
  {https://arxiv.org/abs/hep-ph/9410205} {arXiv:hep-ph/9410205} \BibitemShut
  {NoStop}%
\bibitem [{\citenamefont {Viaux}\ \emph {et~al.}(2013)\citenamefont {Viaux},
  \citenamefont {Catelan}, \citenamefont {Stetson}, \citenamefont {Raffelt},
  \citenamefont {Redondo}, \citenamefont {Valcarce},\ and\ \citenamefont
  {Weiss}}]{Viaux:2013lha}%
  \BibitemOpen
  \bibfield  {author} {\bibinfo {author} {\bibfnamefont {N.}~\bibnamefont
  {Viaux}}, \bibinfo {author} {\bibfnamefont {M.}~\bibnamefont {Catelan}},
  \bibinfo {author} {\bibfnamefont {P.~B.}\ \bibnamefont {Stetson}}, \bibinfo
  {author} {\bibfnamefont {G.}~\bibnamefont {Raffelt}}, \bibinfo {author}
  {\bibfnamefont {J.}~\bibnamefont {Redondo}}, \bibinfo {author} {\bibfnamefont
  {A.~A.~R.}\ \bibnamefont {Valcarce}}, and\ \bibinfo {author} {\bibfnamefont
  {A.}~\bibnamefont {Weiss}},\ }\bibfield  {title} {\bibinfo {title} {{Neutrino
  and axion bounds from the globular cluster M5 (NGC 5904)}},\ }\href
  {https://doi.org/10.1103/PhysRevLett.111.231301} {\bibfield  {journal}
  {\bibinfo  {journal} {Phys. Rev. Lett.}\ }\textbf {\bibinfo {volume} {111}},\
  \bibinfo {pages} {231301} (\bibinfo {year} {2013})},\ \Eprint
  {https://arxiv.org/abs/1311.1669} {arXiv:1311.1669 [astro-ph.SR]}
  \BibitemShut {NoStop}%
\bibitem [{\citenamefont {Miller~Bertolami}\ \emph {et~al.}(2014)\citenamefont
  {Miller~Bertolami}, \citenamefont {Melendez}, \citenamefont {Althaus},\ and\
  \citenamefont {Isern}}]{Bertolami:2014wua}%
  \BibitemOpen
  \bibfield  {author} {\bibinfo {author} {\bibfnamefont {M.~M.}\ \bibnamefont
  {Miller~Bertolami}}, \bibinfo {author} {\bibfnamefont {B.~E.}\ \bibnamefont
  {Melendez}}, \bibinfo {author} {\bibfnamefont {L.~G.}\ \bibnamefont
  {Althaus}}, and\ \bibinfo {author} {\bibfnamefont {J.}~\bibnamefont
  {Isern}},\ }\bibfield  {title} {\bibinfo {title} {{Revisiting the axion
  bounds from the Galactic white dwarf luminosity function}},\ }\href
  {https://doi.org/10.1088/1475-7516/2014/10/069} {\bibfield  {journal}
  {\bibinfo  {journal} {JCAP}\ }\textbf {\bibinfo {volume} {10}},\ \bibinfo
  {pages} {069}},\ \Eprint {https://arxiv.org/abs/1406.7712} {arXiv:1406.7712
  [hep-ph]} \BibitemShut {NoStop}%
\bibitem [{\citenamefont {Randall}(1992)}]{Randall:1992ut}%
  \BibitemOpen
  \bibfield  {author} {\bibinfo {author} {\bibfnamefont {L.}~\bibnamefont
  {Randall}},\ }\bibfield  {title} {\bibinfo {title} {{Composite axion models
  and Planck scale physics}},\ }\href
  {https://doi.org/10.1016/0370-2693(92)91928-3} {\bibfield  {journal}
  {\bibinfo  {journal} {Phys. Lett. B}\ }\textbf {\bibinfo {volume} {284}},\
  \bibinfo {pages} {77} (\bibinfo {year} {1992})}\BibitemShut {NoStop}%
\bibitem [{\citenamefont {{M. Redi and R. Sato}}(2016)}]{Redi:2016esr}%
  \BibitemOpen
  \bibfield  {author} {\bibinfo {author} {\bibnamefont {{M. Redi and R.
  Sato}}},\ }\bibfield  {title} {\bibinfo {title} {{Composite Accidental
  Axions}},\ }\href {https://doi.org/10.1007/JHEP05(2016)104} {\bibfield
  {journal} {\bibinfo  {journal} {JHEP}\ }\textbf {\bibinfo {volume} {05}},\
  \bibinfo {pages} {104}},\ \Eprint {https://arxiv.org/abs/1602.05427}
  {arXiv:1602.05427 [hep-ph]} \BibitemShut {NoStop}%
\bibitem [{\citenamefont {Di~Luzio}\ \emph {et~al.}(2017)\citenamefont
  {Di~Luzio}, \citenamefont {Nardi},\ and\ \citenamefont
  {Ubaldi}}]{DiLuzio:2017tjx}%
  \BibitemOpen
  \bibfield  {author} {\bibinfo {author} {\bibfnamefont {L.}~\bibnamefont
  {Di~Luzio}}, \bibinfo {author} {\bibfnamefont {E.}~\bibnamefont {Nardi}},
  and\ \bibinfo {author} {\bibfnamefont {L.}~\bibnamefont {Ubaldi}},\
  }\bibfield  {title} {\bibinfo {title} {{Accidental Peccei-Quinn symmetry
  protected to arbitrary order}},\ }\href
  {https://doi.org/10.1103/PhysRevLett.119.011801} {\bibfield  {journal}
  {\bibinfo  {journal} {Phys. Rev. Lett.}\ }\textbf {\bibinfo {volume} {119}},\
  \bibinfo {pages} {011801} (\bibinfo {year} {2017})},\ \Eprint
  {https://arxiv.org/abs/1704.01122} {arXiv:1704.01122 [hep-ph]} \BibitemShut
  {NoStop}%
\bibitem [{\citenamefont {Duerr}\ \emph {et~al.}(2018)\citenamefont {Duerr},
  \citenamefont {Schmidt-Hoberg},\ and\ \citenamefont {Unwin}}]{Duerr:2017amf}%
  \BibitemOpen
  \bibfield  {author} {\bibinfo {author} {\bibfnamefont {M.}~\bibnamefont
  {Duerr}}, \bibinfo {author} {\bibfnamefont {K.}~\bibnamefont
  {Schmidt-Hoberg}}, and\ \bibinfo {author} {\bibfnamefont {J.}~\bibnamefont
  {Unwin}},\ }\bibfield  {title} {\bibinfo {title} {{Protecting the Axion with
  Local Baryon Number}},\ }\href
  {https://doi.org/10.1016/j.physletb.2018.03.054} {\bibfield  {journal}
  {\bibinfo  {journal} {Phys. Lett. B}\ }\textbf {\bibinfo {volume} {780}},\
  \bibinfo {pages} {553} (\bibinfo {year} {2018})},\ \Eprint
  {https://arxiv.org/abs/1712.01841} {arXiv:1712.01841 [hep-ph]} \BibitemShut
  {NoStop}%
\bibitem [{\citenamefont {Rubakov}(1997)}]{Rubakov:1997vp}%
  \BibitemOpen
  \bibfield  {author} {\bibinfo {author} {\bibfnamefont {V.}~\bibnamefont
  {Rubakov}},\ }\bibfield  {title} {\bibinfo {title} {{Grand unification and
  heavy axion}},\ }\href {https://doi.org/10.1134/1.567390} {\bibfield
  {journal} {\bibinfo  {journal} {JETP Lett.}\ }\textbf {\bibinfo {volume}
  {65}},\ \bibinfo {pages} {621} (\bibinfo {year} {1997})},\ \Eprint
  {https://arxiv.org/abs/hep-ph/9703409} {arXiv:hep-ph/9703409} \BibitemShut
  {NoStop}%
\bibitem [{\citenamefont {Berezhiani}\ \emph {et~al.}(2001)\citenamefont
  {Berezhiani}, \citenamefont {Gianfagna},\ and\ \citenamefont
  {Giannotti}}]{Berezhiani:2000gh}%
  \BibitemOpen
  \bibfield  {author} {\bibinfo {author} {\bibfnamefont {Z.}~\bibnamefont
  {Berezhiani}}, \bibinfo {author} {\bibfnamefont {L.}~\bibnamefont
  {Gianfagna}}, and\ \bibinfo {author} {\bibfnamefont {M.}~\bibnamefont
  {Giannotti}},\ }\bibfield  {title} {\bibinfo {title} {{Strong CP problem and
  mirror world: The Weinberg-Wilczek axion revisited}},\ }\href
  {https://doi.org/10.1016/S0370-2693(00)01392-7} {\bibfield  {journal}
  {\bibinfo  {journal} {Phys. Lett. B}\ }\textbf {\bibinfo {volume} {500}},\
  \bibinfo {pages} {286} (\bibinfo {year} {2001})},\ \Eprint
  {https://arxiv.org/abs/hep-ph/0009290} {arXiv:hep-ph/0009290} \BibitemShut
  {NoStop}%
\bibitem [{\citenamefont {Hook}(2015)}]{Hook:2014cda}%
  \BibitemOpen
  \bibfield  {author} {\bibinfo {author} {\bibfnamefont {A.}~\bibnamefont
  {Hook}},\ }\bibfield  {title} {\bibinfo {title} {{Anomalous solutions to the
  strong CP problem}},\ }\href {https://doi.org/10.1103/PhysRevLett.114.141801}
  {\bibfield  {journal} {\bibinfo  {journal} {Phys. Rev. Lett.}\ }\textbf
  {\bibinfo {volume} {114}},\ \bibinfo {pages} {141801} (\bibinfo {year}
  {2015})},\ \Eprint {https://arxiv.org/abs/1411.3325} {arXiv:1411.3325
  [hep-ph]} \BibitemShut {NoStop}%
\bibitem [{\citenamefont {Fukuda}\ \emph {et~al.}(2015)\citenamefont {Fukuda},
  \citenamefont {Harigaya}, \citenamefont {Ibe},\ and\ \citenamefont
  {Yanagida}}]{Fukuda:2015ana}%
  \BibitemOpen
  \bibfield  {author} {\bibinfo {author} {\bibfnamefont {H.}~\bibnamefont
  {Fukuda}}, \bibinfo {author} {\bibfnamefont {K.}~\bibnamefont {Harigaya}},
  \bibinfo {author} {\bibfnamefont {M.}~\bibnamefont {Ibe}}, and\ \bibinfo
  {author} {\bibfnamefont {T.~T.}\ \bibnamefont {Yanagida}},\ }\bibfield
  {title} {\bibinfo {title} {{Model of visible QCD axion}},\ }\href
  {https://doi.org/10.1103/PhysRevD.92.015021} {\bibfield  {journal} {\bibinfo
  {journal} {Phys. Rev. D}\ }\textbf {\bibinfo {volume} {92}},\ \bibinfo
  {pages} {015021} (\bibinfo {year} {2015})},\ \Eprint
  {https://arxiv.org/abs/1504.06084} {arXiv:1504.06084 [hep-ph]} \BibitemShut
  {NoStop}%
\bibitem [{\citenamefont {Dimopoulos}\ \emph {et~al.}(2016)\citenamefont
  {Dimopoulos}, \citenamefont {Hook}, \citenamefont {Huang},\ and\
  \citenamefont {Marques-Tavares}}]{Dimopoulos:2016lvn}%
  \BibitemOpen
  \bibfield  {author} {\bibinfo {author} {\bibfnamefont {S.}~\bibnamefont
  {Dimopoulos}}, \bibinfo {author} {\bibfnamefont {A.}~\bibnamefont {Hook}},
  \bibinfo {author} {\bibfnamefont {J.}~\bibnamefont {Huang}}, and\ \bibinfo
  {author} {\bibfnamefont {G.}~\bibnamefont {Marques-Tavares}},\ }\bibfield
  {title} {\bibinfo {title} {{A collider observable QCD axion}},\ }\href
  {https://doi.org/10.1007/JHEP11(2016)052} {\bibfield  {journal} {\bibinfo
  {journal} {JHEP}\ }\textbf {\bibinfo {volume} {11}},\ \bibinfo {pages}
  {052}},\ \Eprint {https://arxiv.org/abs/1606.03097} {arXiv:1606.03097
  [hep-ph]} \BibitemShut {NoStop}%
\bibitem [{\citenamefont {{B. Holdom and M. E. Peskin}}(1982)}]{Holdom:1982ex}%
  \BibitemOpen
  \bibfield  {author} {\bibinfo {author} {\bibnamefont {{B. Holdom and M. E.
  Peskin}}},\ }\bibfield  {title} {\bibinfo {title} {{Raising the Axion
  Mass}},\ }\href {https://doi.org/10.1016/0550-3213(82)90228-0} {\bibfield
  {journal} {\bibinfo  {journal} {Nucl. Phys. B}\ }\textbf {\bibinfo {volume}
  {208}},\ \bibinfo {pages} {397} (\bibinfo {year} {1982})}\BibitemShut
  {NoStop}%
\bibitem [{\citenamefont {Choi}\ \emph {et~al.}(1988)\citenamefont {Choi},
  \citenamefont {Kim},\ and\ \citenamefont {Sze}}]{Choi:1988sy}%
  \BibitemOpen
  \bibfield  {author} {\bibinfo {author} {\bibfnamefont {K.}~\bibnamefont
  {Choi}}, \bibinfo {author} {\bibfnamefont {C.}~\bibnamefont {Kim}}, and\
  \bibinfo {author} {\bibfnamefont {W.}~\bibnamefont {Sze}},\ }\bibfield
  {title} {\bibinfo {title} {{Mass Renormalization by Instantons and the Strong
  CP Problem}},\ }\href {https://doi.org/10.1103/PhysRevLett.61.794} {\bibfield
   {journal} {\bibinfo  {journal} {Phys. Rev. Lett.}\ }\textbf {\bibinfo
  {volume} {61}},\ \bibinfo {pages} {794} (\bibinfo {year} {1988})}\BibitemShut
  {NoStop}%
\bibitem [{\citenamefont {Holdom}(1985)}]{Holdom:1985vx}%
  \BibitemOpen
  \bibfield  {author} {\bibinfo {author} {\bibfnamefont {B.}~\bibnamefont
  {Holdom}},\ }\bibfield  {title} {\bibinfo {title} {{Strong QCD at
  High-energies and a Heavy Axion}},\ }\href
  {https://doi.org/10.1016/0370-2693(85)90371-5} {\bibfield  {journal}
  {\bibinfo  {journal} {Phys. Lett. B}\ }\textbf {\bibinfo {volume} {154}},\
  \bibinfo {pages} {316} (\bibinfo {year} {1985})},\ \bibinfo {note} {[Erratum:
  Phys.Lett.B 156, 452 (1985)]}\BibitemShut {NoStop}%
\bibitem [{\citenamefont {{M. Dine and N. Seiberg}}(1986)}]{Dine:1986bg}%
  \BibitemOpen
  \bibfield  {author} {\bibinfo {author} {\bibnamefont {{M. Dine and N.
  Seiberg}}},\ }\bibfield  {title} {\bibinfo {title} {{String Theory and the
  Strong CP Problem}},\ }\href {https://doi.org/10.1016/0550-3213(86)90043-X}
  {\bibfield  {journal} {\bibinfo  {journal} {Nucl. Phys. B}\ }\textbf
  {\bibinfo {volume} {273}},\ \bibinfo {pages} {109} (\bibinfo {year}
  {1986})}\BibitemShut {NoStop}%
\bibitem [{\citenamefont {{J. M. Flynn and L. Randall}}(1987)}]{Flynn:1987rs}%
  \BibitemOpen
  \bibfield  {author} {\bibinfo {author} {\bibnamefont {{J. M. Flynn and L.
  Randall}}},\ }\bibfield  {title} {\bibinfo {title} {{A Computation of the
  Small Instanton Contribution to the Axion Potential}},\ }\href
  {https://doi.org/10.1016/0550-3213(87)90089-7} {\bibfield  {journal}
  {\bibinfo  {journal} {Nucl. Phys. B}\ }\textbf {\bibinfo {volume} {293}},\
  \bibinfo {pages} {731} (\bibinfo {year} {1987})}\BibitemShut {NoStop}%
\bibitem [{\citenamefont {{K. Choi and H. D. Kim}}(1999)}]{Choi:1998ep}%
  \BibitemOpen
  \bibfield  {author} {\bibinfo {author} {\bibnamefont {{K. Choi and H. D.
  Kim}}},\ }\bibfield  {title} {\bibinfo {title} {{Small instanton contribution
  to the axion potential in supersymmetric models}},\ }\href
  {https://doi.org/10.1103/PhysRevD.59.072001} {\bibfield  {journal} {\bibinfo
  {journal} {Phys. Rev. D}\ }\textbf {\bibinfo {volume} {59}},\ \bibinfo
  {pages} {072001} (\bibinfo {year} {1999})},\ \Eprint
  {https://arxiv.org/abs/hep-ph/9809286} {arXiv:hep-ph/9809286} \BibitemShut
  {NoStop}%
\bibitem [{\citenamefont {{P. Agrawal and K. Howe}}(2018)}]{Agrawal:2017ksf}%
  \BibitemOpen
  \bibfield  {author} {\bibinfo {author} {\bibnamefont {{P. Agrawal and K.
  Howe}}},\ }\bibfield  {title} {\bibinfo {title} {{Factoring the Strong CP
  Problem}},\ }\href {https://doi.org/10.1007/JHEP12(2018)029} {\bibfield
  {journal} {\bibinfo  {journal} {JHEP}\ }\textbf {\bibinfo {volume} {12}},\
  \bibinfo {pages} {029}},\ \Eprint {https://arxiv.org/abs/1710.04213}
  {arXiv:1710.04213 [hep-ph]} \BibitemShut {NoStop}%
\bibitem [{\citenamefont {Krnjaic}(2016)}]{Krnjaic:2015mbs}%
  \BibitemOpen
  \bibfield  {author} {\bibinfo {author} {\bibfnamefont {G.}~\bibnamefont
  {Krnjaic}},\ }\bibfield  {title} {\bibinfo {title} {{Probing Light Thermal
  Dark-Matter With a Higgs Portal Mediator}},\ }\href
  {https://doi.org/10.1103/PhysRevD.94.073009} {\bibfield  {journal} {\bibinfo
  {journal} {Phys. Rev. D}\ }\textbf {\bibinfo {volume} {94}},\ \bibinfo
  {pages} {073009} (\bibinfo {year} {2016})},\ \Eprint
  {https://arxiv.org/abs/1512.04119} {arXiv:1512.04119 [hep-ph]} \BibitemShut
  {NoStop}%
\bibitem [{\citenamefont {Matsumoto}\ \emph {et~al.}(2019)\citenamefont
  {Matsumoto}, \citenamefont {Tsai},\ and\ \citenamefont
  {Tseng}}]{Matsumoto:2018acr}%
  \BibitemOpen
  \bibfield  {author} {\bibinfo {author} {\bibfnamefont {S.}~\bibnamefont
  {Matsumoto}}, \bibinfo {author} {\bibfnamefont {Y.-L.~S.}\ \bibnamefont
  {Tsai}}, and\ \bibinfo {author} {\bibfnamefont {P.-Y.}\ \bibnamefont
  {Tseng}},\ }\bibfield  {title} {\bibinfo {title} {{Light Fermionic WIMP Dark
  Matter with Light Scalar Mediator}},\ }\href
  {https://doi.org/10.1007/JHEP07(2019)050} {\bibfield  {journal} {\bibinfo
  {journal} {JHEP}\ }\textbf {\bibinfo {volume} {07}},\ \bibinfo {pages}
  {050}},\ \Eprint {https://arxiv.org/abs/1811.03292} {arXiv:1811.03292
  [hep-ph]} \BibitemShut {NoStop}%
\bibitem [{\citenamefont {D'Agnolo}\ \emph {et~al.}(2018)\citenamefont
  {D'Agnolo}, \citenamefont {Mondino}, \citenamefont {Ruderman},\ and\
  \citenamefont {Wang}}]{DAgnolo:2018wcn}%
  \BibitemOpen
  \bibfield  {author} {\bibinfo {author} {\bibfnamefont {R.~T.}\ \bibnamefont
  {D'Agnolo}}, \bibinfo {author} {\bibfnamefont {C.}~\bibnamefont {Mondino}},
  \bibinfo {author} {\bibfnamefont {J.~T.}\ \bibnamefont {Ruderman}}, and\
  \bibinfo {author} {\bibfnamefont {P.-J.}\ \bibnamefont {Wang}},\ }\bibfield
  {title} {\bibinfo {title} {{Exponentially Light Dark Matter from
  Coannihilation}},\ }\href {https://doi.org/10.1007/JHEP08(2018)079}
  {\bibfield  {journal} {\bibinfo  {journal} {JHEP}\ }\textbf {\bibinfo
  {volume} {08}},\ \bibinfo {pages} {079}},\ \Eprint
  {https://arxiv.org/abs/1803.02901} {arXiv:1803.02901 [hep-ph]} \BibitemShut
  {NoStop}%
\bibitem [{\citenamefont {D'Agnolo}\ \emph {et~al.}(2020)\citenamefont
  {D'Agnolo}, \citenamefont {Pappadopulo}, \citenamefont {Ruderman},\ and\
  \citenamefont {Wang}}]{DAgnolo:2019zkf}%
  \BibitemOpen
  \bibfield  {author} {\bibinfo {author} {\bibfnamefont {R.~T.}\ \bibnamefont
  {D'Agnolo}}, \bibinfo {author} {\bibfnamefont {D.}~\bibnamefont
  {Pappadopulo}}, \bibinfo {author} {\bibfnamefont {J.~T.}\ \bibnamefont
  {Ruderman}}, and\ \bibinfo {author} {\bibfnamefont {P.-J.}\ \bibnamefont
  {Wang}},\ }\bibfield  {title} {\bibinfo {title} {{Thermal Relic Targets with
  Exponentially Small Couplings}},\ }\href
  {https://doi.org/10.1103/PhysRevLett.124.151801} {\bibfield  {journal}
  {\bibinfo  {journal} {Phys. Rev. Lett.}\ }\textbf {\bibinfo {volume} {124}},\
  \bibinfo {pages} {151801} (\bibinfo {year} {2020})},\ \Eprint
  {https://arxiv.org/abs/1906.09269} {arXiv:1906.09269 [hep-ph]} \BibitemShut
  {NoStop}%
\bibitem [{\citenamefont {{Y. Cui and R. Sundrum}}(2013)}]{Cui:2012jh}%
  \BibitemOpen
  \bibfield  {author} {\bibinfo {author} {\bibnamefont {{Y. Cui and R.
  Sundrum}}},\ }\bibfield  {title} {\bibinfo {title} {{Baryogenesis for weakly
  interacting massive particles}},\ }\href
  {https://doi.org/10.1103/PhysRevD.87.116013} {\bibfield  {journal} {\bibinfo
  {journal} {Phys. Rev.}\ }\textbf {\bibinfo {volume} {D87}},\ \bibinfo {pages}
  {116013} (\bibinfo {year} {2013})},\ \Eprint
  {https://arxiv.org/abs/1212.2973} {arXiv:1212.2973 [hep-ph]} \BibitemShut
  {NoStop}%
\bibitem [{\citenamefont {{Y. Cui and B. Shuve}}(2015)}]{Cui:2014twa}%
  \BibitemOpen
  \bibfield  {author} {\bibinfo {author} {\bibnamefont {{Y. Cui and B.
  Shuve}}},\ }\bibfield  {title} {\bibinfo {title} {{Probing Baryogenesis with
  Displaced Vertices at the LHC}},\ }\href
  {https://doi.org/10.1007/JHEP02(2015)049} {\bibfield  {journal} {\bibinfo
  {journal} {JHEP}\ }\textbf {\bibinfo {volume} {02}},\ \bibinfo {pages}
  {049}},\ \Eprint {https://arxiv.org/abs/1409.6729} {arXiv:1409.6729 [hep-ph]}
  \BibitemShut {NoStop}%
\bibitem [{\citenamefont {{D. McKeen and A. E.
  Nelson}}(2016)}]{McKeen:2015cuz}%
  \BibitemOpen
  \bibfield  {author} {\bibinfo {author} {\bibnamefont {{D. McKeen and A. E.
  Nelson}}},\ }\bibfield  {title} {\bibinfo {title} {{CP Violating Baryon
  Oscillations}},\ }\href {https://doi.org/10.1103/PhysRevD.94.076002}
  {\bibfield  {journal} {\bibinfo  {journal} {Phys. Rev.}\ }\textbf {\bibinfo
  {volume} {D94}},\ \bibinfo {pages} {076002} (\bibinfo {year} {2016})},\
  \Eprint {https://arxiv.org/abs/1512.05359} {arXiv:1512.05359 [hep-ph]}
  \BibitemShut {NoStop}%
\bibitem [{\citenamefont {Aitken}\ \emph {et~al.}(2017)\citenamefont {Aitken},
  \citenamefont {McKeen}, \citenamefont {Neder},\ and\ \citenamefont
  {Nelson}}]{Aitken:2017wie}%
  \BibitemOpen
  \bibfield  {author} {\bibinfo {author} {\bibfnamefont {K.}~\bibnamefont
  {Aitken}}, \bibinfo {author} {\bibfnamefont {D.}~\bibnamefont {McKeen}},
  \bibinfo {author} {\bibfnamefont {T.}~\bibnamefont {Neder}}, and\ \bibinfo
  {author} {\bibfnamefont {A.~E.}\ \bibnamefont {Nelson}},\ }\bibfield  {title}
  {\bibinfo {title} {{Baryogenesis from Oscillations of Charmed or Beautiful
  Baryons}},\ }\href {https://doi.org/10.1103/PhysRevD.96.075009} {\bibfield
  {journal} {\bibinfo  {journal} {Phys. Rev.}\ }\textbf {\bibinfo {volume}
  {D96}},\ \bibinfo {pages} {075009} (\bibinfo {year} {2017})},\ \Eprint
  {https://arxiv.org/abs/1708.01259} {arXiv:1708.01259 [hep-ph]} \BibitemShut
  {NoStop}%
\bibitem [{\citenamefont {Elor}\ \emph {et~al.}(2019)\citenamefont {Elor},
  \citenamefont {Escudero},\ and\ \citenamefont {Nelson}}]{Elor:2018twp}%
  \BibitemOpen
  \bibfield  {author} {\bibinfo {author} {\bibfnamefont {G.}~\bibnamefont
  {Elor}}, \bibinfo {author} {\bibfnamefont {M.}~\bibnamefont {Escudero}}, and\
  \bibinfo {author} {\bibfnamefont {A.}~\bibnamefont {Nelson}},\ }\bibfield
  {title} {\bibinfo {title} {{Baryogenesis and Dark Matter from $B$ Mesons}},\
  }\href {https://doi.org/10.1103/PhysRevD.99.035031} {\bibfield  {journal}
  {\bibinfo  {journal} {Phys. Rev.}\ }\textbf {\bibinfo {volume} {D99}},\
  \bibinfo {pages} {035031} (\bibinfo {year} {2019})},\ \Eprint
  {https://arxiv.org/abs/1810.00880} {arXiv:1810.00880 [hep-ph]} \BibitemShut
  {NoStop}%
\bibitem [{\citenamefont {{A. E. Nelson and H. Xiao}}(2019)}]{Nelson:2019fln}%
  \BibitemOpen
  \bibfield  {author} {\bibinfo {author} {\bibnamefont {{A. E. Nelson and H.
  Xiao}}},\ }\bibfield  {title} {\bibinfo {title} {{Baryogenesis from B Meson
  Oscillations}},\ }\href {https://doi.org/10.1103/PhysRevD.100.075002}
  {\bibfield  {journal} {\bibinfo  {journal} {Phys. Rev. D}\ }\textbf {\bibinfo
  {volume} {100}},\ \bibinfo {pages} {075002} (\bibinfo {year} {2019})},\
  \Eprint {https://arxiv.org/abs/1901.08141} {arXiv:1901.08141 [hep-ph]}
  \BibitemShut {NoStop}%
\bibitem [{\citenamefont {Alonso-Álvarez}\ \emph {et~al.}(2020)\citenamefont
  {Alonso-Álvarez}, \citenamefont {Elor}, \citenamefont {Nelson},\ and\
  \citenamefont {Xiao}}]{Alonso-Alvarez:2019fym}%
  \BibitemOpen
  \bibfield  {author} {\bibinfo {author} {\bibfnamefont {G.}~\bibnamefont
  {Alonso-Álvarez}}, \bibinfo {author} {\bibfnamefont {G.}~\bibnamefont
  {Elor}}, \bibinfo {author} {\bibfnamefont {A.~E.}\ \bibnamefont {Nelson}},
  and\ \bibinfo {author} {\bibfnamefont {H.}~\bibnamefont {Xiao}},\ }\bibfield
  {title} {\bibinfo {title} {{A Supersymmetric Theory of Baryogenesis and
  Sterile Sneutrino Dark Matter from $B$ Mesons}},\ }\href
  {https://doi.org/10.1007/JHEP03(2020)046} {\bibfield  {journal} {\bibinfo
  {journal} {JHEP}\ }\textbf {\bibinfo {volume} {03}},\ \bibinfo {pages}
  {046}},\ \Eprint {https://arxiv.org/abs/1907.10612} {arXiv:1907.10612
  [hep-ph]} \BibitemShut {NoStop}%
\bibitem [{CER(2020)}]{CERN-LHCC-2020-004}%
  \BibitemOpen
  \href {https://cds.cern.ch/record/2714892} {\emph {\bibinfo {title} {{The
  Phase-2 Upgrade of the CMS Level-1 Trigger}}}},\ \bibinfo {type} {Tech.
  Rep.}\ \bibinfo {number} {CERN-LHCC-2020-004. CMS-TDR-021}\ (\bibinfo
  {institution} {CERN},\ \bibinfo {address} {Geneva},\ \bibinfo {year} {2020})\
  \bibinfo {note} {final version}\BibitemShut {NoStop}%
\bibitem [{\citenamefont {Klein}\ \emph {et~al.}(2017)\citenamefont {Klein}
  \emph {et~al.}}]{Collaboration:2272264}%
  \BibitemOpen
  \bibfield  {author} {\bibinfo {author} {\bibfnamefont {K.}~\bibnamefont
  {Klein}} \emph {et~al.} (\bibinfo {collaboration} {CMS collaboration}),\
  }\href {https://cds.cern.ch/record/2272264} {\emph {\bibinfo {title} {{The
  Phase-2 Upgrade of the CMS Tracker}}}},\ \bibinfo {type} {Tech. Rep.}\
  \bibinfo {number} {CERN-LHCC-2017-009. CMS-TDR-014}\ (\bibinfo  {institution}
  {CERN},\ \bibinfo {address} {Geneva},\ \bibinfo {year} {2017})\BibitemShut
  {NoStop}%
\bibitem [{\citenamefont {Tanabashi}\ \emph {et~al.}(2018)\citenamefont
  {Tanabashi} \emph {et~al.}}]{Tanabashi:2018oca}%
  \BibitemOpen
  \bibfield  {author} {\bibinfo {author} {\bibfnamefont {M.}~\bibnamefont
  {Tanabashi}} \emph {et~al.} (\bibinfo {collaboration} {Particle Data
  Group}),\ }\bibfield  {title} {\bibinfo {title} {{Review of Particle Physics,
  section 33.3}},\ }\href {https://doi.org/10.1103/PhysRevD.98.030001}
  {\bibfield  {journal} {\bibinfo  {journal} {Phys. Rev.}\ }\textbf {\bibinfo
  {volume} {D98}},\ \bibinfo {pages} {030001} (\bibinfo {year}
  {2018})}\BibitemShut {NoStop}%
\bibitem [{\citenamefont {{CMS Collaboration}}(2010)}]{CMS-PAS-TRK-10-003}%
  \BibitemOpen
  \bibfield  {author} {\bibinfo {author} {\bibnamefont {{CMS Collaboration}}},\
  }\href {https://cds.cern.ch/record/1279138} {\emph {\bibinfo {title}
  {{Studies of Tracker Material}}}},\ \bibinfo {type} {Tech. Rep.}\ \bibinfo
  {number} {CMS-PAS-TRK-10-003}\ (\bibinfo {year} {2010})\BibitemShut {NoStop}%
\bibitem [{\citenamefont {Agostinelli}\ \emph {et~al.}(2003)\citenamefont
  {Agostinelli} \emph {et~al.}}]{AGOSTINELLI2003250}%
  \BibitemOpen
  \bibfield  {author} {\bibinfo {author} {\bibfnamefont {S.}~\bibnamefont
  {Agostinelli}} \emph {et~al.},\ }\bibfield  {title} {\bibinfo {title}
  {Geant4—a simulation toolkit},\ }\href
  {https://doi.org/https://doi.org/10.1016/S0168-9002(03)01368-8} {\bibfield
  {journal} {\bibinfo  {journal} {Nuclear Instruments and Methods in Physics
  Research Section A: Accelerators, Spectrometers, Detectors and Associated
  Equipment}\ }\textbf {\bibinfo {volume} {506}},\ \bibinfo {pages} {250 }
  (\bibinfo {year} {2003})}\BibitemShut {NoStop}%
\bibitem [{\citenamefont {Andersson}\ \emph {et~al.}(1987)\citenamefont
  {Andersson}, \citenamefont {Gustafson},\ and\ \citenamefont
  {Nilsson-Almqvist}}]{ANDERSSON1987289}%
  \BibitemOpen
  \bibfield  {author} {\bibinfo {author} {\bibfnamefont {B.}~\bibnamefont
  {Andersson}}, \bibinfo {author} {\bibfnamefont {G.}~\bibnamefont
  {Gustafson}}, and\ \bibinfo {author} {\bibfnamefont {B.}~\bibnamefont
  {Nilsson-Almqvist}},\ }\bibfield  {title} {\bibinfo {title} {A model for
  low-pt hadronic reactions with generalizations to hadron-nucleus and
  nucleus-nucleus collisions},\ }\href
  {https://doi.org/http://dx.doi.org/10.1016/0550-3213(87)90257-4} {\bibfield
  {journal} {\bibinfo  {journal} {Nuclear Physics B}\ }\textbf {\bibinfo
  {volume} {281}},\ \bibinfo {pages} {289 } (\bibinfo {year}
  {1987})}\BibitemShut {NoStop}%
\bibitem [{\citenamefont {Andersson}\ \emph {et~al.}(1996)\citenamefont
  {Andersson}, \citenamefont {Tai},\ and\ \citenamefont {Sa}}]{Andersson1996}%
  \BibitemOpen
  \bibfield  {author} {\bibinfo {author} {\bibfnamefont {B.}~\bibnamefont
  {Andersson}}, \bibinfo {author} {\bibfnamefont {A.}~\bibnamefont {Tai}}, and\
  \bibinfo {author} {\bibfnamefont {B.-H.}\ \bibnamefont {Sa}},\ }\bibfield
  {title} {\bibinfo {title} {Final state interactions in the (nuclear) fritiof
  string interaction scenario},\ }\href {https://doi.org/10.1007/s002880050127}
  {\bibfield  {journal} {\bibinfo  {journal} {Zeitschrift f{\"u}r Physik C
  Particles and Fields}\ }\textbf {\bibinfo {volume} {70}},\ \bibinfo {pages}
  {499} (\bibinfo {year} {1996})}\BibitemShut {NoStop}%
\bibitem [{\citenamefont {{B. Nilsson-Almqvist and E.
  Stenlund}}(1987)}]{NilssonAlmqvist:1986rx}%
  \BibitemOpen
  \bibfield  {author} {\bibinfo {author} {\bibnamefont {{B. Nilsson-Almqvist
  and E. Stenlund}}},\ }\bibfield  {title} {\bibinfo {title} {{Interactions
  Between Hadrons and Nuclei: The Lund Monte Carlo, Fritiof Version 1.6}},\
  }\href {https://doi.org/10.1016/0010-4655(87)90056-7} {\bibfield  {journal}
  {\bibinfo  {journal} {Comput. Phys. Commun.}\ }\textbf {\bibinfo {volume}
  {43}},\ \bibinfo {pages} {387} (\bibinfo {year} {1987})}\BibitemShut
  {NoStop}%
\bibitem [{\citenamefont {{B. Ganhuyag and V.
  Uzhinsky}}(1997)}]{Ganhuyag:1997gz}%
  \BibitemOpen
  \bibfield  {author} {\bibinfo {author} {\bibnamefont {{B. Ganhuyag and V.
  Uzhinsky}}},\ }\bibfield  {title} {\bibinfo {title} {{Modified FRITIOF code:
  Negative charged particle production in high energy nucleus nucleus
  interactions}},\ }\bibfield  {booktitle} {\emph {\bibinfo {booktitle}
  {{Relativistic heavy-ion physics. Proceedings, International School/Workshop
  for Young Physicists, Prague, Czech Republic, September 2-6, 1996}}},\ }\href
  {https://doi.org/10.1023/A:1021296114786} {\bibfield  {journal} {\bibinfo
  {journal} {Czech. J. Phys.}\ }\textbf {\bibinfo {volume} {47}},\ \bibinfo
  {pages} {913} (\bibinfo {year} {1997})}\BibitemShut {NoStop}%
\bibitem [{\citenamefont {Guthrie}\ \emph {et~al.}(1968)\citenamefont
  {Guthrie}, \citenamefont {Alsmiller},\ and\ \citenamefont
  {Bertini}}]{Guthrie:1968ue}%
  \BibitemOpen
  \bibfield  {author} {\bibinfo {author} {\bibfnamefont {M.~P.}\ \bibnamefont
  {Guthrie}}, \bibinfo {author} {\bibfnamefont {R.~G.}\ \bibnamefont
  {Alsmiller}}, and\ \bibinfo {author} {\bibfnamefont {H.~W.}\ \bibnamefont
  {Bertini}},\ }\bibfield  {title} {\bibinfo {title} {{Calculation of the
  capture of negative pions in light elements and comparison with experiments
  pertaining to cancer radiotherapy}},\ }\href
  {https://doi.org/10.1016/0029-554X(68)90054-2} {\bibfield  {journal}
  {\bibinfo  {journal} {Nucl. Instrum. Meth.}\ }\textbf {\bibinfo {volume}
  {66}},\ \bibinfo {pages} {29} (\bibinfo {year} {1968})}\BibitemShut {NoStop}%
\bibitem [{\citenamefont {{H. W. Bertini and M.P.
  Guthrie}}(1971)}]{Bertini:1971xb}%
  \BibitemOpen
  \bibfield  {author} {\bibinfo {author} {\bibnamefont {{H. W. Bertini and M.P.
  Guthrie}}},\ }\bibfield  {title} {\bibinfo {title} {{News item results from
  medium-energy intranuclear-cascade calculation}},\ }\href
  {https://doi.org/10.1016/0375-9474(71)90710-X} {\bibfield  {journal}
  {\bibinfo  {journal} {Nucl. Phys.}\ }\textbf {\bibinfo {volume} {A169}},\
  \bibinfo {pages} {670} (\bibinfo {year} {1971})}\BibitemShut {NoStop}%
\bibitem [{\citenamefont {Karmanov}(1980)}]{Karmanov:1979if}%
  \BibitemOpen
  \bibfield  {author} {\bibinfo {author} {\bibfnamefont {V.~A.}\ \bibnamefont
  {Karmanov}},\ }\bibfield  {title} {\bibinfo {title} {{Light front wave
  function of relativistic composite system in explicitly solvable model}},\
  }\href {https://doi.org/10.1016/0550-3213(80)90204-7} {\bibfield  {journal}
  {\bibinfo  {journal} {Nucl. Phys.}\ }\textbf {\bibinfo {volume} {B166}},\
  \bibinfo {pages} {378} (\bibinfo {year} {1980})}\BibitemShut {NoStop}%
\bibitem [{\citenamefont {Sjostrand}\ \emph {et~al.}(2006)\citenamefont
  {Sjostrand}, \citenamefont {Mrenna},\ and\ \citenamefont
  {Skands}}]{Sjostrand:2006za}%
  \BibitemOpen
  \bibfield  {author} {\bibinfo {author} {\bibfnamefont {T.}~\bibnamefont
  {Sjostrand}}, \bibinfo {author} {\bibfnamefont {S.}~\bibnamefont {Mrenna}},
  and\ \bibinfo {author} {\bibfnamefont {P.~Z.}\ \bibnamefont {Skands}},\
  }\bibfield  {title} {\bibinfo {title} {{PYTHIA 6.4 Physics and Manual}},\
  }\href {https://doi.org/10.1088/1126-6708/2006/05/026} {\bibfield  {journal}
  {\bibinfo  {journal} {JHEP}\ }\textbf {\bibinfo {volume} {05}},\ \bibinfo
  {pages} {026}},\ \Eprint {https://arxiv.org/abs/hep-ph/0603175}
  {arXiv:hep-ph/0603175 [hep-ph]} \BibitemShut {NoStop}%
\bibitem [{\citenamefont {Aaboud}\ \emph {et~al.}(2016)\citenamefont {Aaboud}
  \emph {et~al.}}]{Aaboud:2016mmw}%
  \BibitemOpen
  \bibfield  {author} {\bibinfo {author} {\bibfnamefont {M.}~\bibnamefont
  {Aaboud}} \emph {et~al.} (\bibinfo {collaboration} {ATLAS}),\ }\bibfield
  {title} {\bibinfo {title} {{Measurement of the Inelastic Proton-Proton Cross
  Section at $\sqrt{s} = 13$ TeV with the ATLAS Detector at the LHC}},\ }\href
  {https://doi.org/10.1103/PhysRevLett.117.182002} {\bibfield  {journal}
  {\bibinfo  {journal} {Phys. Rev. Lett.}\ }\textbf {\bibinfo {volume} {117}},\
  \bibinfo {pages} {182002} (\bibinfo {year} {2016})},\ \Eprint
  {https://arxiv.org/abs/1606.02625} {arXiv:1606.02625 [hep-ex]} \BibitemShut
  {NoStop}%
\bibitem [{\citenamefont {Zyla}\ \emph {et~al.}(2020)\citenamefont {Zyla} \emph
  {et~al.}}]{10.1093/ptep/ptaa104}%
  \BibitemOpen
  \bibfield  {author} {\bibinfo {author} {\bibfnamefont {P.~A.}\ \bibnamefont
  {Zyla}} \emph {et~al.},\ }\bibfield  {title} {\bibinfo {title} {{Review of
  Particle Physics}},\ }\bibfield  {journal} {\bibinfo  {journal} {Progress of
  Theoretical and Experimental Physics}\ }\textbf {\bibinfo {volume} {2020}},\
  \href {https://doi.org/10.1093/ptep/ptaa104} {10.1093/ptep/ptaa104} (\bibinfo
  {year} {2020}),\ \bibinfo {note} {083C01}\BibitemShut {NoStop}%
\bibitem [{\citenamefont {Cacciari}\ \emph {et~al.}(1998)\citenamefont
  {Cacciari}, \citenamefont {Greco},\ and\ \citenamefont
  {Nason}}]{Cacciari:1998it}%
  \BibitemOpen
  \bibfield  {author} {\bibinfo {author} {\bibfnamefont {M.}~\bibnamefont
  {Cacciari}}, \bibinfo {author} {\bibfnamefont {M.}~\bibnamefont {Greco}},
  and\ \bibinfo {author} {\bibfnamefont {P.}~\bibnamefont {Nason}},\ }\bibfield
   {title} {\bibinfo {title} {{The P(T) spectrum in heavy flavor
  hadroproduction}},\ }\href {https://doi.org/10.1088/1126-6708/1998/05/007}
  {\bibfield  {journal} {\bibinfo  {journal} {JHEP}\ }\textbf {\bibinfo
  {volume} {05}},\ \bibinfo {pages} {007}},\ \Eprint
  {https://arxiv.org/abs/hep-ph/9803400} {arXiv:hep-ph/9803400} \BibitemShut
  {NoStop}%
\bibitem [{\citenamefont {Cacciari}\ \emph {et~al.}(2001)\citenamefont
  {Cacciari}, \citenamefont {Frixione},\ and\ \citenamefont
  {Nason}}]{Cacciari:2001td}%
  \BibitemOpen
  \bibfield  {author} {\bibinfo {author} {\bibfnamefont {M.}~\bibnamefont
  {Cacciari}}, \bibinfo {author} {\bibfnamefont {S.}~\bibnamefont {Frixione}},
  and\ \bibinfo {author} {\bibfnamefont {P.}~\bibnamefont {Nason}},\ }\bibfield
   {title} {\bibinfo {title} {{The p(T) spectrum in heavy flavor
  photoproduction}},\ }\href {https://doi.org/10.1088/1126-6708/2001/03/006}
  {\bibfield  {journal} {\bibinfo  {journal} {JHEP}\ }\textbf {\bibinfo
  {volume} {03}},\ \bibinfo {pages} {006}},\ \Eprint
  {https://arxiv.org/abs/hep-ph/0102134} {arXiv:hep-ph/0102134} \BibitemShut
  {NoStop}%
\bibitem [{\citenamefont {Cacciari}\ \emph {et~al.}(2012)\citenamefont
  {Cacciari}, \citenamefont {Frixione}, \citenamefont {Houdeau}, \citenamefont
  {Mangano}, \citenamefont {Nason},\ and\ \citenamefont
  {Ridolfi}}]{Cacciari:2012ny}%
  \BibitemOpen
  \bibfield  {author} {\bibinfo {author} {\bibfnamefont {M.}~\bibnamefont
  {Cacciari}}, \bibinfo {author} {\bibfnamefont {S.}~\bibnamefont {Frixione}},
  \bibinfo {author} {\bibfnamefont {N.}~\bibnamefont {Houdeau}}, \bibinfo
  {author} {\bibfnamefont {M.~L.}\ \bibnamefont {Mangano}}, \bibinfo {author}
  {\bibfnamefont {P.}~\bibnamefont {Nason}}, and\ \bibinfo {author}
  {\bibfnamefont {G.}~\bibnamefont {Ridolfi}},\ }\bibfield  {title} {\bibinfo
  {title} {{Theoretical predictions for charm and bottom production at the
  LHC}},\ }\href {https://doi.org/10.1007/JHEP10(2012)137} {\bibfield
  {journal} {\bibinfo  {journal} {JHEP}\ }\textbf {\bibinfo {volume} {10}},\
  \bibinfo {pages} {137}},\ \Eprint {https://arxiv.org/abs/1205.6344}
  {arXiv:1205.6344 [hep-ph]} \BibitemShut {NoStop}%
\bibitem [{\citenamefont {Cacciari}\ \emph {et~al.}(2015)\citenamefont
  {Cacciari}, \citenamefont {Mangano},\ and\ \citenamefont
  {Nason}}]{Cacciari:2015fta}%
  \BibitemOpen
  \bibfield  {author} {\bibinfo {author} {\bibfnamefont {M.}~\bibnamefont
  {Cacciari}}, \bibinfo {author} {\bibfnamefont {M.~L.}\ \bibnamefont
  {Mangano}}, and\ \bibinfo {author} {\bibfnamefont {P.}~\bibnamefont
  {Nason}},\ }\bibfield  {title} {\bibinfo {title} {{Gluon PDF constraints from
  the ratio of forward heavy-quark production at the LHC at $\sqrt{S}=7$ and 13
  TeV}},\ }\href {https://doi.org/10.1140/epjc/s10052-015-3814-x} {\bibfield
  {journal} {\bibinfo  {journal} {Eur.\ Phys.\ J.\ C}\ }\textbf {\bibinfo
  {volume} {75}},\ \bibinfo {pages} {610} (\bibinfo {year} {2015})},\ \Eprint
  {https://arxiv.org/abs/1507.06197} {arXiv:1507.06197 [hep-ph]} \BibitemShut
  {NoStop}%
\bibitem [{\citenamefont {Contardo}\ \emph {et~al.}(2015)\citenamefont
  {Contardo}, \citenamefont {Klute}, \citenamefont {Mans}, \citenamefont
  {Silvestris},\ and\ \citenamefont {Butler}}]{CMSCollaboration:2015zni}%
  \BibitemOpen
  \bibfield  {author} {\bibinfo {author} {\bibfnamefont {D.}~\bibnamefont
  {Contardo}}, \bibinfo {author} {\bibfnamefont {M.}~\bibnamefont {Klute}},
  \bibinfo {author} {\bibfnamefont {J.}~\bibnamefont {Mans}}, \bibinfo {author}
  {\bibfnamefont {L.}~\bibnamefont {Silvestris}}, and\ \bibinfo {author}
  {\bibfnamefont {J.}~\bibnamefont {Butler}},\ }\href
  {https://cds.cern.ch/record/2020886?ln=en} {\emph {\bibinfo {title}
  {{Technical Proposal for the Phase-II Upgrade of the CMS Detector}}}},\
  \bibinfo {type} {Tech. Rep.}\ \bibinfo {number} {CMS-TDR-15-02}\ (\bibinfo
  {year} {2015})\BibitemShut {NoStop}%
\bibitem [{\citenamefont {Gershtein}\ \emph {et~al.}(2021)\citenamefont
  {Gershtein}, \citenamefont {Junius}, \citenamefont {Knapen}, \citenamefont
  {Mariotti},\ and\ \citenamefont {Redigolo}}]{futurework}%
  \BibitemOpen
  \bibfield  {author} {\bibinfo {author} {\bibfnamefont {Y.}~\bibnamefont
  {Gershtein}}, \bibinfo {author} {\bibfnamefont {S.}~\bibnamefont {Junius}},
  \bibinfo {author} {\bibfnamefont {S.}~\bibnamefont {Knapen}}, \bibinfo
  {author} {\bibfnamefont {A.}~\bibnamefont {Mariotti}}, and\ \bibinfo {author}
  {\bibfnamefont {D.}~\bibnamefont {Redigolo}},\ }\href@noop {} {\bibfield
  {journal} {\bibinfo  {journal} {in progress}\ } (\bibinfo {year}
  {2021})}\BibitemShut {NoStop}%
\bibitem [{\citenamefont {Izaguirre}\ \emph {et~al.}(2016)\citenamefont
  {Izaguirre}, \citenamefont {Krnjaic},\ and\ \citenamefont
  {Shuve}}]{Izaguirre:2015zva}%
  \BibitemOpen
  \bibfield  {author} {\bibinfo {author} {\bibfnamefont {E.}~\bibnamefont
  {Izaguirre}}, \bibinfo {author} {\bibfnamefont {G.}~\bibnamefont {Krnjaic}},
  and\ \bibinfo {author} {\bibfnamefont {B.}~\bibnamefont {Shuve}},\ }\bibfield
   {title} {\bibinfo {title} {{Discovering Inelastic Thermal-Relic Dark Matter
  at Colliders}},\ }\href {https://doi.org/10.1103/PhysRevD.93.063523}
  {\bibfield  {journal} {\bibinfo  {journal} {Phys. Rev.}\ }\textbf {\bibinfo
  {volume} {D93}},\ \bibinfo {pages} {063523} (\bibinfo {year} {2016})},\
  \Eprint {https://arxiv.org/abs/1508.03050} {arXiv:1508.03050 [hep-ph]}
  \BibitemShut {NoStop}%
\bibitem [{\citenamefont {{A. Berlin and F. Kling}}(2019)}]{Berlin:2018jbm}%
  \BibitemOpen
  \bibfield  {author} {\bibinfo {author} {\bibnamefont {{A. Berlin and F.
  Kling}}},\ }\bibfield  {title} {\bibinfo {title} {{Inelastic Dark Matter at
  the LHC Lifetime Frontier: ATLAS, CMS, LHCb, CODEX-b, FASER, and MATHUSLA}},\
  }\href {https://doi.org/10.1103/PhysRevD.99.015021} {\bibfield  {journal}
  {\bibinfo  {journal} {Phys. Rev.}\ }\textbf {\bibinfo {volume} {D99}},\
  \bibinfo {pages} {015021} (\bibinfo {year} {2019})},\ \Eprint
  {https://arxiv.org/abs/1810.01879} {arXiv:1810.01879 [hep-ph]} \BibitemShut
  {NoStop}%
\end{thebibliography}%

\end{document}